\documentclass[fleqn,usenatbib]{mnras}

\usepackage{adjustbox}
\usepackage{multirow}
\usepackage{threeparttable}

\usepackage{graphicx}	
\usepackage{amsmath}	
\usepackage{amssymb}	
\usepackage{multicol}        
\usepackage{bm}		
\usepackage{pdflscape}	

\usepackage[T1]{fontenc}
\usepackage{ae,aecompl}

\usepackage{newtxtext,newtxmath}

\title[Shocked gas in Sgr B2]{Structure and kinematics of shocked gas in Sgr B2: further evidence of a cloud-cloud collision from SiO emission maps}

\author[J. Armijos-Abenda\~no et al.]{J. Armijos-Abenda\~no,$^{1}$\thanks{E-mail: jairo.armijos@epn.edu.ec}\thanks{These two authors contributed equally to this paper.}
W. E. Banda-Barrag\'an,$^{2}$\footnotemark[2]
J. Mart\'in-Pintado,$^{3}$
H. D\'enes,$^{4}$
\newauthor
C. Federrath,$^{5}$
and M. A. Requena-Torres$^{6,7}$
\\
$^{1}$Observatorio Astron\'omico de Quito, Escuela Polit\'ecnica Nacional, Interior del Parque La Alameda, 170136, Quito, Ecuador\\
$^{2}$Hamburger Sternwarte, Universit\"{a}t Hamburg, Gojenbergsweg 112, D-21029 Hamburg, Germany\\
$^{3}$Centro de Astrobiolog\'ia (CSIC, INTA), Ctra a Ajalvir, km 4, 28850, Torrej\'on de Ardoz, Madrid, Spain\\
$^{4}$ASTRON - The Netherlands Institute for Radio Astronomy, NL-7991 PD Dwingeloo, The Netherlands\\
$^{5}$Research School of Astronomy and Astrophysics, Australian National University, Canberra, ACT 2611, Australia\\
$^{6}$Department of Astronomy, University of Maryland, College Park, MD 20742, USA\\
$^{7}$Department of Physics, Astronomy, and Geosciences, Towson University, Towson, MD 21252, USA\\
}

\date{Accepted XXX. Received YYY; in original form ZZZ}

\pubyear{2020}

\begin{document}
\label{firstpage}
\pagerange{\pageref{firstpage}--\pageref{lastpage}}
\maketitle

\begin{abstract}
We present SiO J=2-1 maps of the Sgr B2 molecular cloud, which show shocked gas with a turbulent substructure comprising at least three cavities at velocities of $[10,40]\,\rm km\,s^{-1}$ and an arc at velocities of {\bf $[-20,10]\,\rm km\,s^{-1}$}. The spatial anti-correlation of shocked gas at low and high velocities, and the presence of bridging features in position-velocity diagrams suggest that these structures formed in a cloud-cloud collision. Some of the known compact HII regions spatially overlap with sites of strong SiO emission at velocities of $[40,85]\,\rm km\,s^{-1}$, and are between or along the edges of SiO gas features at $[100,120]\,\rm km\,s^{-1}$, suggesting that the stars responsible for ionizing the compact HII regions formed in compressed gas due to this collision. We find gas densities and kinetic temperatures of the order of $n_{\rm H_2}\sim10^5\,\rm cm^{-3}$ and $\sim 30\,\rm K$, respectively, towards three positions of Sgr B2. The average values of the SiO relative abundances, integrated line intensities, and line widths are $\sim$10$^{-9}$, $\sim11\,\rm K\,km\,s^{-1}$, and $\sim31\,\rm km\,s^{-1}$, respectively. These values agree with those obtained with chemical models that mimic grain sputtering by C-type shocks. A comparison of our observations with hydrodynamical simulations shows that a cloud-cloud collision that took place $\lesssim 0.5\,\rm Myr$ ago can explain the density distribution with a mean column density of $\bar{N}_{\rm H_2}\gtrsim 5\times10^{22}\,\rm cm^{-2}$, and the morphology and kinematics of shocked gas in different velocity channels. Colliding clouds are efficient at producing internal shocks with velocities $\sim 5-50\,\rm km\,s^{-1}$. High-velocity shocks are produced during the early stages of the collision and can readily ignite star formation, while moderate- and low-velocity shocks are important over longer timescales and can explain the widespread SiO emission in Sgr B2.
\end{abstract}




\section{Introduction}
The Sgr B2 cloud (hereafter Sgr B2) with a mass of $\sim$10$^6$ M$_{\odot}$ is one of the most massive clouds in the Galactic Center (GC) at a distance of 7.9 kpc \citep{Boehle2016}. Sgr B2 is located at a projected distance of around 100 pc from the central supermassive black hole Sgr A*. Sgr B2 harbors three main star-forming cores known as Sgr B2N, Sgr B2M and Sgr B2S, enclosing compact and ultra-compact HII regions \citep{Gaume1990}. A burst of star formation is taking place in the extended envelope of Sgr B2 as shown by a large population of high-mass protostellar cores found along the cloud and not just in the three cores \citep{Ginsburg2018}.\par

The Sgr B2N and Sgr B2M cores have H$_2$ densities of 2$\times$10$^7$ cm$^{-3}$ and 4$\times$10$^6$ cm$^{-3}$, respectively \citep{Goldsmith1990}. These cores are embedded in two envelopes, one of moderate density ($\sim$10$^5$ cm$^{-3}$) with a size of 2.5 pc$\times$5.0 pc, and a kinetic temperature of $\sim$100 K, and another one with a low density $<$5$\times$10$^3$ cm$^{-3}$, a diameter of 38 pc \citep{Schmiedeke2016}, and a kinetic temperature $>$200 K \citep{Huttemeister1995}.
A warm CO gas component ($\sim$50--100 K) associated with the extended envelope of Sgr B2 is likely heated by UV photons, while shocks and UV photons may be responsible for the heating of the hot CO gas ($>$200 K) in the Sgr B2N and Sgr B2M cores \citep{Etzaluze2013}.\par

Rings, arcs, and filaments have been discovered towards the Sgr B2 envelope. The rings with sizes of around 1-3 pc show kinetic temperatures of 40-70 K and their radial velocity gradients agree with those of three-dimensional shells, whose formation may be explained by a wind bubble driven by massive stars \citep{Martin1999}. \cite{Hasegawa1994} studied the $^{13}$CO gas kinematics of Sgr B2, finding kinematical features that are consistent with a scenario of cloud-cloud collision. \cite{Sato00} found a correlation between a cavity observed at gas velocities \mbox{of [40,50] km s$^{-1}$} and a clump associated with Sgr B2N and Sgr B2M at [70,80] km s$^{-1}$, supporting the cloud collision scenario, in which the clump would have collided with another cloud with a velocity of $\sim$30 km s$^{-1}$. They also found H$_2$CO and CH$_3$OH masers with velocities lower than 65 km s$^{-1}$ located along the eastern edge of the cavity, which would be the interface between the two colliding clouds. On the other hand, OH, H$_2$CO, CH$_3$OH and SiO masers with velocities higher than 65 km s$^{-1}$ and associated compact HII regions are observed towards the clump \citep{Sato00}. In addition, a group of six shell-like structures have been found in a southeastern extension of Sgr B2 \citep{Tsuboi15}. Position-velocity maps of Sgr B2 show that these structures could be expanding shells \citep{Tsuboi15}. Integrated emission maps and position-velocity diagrams of HNCO, SiO, CH$_3$OH, and HNCO towards the G$+$0.693 region in Sgr B2 have shown observational characteristics of a cloud-cloud collision \citep{Zeng_2020}.\par

A current model is that Sgr B2 is located at the projected extrema of a 100 pc ring, along which gas and dust rotate around the nucleus of the Galaxy at a constant orbital speed of $\sim 80\,\rm km\,s^{-1}$ \citep{Molinari2011}. This ring could be associated with a system of orbits called x$_2$, which would be enclosed within another system of orbits called x$_1$, aligned along the Galactic bar (\citealt{1991MNRAS.252..210B}). Clouds moving on x$_1$ and x$_2$ orbits may be colliding at the location of Sgr B2, thus triggering star formation \citep{Molinari2011}. An alternative model to the one of the 100 pc ring was proposed by \cite{2015MNRAS.447.1059K}, who considered four gas streams of molecular gas orbiting the GC. In this scenario the gas follows an open orbit and has a varying orbital velocity in the range of $\sim 100-200\,\rm km\,s^{-1}$, which can explain the observed kinematics of cloud complexes in the region (including Sgr B2). In addition, using hydrodynamical simulations, \cite{Sormani18} showed that corrugation and thermal instabilities, as well as bombardment of the Central Molecular Zone (CMZ) from the Galactic bar shocks, can explain the asymmetry observed in the molecular gas distribution in the CMZ and the potentially episodic star formation (e.g., see \citealt{2015MNRAS.453..739K}).\par

Cloud-cloud collisions are not uncommon in the interstellar medium, and they are believed to trigger star formation. Indeed, it is thought that around 10\% of the star formation occurring in the Galaxy is driven by cloud-cloud collision processes of mainly massive giant molecular clouds with masses of $\gtrsim$10$^{5.5}$ M$\odot$ \citep{Kobayashi18}. \cite{Fukui16} have suggested that multiple O stars have been formed towards the super star cluster RCW 38 due to the collision of two clouds. \cite{Torii17} showed that the collision of two molecular clouds $\sim$0.3 Myr ago is likely what triggered the formation of the O star ionizing the Trifid Nebula M20. They also found that the spatially complementary distribution shown by the CO J=1-0 and 3-2 emission of both clouds is reproduced by cloud-cloud collision simulations. Studies of the emission distribution of molecular gas towards other sources in the Galaxy also support the idea that cloud-cloud collisions may trigger star formation (e.g., see \citealt{Hayashi18,Ohama18,Enokiya19}).\par

In this paper, we present large-scale maps of SiO, which traces shocks (e.g., see \citealt{MartinP1992,2016A&A...595A.122L}), to study the large-scale kinematics of Sgr B2 and to investigate whether the scenario of a cloud-cloud collision could have taken place in this outstanding site of star formation in the Galaxy. We derive physical properties for the entire Sgr B2 cloud and also for selected positions. The kinematical features shown by the SiO emission towards Sgr B2 and their derived properties are compared with those obtained by hydrodynamical simulations of cloud-cloud collisions.\par

\section{Observation and data reduction}\label{observation}
The data were observed with the IRAM 30-m telescope in August 2006. We used the On-The Fly (OTF) observing mode to map an area of around 15$\arcmin\times$15$\arcmin$ (36$\times$36 pc$^2$ at the GC distance) centered on the Sgr B2M hot core with $\rmn{RA}(J2000)=17^{\rmn{h}} 47^{\rmn{m}} 20\fs4$ and $\rmn{Dec.}(J2000)~-28\degr 23\arcmin 07\farcs 2$. The data were calibrated by using hot and cold loads. We used the A100 receiver for the vertical polarization and the B100 receiver for the horizontal polarization. Both receivers covered the frequencies of the SiO J=2-1, C$^{18}$O J=1-0 and $^{13}$CO J=1-0 transitions listed in Table \ref{Obs_mol}.
The 1 MHz filter bank was used as spectrometer during our observations, providing a bandwidth of 512 MHz.

\begin{table}
    \centering
    \caption{Observed molecular lines}\label{Obs_mol}
    \begin{threeparttable}
    \begin{tabular}{cccc}
    \hline
    Line & $\nu$ & E$_{\rm up}$\tnote{(a)} & HPBW\tnote{(b)}\\
         & (GHz) & (K)    & (\arcsec)\\
   \hline
    SiO J=2-1 & 86.8 & 6.25 & 28.3\\
    C$^{18}$O J=1-0 & 109.8 & 5.27 & 22.4 \\
    $^{13}$CO J=1-0 & 110.2 & 5.29 & 22.3 \\
    OCS J=7-6 & 85.1 & 16.34 & 28.9\\
    OCS J=21-20 & 255.4 & 134.83 & 9.6 \\
   \hline
    \end{tabular}
    \begin{tablenotes}
	\item[(a)] Upper state energy of the transition.
	\item[(b)] Half-power beam width (HPBW) of the IRAM 30-m telescope.
	\end{tablenotes}
    \end{threeparttable}
\end{table}

In our study, we also used data observed with the IRAM 30-m telescope in December 2014. The EMIR receivers operating in the E090 and E230 frequency bands were connected to the Fast Fourier Transformed Spectrometer (FTS) to observe the J=7-6 and 21-20 transitions of OCS towards three selected positions of Sgr B2, which we use to study the H$_2$ density in Section \ref{SiOabundances}. Spectra were calibrated by using ambient temperature loads, which provided a calibration accuracy around 10\%. Frequencies and the upper state energy of the observed molecular transitions are given in Table \ref{Obs_mol}.\par

\begin{figure*}
    \includegraphics[width=17.8cm]{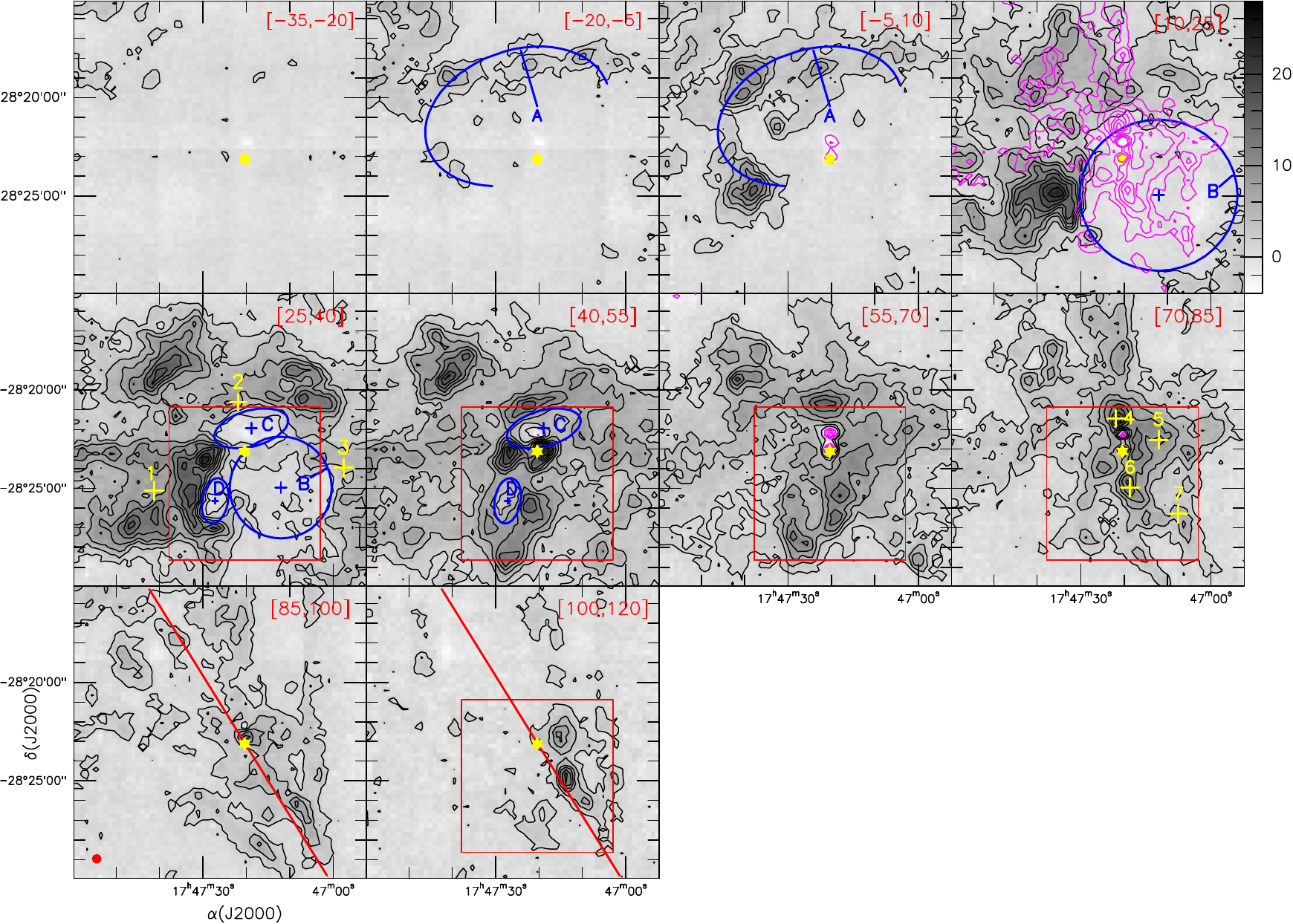}\\
    \includegraphics[width=17.7cm]{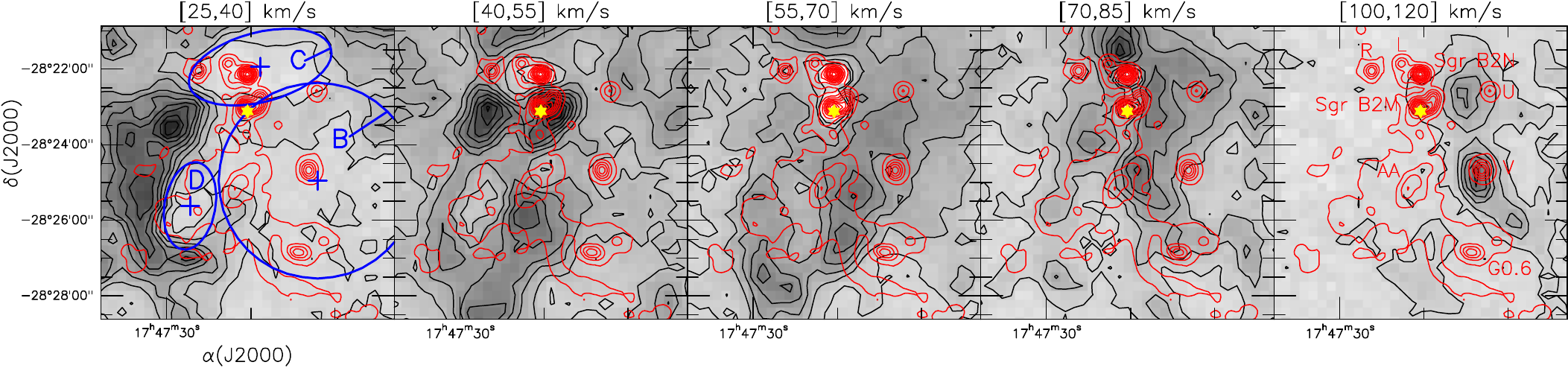}
    \caption{\textbf{First row to third row:} Maps of the SiO J=2-1 line integrated over ten velocity ranges in km s$^{-1}$ indicated at the top-right side of each panel. Contour levels start at 3$\sigma$ (black contours) for the emission and -3$\sigma$ (magenta contours) for the absorption (observed only towards the Sgr B2M and Sgr B2N hot cores at velocities of [-5,10] and [55,70] km s$^{-1}$). Contours increase in 5$\sigma$ and -5$\sigma$ steps for the emission and absorption, respectively. $\sigma$ $\approx$0.5 K km s$^{-1}$ for all maps. The wedge at the right side of the panel of [10,25] km s$^{-1}$ shows the SiO J=2-1 integrated intensity scale given in K km s$^{-1}$ for all maps. The Sgr B2M position is shown with a yellow star. An arc labeled as A and cavities labeled as B, C and D are outlined by a semi-ellipse and ellipses, respectively, on some of the panels. The blue crosses are the centers of the ellipses.
    Yellow crosses and their numbers in the panels of [25,40] and [70,85] km s$^{-1}$ show positions selected for studying SiO relative abundances (see Section \ref{SiOabundances}). The magenta contours in the panel of [10,25] km s$^{-1}$ represent the SiO J=2-1 line emission at 6$\sigma$ integrated over the velocity range of [70,85] km s$^{-1}$.
    The red line in the two panels with velocities higher than 85 km s$^{-1}$ represent the Galactic plane. The 28 arcsec beam size at 87 GHz of the IRAM 30-m telescope is shown in the left-bottom corner of the panel of [85,100] km s$^{-1}$. The red squares on the panels with velocities higher than 25 km s$^{-1}$ show regions enlarged on the fourth row of this figure.
    \textbf{Fourth row:} Enlarged regions shown with red squares on the maps of the second and third rows. The red contours show the 20 cm radio continuum emission obtained by \protect\cite{Yusef2004}. The main compact HII regions of Sgr B2 \protect\citep{Mehringer93} are labeled on the map of [100,120] km s$^{-1}$. The position of the compact HII regions spatially coincides with regions of high SiO J=2-1 emission in the velocity ranges of [40,85] km s$^{-1}$ (except at [55,70] km s$^{-1}$, where the HII regions of Sgr B2M and Sgr B2N spatially coincide with SiO absorption), while there is a spatial anticorrelation with the SiO J=2-1 emission in the other channels.}
    \label{fig:SiO_maps}
\end{figure*}

The baseline correction of the 2006 data was done using the GILDAS software\footnote{http://www.iram.fr/IRAMFR/GILDAS}, which was also used to build the SiO J=2-1 and $^{13}$CO J=1-0 data cubes with half-power beamwidths (HPBW) of 28 and 23 arcsec, respectively. Both data cubes have a velocity resolution of $\sim$5 km s$^{-1}$ appropriate to resolve the typical linewidths of 20 km s$^{-1}$ observed in the GC. The data observed in 2014 were imported in the MADCUBA software \citep{Martin19} to apply the baseline subtraction and spectrum averaging. Then, the spectra were smoothed to a velocity resolution of $\sim$5 km s$^{-1}$ as in the case of the OTF data. The line intensity of the spectra is given in the antenna temperature scale (T$\rm _a^*$) as the emission is extended and fills the telescope beam of the IRAM 30-m telescope (see Figure~\ref{fig:SiO_maps}).\par

\section{Integrated intensity maps of SiO J=2-1}\label{Maps_sio}
\begin{figure*}
	\includegraphics[width=14.3cm]{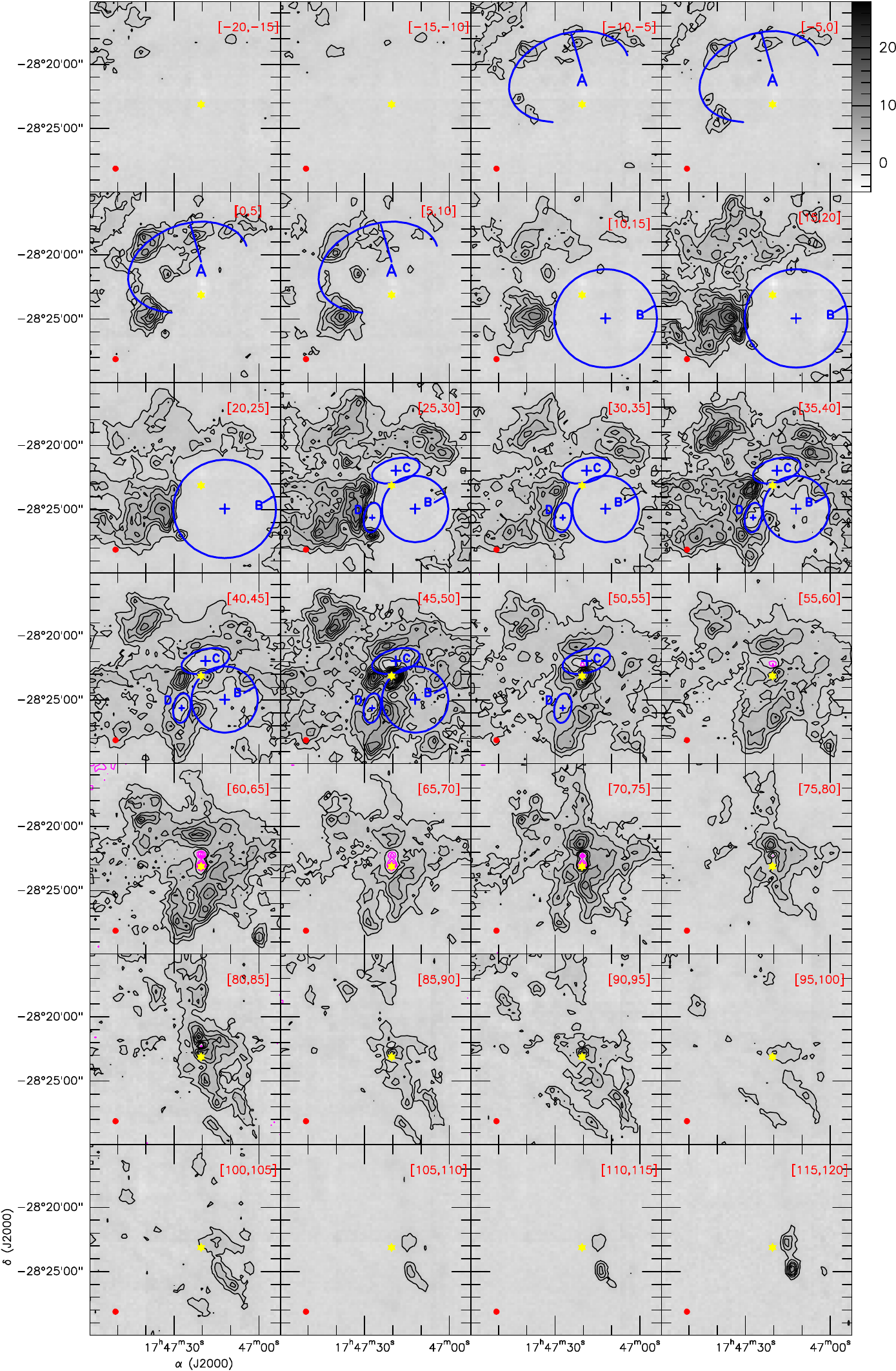}
    \caption{Channel maps of the SiO J=2-1 line towards Sgr B2. Contours in black and magenta start at 3$\sigma$ (for the emission) and -3$\sigma$ (for the absorption), respectively. $\sigma$ $\approx$0.5 K km s$^{-1}$ for all maps. Contours increase in steps of 1.5 K km s$^{-1}$ and -1.5 K km s$^{-1}$ for the emission and absorption, respectively. Channel maps range from -20 km s$^{-1}$ to 120 km s$^{-1}$ in steps of 5 km s$^{-1}$. The velocity range in km s$^{-1}$ is indicated on the right-top side of each map. The wedge at the right side of the [-5,0] km s$^{-1}$ map shows the intensity scale in K km s$^{-1}$. Arc A and cavities B, C, and D are outlined by a semi-ellipse and ellipses, respectively, on the maps with velocities within [-10,55] km s$^{-1}$. The blue crosses represent the centers of the ellipses. The yellow star represents the position of Sgr B2M. The beam size of 28 arcsec of the IRAM 30-m telescope at 86 GHz is indicated by a filled circle in the left-bottom side of each map.}
    \label{fig:channel_maps}
\end{figure*}

To study the SiO kinematics of Sgr B2, Figure~\ref{fig:SiO_maps} shows ten different maps of the SiO J=2-1 line integrated in different velocity ranges. As seen in this figure, the SiO J=2-1 emission from Sgr B2 is extended and shows velocities from -20 to 120 km s$^{-1}$. SiO J=2-1 absorption is seen in the velocity ranges of [-5,10] and [55,70] km s$^{-1}$ towards the Sgr B2M and Sgr B2N hot cores. The SiO J=2-1 emission in Figure~\ref{fig:SiO_maps} shows shocked gas with a turbulent substructure featuring an arc labeled as A, as well as at least three cavities labeled as B, C, and D. 
The SiO J=2-1 emission with velocities of [40,55] km s$^{-1}$ fills almost the entire mapped region of Sgr B2. The four features were identified by eye at different gas velocities in our SiO J=2-1 maps. Arc A is outlined by a semi-ellipse while cavities B, C, and D by ellipses drawn on the first and second rows of Figure~\ref{fig:SiO_maps}. The centers, semi-major and semi-minor axes, and position angles of these four features are given in Table \ref{Parameters_features}. Arc A has velocities in the range of [-20,10] km s$^{-1}$. Cavity B appears in the SiO J=2-1 maps at velocities of [10,25] km s$^{-1}$, and disappears at gas velocities of [40,55] km s$^{-1}$. The size of this cavity decreases by a factor of $\sim$1.5 at velocities of [25,40] km s$^{-1}$ compared to its size at velocities of [10,25] km s$^{-1}$. Cavity B was previously identified by \cite{Sato00}. Cavities C and D appear in our maps at velocities of [25,40] km s$^{-1}$, and disappear at velocities higher than $\sim$40 km s$^{-1}$. Both of these features are smaller than cavity B. A high degree of turbulence is typically found in GC molecular clouds (e.g., see \citealt{Bally87,Shetty12}), so such structures can be produced by turbulent stirring (e.g., see \citealt{2009ApJ...692..364F}). Dynamical contribution from the local compact HII regions can also be expected on small scales. At velocities higher than 85 km s$^{-1}$, much of the SiO gas seems to be concentrated along the Galactic plane, which was already pointed out by \cite{Sato00}. Figure~\ref{fig:channel_maps} displays channel maps of the SiO J=2-1 line towards Sgr B2, where arc A and cavities B, C, and D are outlined as in Figure~\ref{fig:SiO_maps}.

\begin{table}
    \centering
    \caption{Parameters of arc A and cavities B, C and D}\label{Parameters_features}
    \begin{threeparttable}
    \begin{adjustbox}{max width=9cm}
    \begin{tabular}{cccccc}
    \hline
    Arc or & \multicolumn{2}{c}{Center} & semimajor axis & semiminor axis & PA\tnote{a}\\
    cavity  & $\alpha$(J2000) & $\delta$(J2000) & ($\arcsec$)&($\arcsec$) & ($\deg$)\\
    \hline
    A\tnote{b} & $17^{\rmn{h}} 47^{\rmn{m}} 24\fs900$ & $-28\degr 20\arcmin 56\farcs 42$ & 290 & 250 & 110\\
    B & $17^{\rmn{h}} 47^{\rmn{m}} 12\fs063$ & $-28\degr 24\arcmin 57\farcs 12$ & 240 & 230 & 0 \\
    C & $17^{\rmn{h}} 47^{\rmn{m}} 18\fs821$ & $-28\degr 21\arcmin 56\farcs 59$ & 115 & 55  & 115 \\
    D & $17^{\rmn{h}} 47^{\rmn{m}} 27\fs268$ & $-28\degr 25\arcmin 37\farcs 23$ & 70 & 40 & 170\\
   \hline
    \end{tabular}
    \end{adjustbox}
    \begin{tablenotes}
	\item[(a)] Position angle of the major axis, which is measured from north through east.
	\item[(b)] Arc A is defined as an arc of an ellipse with its parameters given in this row.
	\end{tablenotes}
    \end{threeparttable}
\end{table}

\begin{figure}
	\includegraphics[width=8cm]{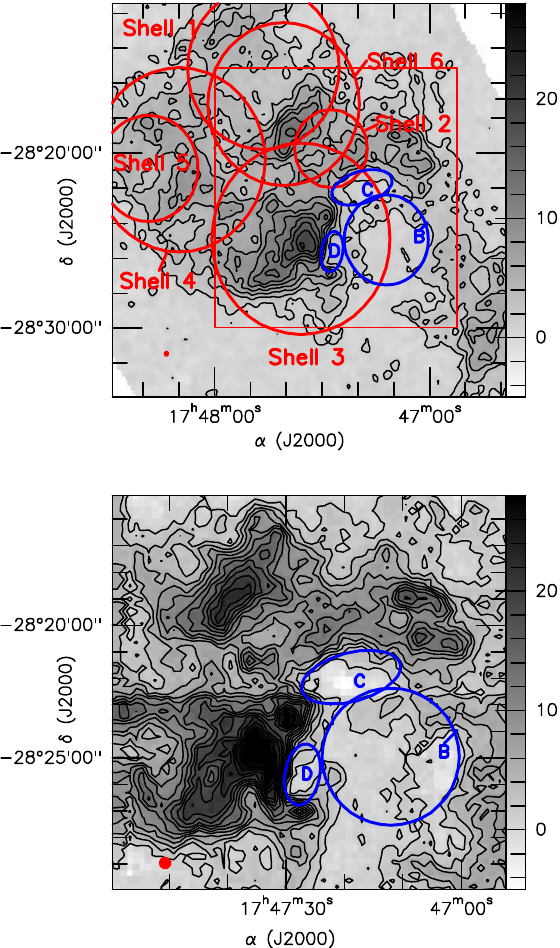}
    \caption{{\bf Top panel}: Sgr B2 map of the SiO J=2-1 emission integrated over the range of 10-40 km s$^{-1}$, which was observed with the 45-m Nobeyama telescope by \protect\cite{Tsuboi15}. Contour levels start at 1.5 K km s$^{-1}$ and increase in 2.0 K km s$^{-1}$. The red ellipses outline six shells identified by \protect\cite{Tsuboi15} (the position and diameter of the shells are shown in their Table 4). Cavities B, C, and D identified in our Figure \ref{fig:SiO_maps} are shown with blue ellipses. The 20 arcsec beam size of the Nobeyama telescope at 86 GHz is indicated by a filled circle on left-bottom side of the map. The red square shows the region observed with the IRAM 30-m telescope displayed on the bottom panel. The right wedge shows the intensity scale in K km s$^{-1}$. {\bf Bottom panel}: the same as on the top panel but the SiO J=2-1 emission map was observed with the IRAM 30-m telescope, the shells studied by \protect\cite{Tsuboi15} are not shown, the filled circle indicates the 28 arcsec beam of the IRAM 30-m telescope at 86 GHz. The mapped region is smaller than that shown on the top panel.}
    \label{fig:Nob_30mteles}
\end{figure}

For comparison, the top panel of Figure~\ref{fig:Nob_30mteles} shows the six shells identified by \cite{Tsuboi15} on their map of the SiO J=2-1 emission observed with the 45-m Nobeyama telescope, whereas the bottom panel of this figure shows our SiO J=2-1 emission map observed with the IRAM 30-m telescope.
In both panels of Figure~\ref{fig:Nob_30mteles} cavities B, C, and D are outlined as in the map of [25,40] km s$^{-1}$ in Figure~\ref{fig:SiO_maps}. As seen in Figure~\ref{fig:Nob_30mteles}, both maps reveal a quite similar emission distribution of the SiO J=2-1 line, but our observed region towards Sgr B2 is smaller than in the map obtained with the Nobeyama telescope. Cavities B, C, and D are located along the western edge of Shell 3 studied in \cite{Tsuboi15} (see the top panel of Figure~\ref{fig:Nob_30mteles}). These three features together with arc A are identified and studied for the first time.\par

Figure~\ref{fig:SiO_maps} also displays the SiO J=2-1 emission in magenta contours (above the 6$\sigma$ level and integrated over the velocity range of [70,85] km s$^{-1}$) overlapped on the SiO J=2-1 emission with velocities within 10-25 km s$^{-1}$. As seen in the overlapped images, the northern and eastern edges of the high-velocity gas lie along cavity B in the low-velocity map, which was already noted in previous studies \citep{Hasegawa1994,Sato00}. This spatial anti-correlation of gas at low and high velocities supports the scenario in which a high-velocity cloud collided with another cloud, in agreement with the work by \cite{Sato00}. This suggests that much of the gas with velocities of 10-25 km s$^{-1}$ is being impacted by the gas with velocities of at least 70-85 km s$^{-1}$. In this scenario, the formation of large-, intermediate-, and small-scale features would be mainly regulated by turbulent stirring and mixing processes resulting from this large-scale cloud-cloud collision, with additional contributions from stellar feedback on small scales.\par

The 20 cm radio continuum emission map obtained by \cite{Yusef2004} is overlapped in red contours on the SiO J=2-1 maps of [25,40], [55,70], [70,85] and [100,120] km s$^{-1}$ in the fourth row of Figure \ref{fig:SiO_maps}. The main compact HII regions of Sgr B2 (Sgr B2M, Sgr B2N, L, R, U, V, AA, and G0.6) studied in detail in \cite{Mehringer93} are labeled on the right panel of the fourth row in Figure \ref{fig:SiO_maps} corresponding to SiO J=2-1 emission at [100,120] km s$^{-1}$. These compact HII regions spatially coincide with strong SiO emission at velocities of [40,85] km s$^{-1}$ (except at [55,70] km s$^{-1}$, where the HII regions of Sgr B2M and Sgr B2N spatially coincide with SiO absorption). This supports the scenario in which star formation could have been triggered in compressed gas due to the collision of clouds with different velocities in Sgr B2 \citep{Hasegawa1994,Sato00,Molinari2011}. We also find that the compact HII regions lie inside or along the cavities B and C as seen in the panel showing the SiO J=2-1 emission at velocities of [25,40] km s$^{-1}$, as well as between or along the edges of the SiO gas features at velocities of [100,120] km s$^{-1}$ (see Figure~\ref{fig:SiO_maps}). This is in agreement with the findings by \cite{Sato00}, who found H$_2$CO and CH$_3$OH masers along the eastern edge of cavity B.


\section{SiO linewidths and velocity components}
We study the velocity distribution and characteristic line widths of the SiO emission by decomposing all of the spatial pixels of the SiO data cube with the automated Gaussian decomposer {\sc GaussPy+} \citep{Riener_2019, Lindner_2015}. {\sc GaussPy+} provides an unbiased way to deconstruct spectra by using computer vision and machine learning algorithms to provide a multi-component Gaussian model. The software first calculates the derivatives of the spectra to provide initial guesses for the parameters of the Gaussian components and then performs an iterative $\chi^{2}$ fit to match the model to the data. After this initial fitting, the software also applies a two-phase spatial coherence fitting, where it optimises the fit based on the Gaussian components of the neighbouring pixels in the data cube. We used the default parameters for the decomposition provided by \cite{Riener_2019}, except for $\alpha_{1}$ and $\alpha_{2}$ for which we used 1.55 and 2.74, respectively. These parameters effectively control the amount of smoothing that is applied during the decomposition and mostly depend on the signal-to-noise and the complexity of the spectra. To determine these parameters, we ran the training module of {\sc GaussPy+} on 100 randomly-selected spectra from the SiO data cube. To investigate the effect of the choice of the $\alpha$ parameters, we also run the decomposition with the default $\alpha$ values, which gave the same result within the uncertainties of the decomposition. We show example spectra from the SiO data cube and results of the decomposition with {\sc GaussPy+} in Figure \ref{fig:SiO_decomposition_gausspy}.\par

\begin{figure}
\begin{center}
  \begin{tabular}{c}
    \resizebox{60mm}{!}{\hspace{-1.2cm}\includegraphics{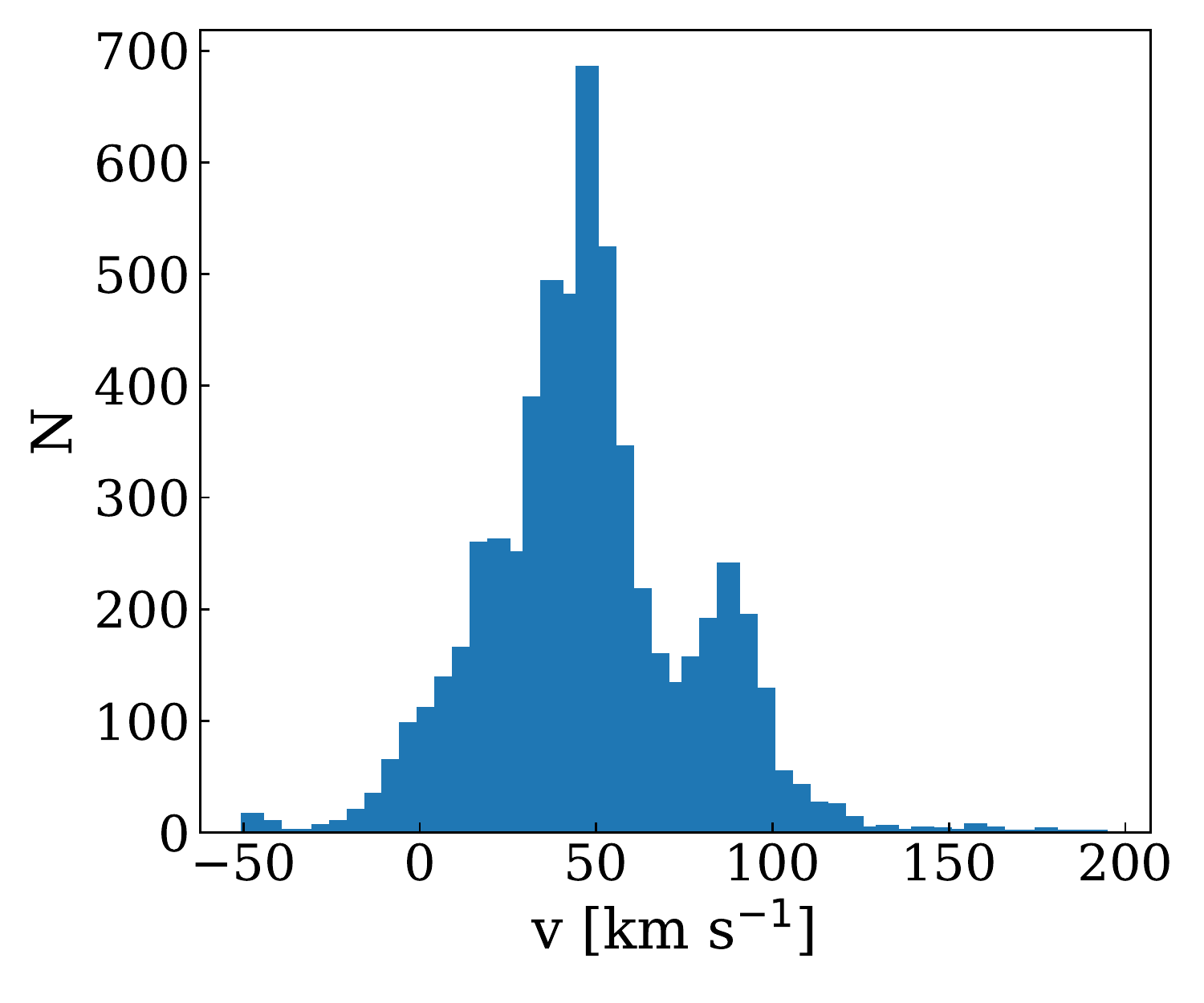}}\\
    \resizebox{60mm}{!}{\hspace{-1.2cm}\includegraphics{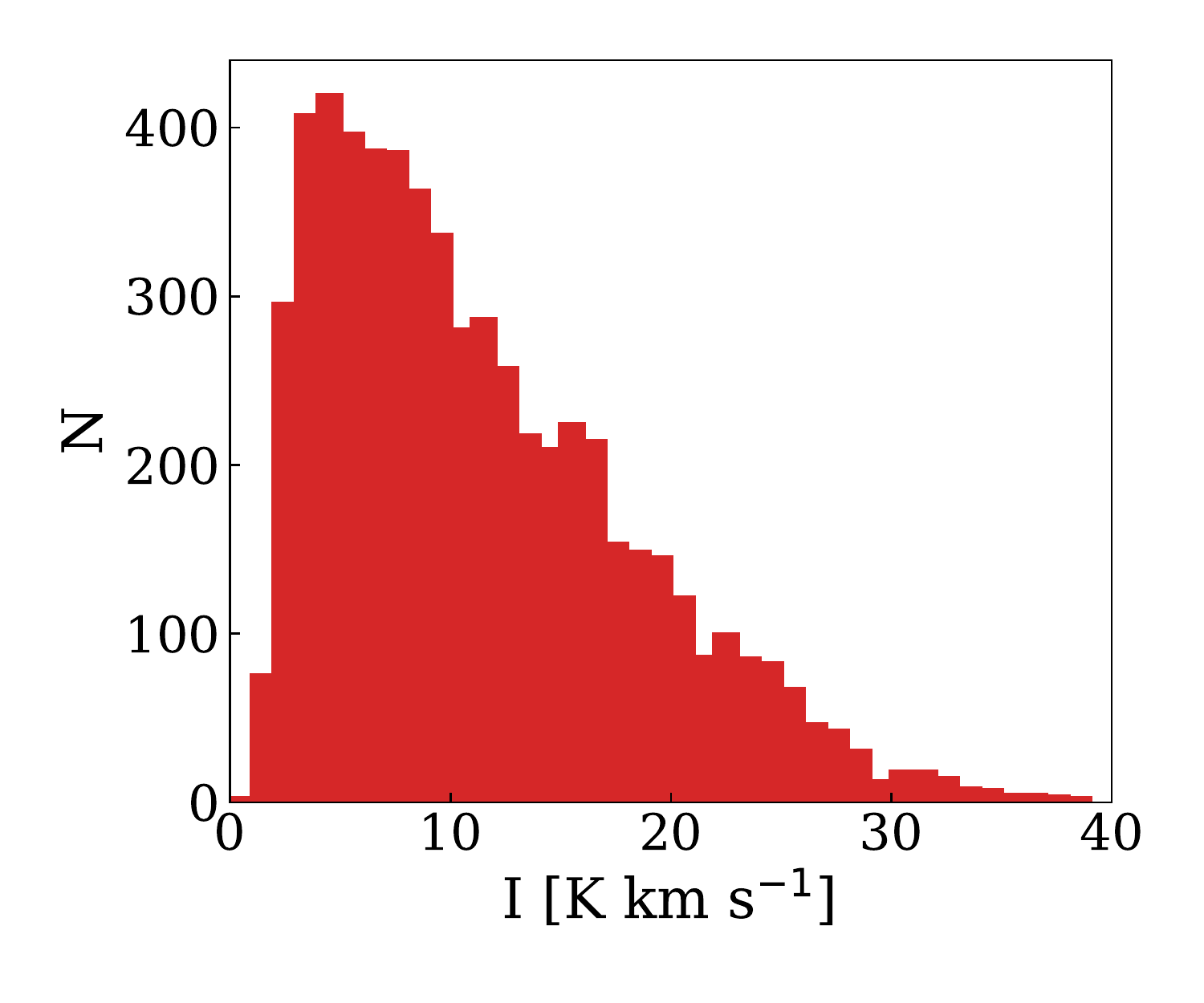}}\\
    \resizebox{60mm}{!}{\hspace{-1.2cm}\includegraphics{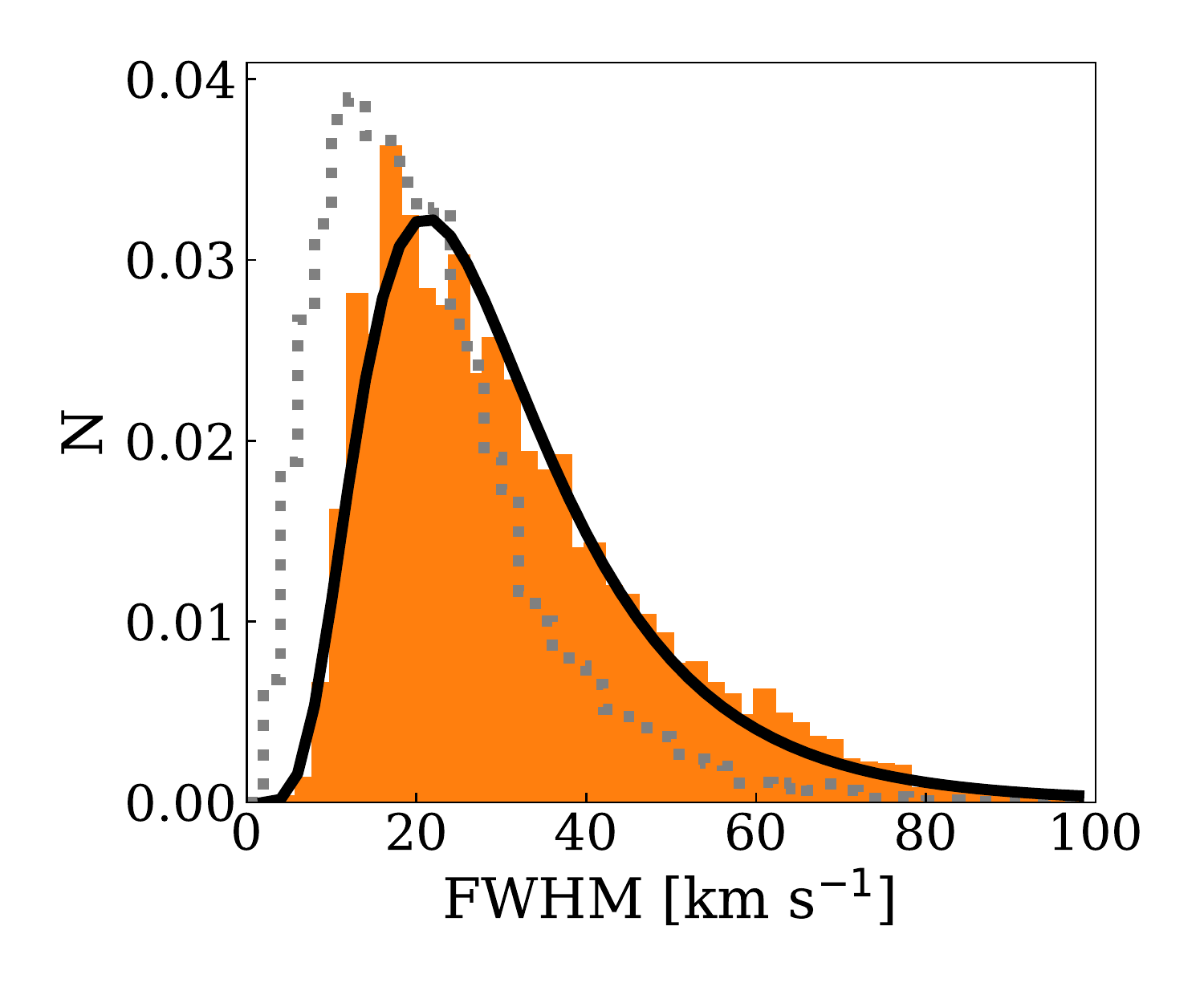}}\\
  \end{tabular}
  \caption{Top panel: velocity distribution of Gaussian components for all SiO image pixels decomposed by {\sc GaussPy+}. Middle panel: Integrated line intensity distribution of SiO Gaussian components, with a mean of 11 $\pm$ 3 K km s$^{-1}$. Bottom panel: normalised line width (FWHM) distribution of Gaussian components for all SiO image pixels. The black line shows a fitted log-normal distribution to the histogram. The mean of the line widths is $31 \pm 5 \rm km\,s^{-1}$. The dashed grey line shows the FWHM histogram of the $^{13}$CO data for comparison. The mean of the $^{13}$CO distribution is 21 $\pm$ 5 km s$^{-1}$.} 
  \label{Histograms}
\end{center}
\end{figure}

The decomposition resulted in 5946 Gaussian components. The top and middle panels of Figure \ref{Histograms} show the SiO velocity distribution, and the total intensity distribution, respectively. The velocity distribution has an average velocity of 47 $\pm$2 km s$^{-1}$ and shows 3 peaks at $v \sim$ 25, 45, and 80 km s$^{-1}$, highlighting the most dominant SiO features in this region. The integrated line intensity peaks at $\sim$ 5 K km s$^{-1}$ and has a mean of 11 $\pm$ 3 K km s$^{-1}$. In the bottom panel of Figure \ref{Histograms} we show the normalised FWHM distribution of the individual Gaussian components of SiO in orange (and, for comparison, of $^{13}$CO in grey). The FWHM histogram of SiO is log-normal with a peak at 21 km s$^{-1}$ and an average of 31 $\pm$ 5 km s$^{-1}$, which are in agreement with the typical values found in the GC (e.g. \citealt{1996ARA&A..34..645M}). To compare the FWHM distribution with that of the ambient medium in Sgr B2, we also decomposed the $^{13}$CO data with {\sc GaussPy+} (for details see Appendix \ref{Sec:AppendixDecomposition}). The mean of the $^{13}$CO FWHM distribution is 21 $\pm$ 2 km s$^{-1}$, which is slightly lower than the mean of the SiO data.\par

The distribution of SiO line widths indicates that `broad' line widths are less common than `narrow' line widths\footnote{Note that if we compare the FWHM of SiO in Sgr B2 with those of regions outside the GC, all these line widths are broader. However, compared to the local ambient gas (e.g. the $^{13}$CO and C$^{18}$O lines in Figures \ref{Histograms} and \ref{fig:SiO_C18O_spectra}), they are not much broader owing to the extreme environment in the GC. Thus, we refer to them as `narrow' if they have similar or lower FWHM than those of the local ambient gas.}. SiO line widths are commonly associated with shock speeds under the assumption that individual 1D shocks are observed face on along individual LOS. However, the actual shock velocities can be higher if projection effects are considered, or lower if the ambient gas is turbulent or if the line is composed of several unresolved shock components (see discussions in \citealt{Gusdorf08a,Gusdorf08b,2009ApJ...695..149J}). In this paper we assume that $v_{\rm shock}\lesssim$ FWHM for our analysis because shocks are inherently associated with supersonic turbulence. Under this assumption, the fact that the bulk of the line widths of SiO emission is between $5-50\,\rm km\,s^{-1}$ indicates that the shocks in the underlying turbulent gas are predominantly moderate- and low-velocity shocks\footnote{Hereafter, we refer to very-low-velocity, low-velocity, moderate-velocity, and high-velocity shocks to those with speeds $v_{\rm shock}<5\,\rm km\,s^{-1}$, $5\,\rm km\,s^{-1}\leq v_{\rm shock} < 20\,\rm km\,s^{-1}$, $20\,\rm km\,s^{-1}\leq v_{\rm shock} < 50\,\rm km\,s^{-1}$, and $v_{\rm shock}\geq 50\,\rm km\,s^{-1}$, respectively.}, and that they are more spatially extended than high-velocity shocks (see also Figure \ref{fig:SiO_velocity_width_decomposition}). This is to some degree similar to what has been reported in recent studies in other regions by \cite{2010MNRAS.406..187J,2013ApJ...773..123S}; and \cite{2016A&A...595A.122L}, who show that the more extended, narrower-line SiO emission can be explained by low-velocity shocks from e.g. colliding flows, while the area covered by broader-line SiO emission is typically much smaller and localised e.g. near proto-stellar outflows, which are known to produce higher-velocity shocks. However, in Sgr B2 we do not see a spatial correlation between regions with broader lines and those with ongoing star formation (see Figure \ref{fig:SiO_velocity_width_decomposition}). Interestingly, we do see a spatial overlap between regions with `broad' SiO emission and gas at $[30,60]\,\rm km\,s^{-1}$ (see Figure \ref{fig:fwhm_vlsr_gausspy}). Thus, in Sgr B2 the SiO line widths are more likely tracing shocks produced in mixed, turbulent gas during the collision, rather than stellar feedback. However, the resolution of our current observations do not allow us to study SiO emission associated with stellar feedback on small scales in detail. Thus, disentangling different SiO velocity components around star-forming regions in Sgr B2 with higher spatial and spectral resolution observations would be interesting for future studies.

\section{SiO abundance}\label{SiOabundances}
\begin{figure}
	\includegraphics[width=7cm]{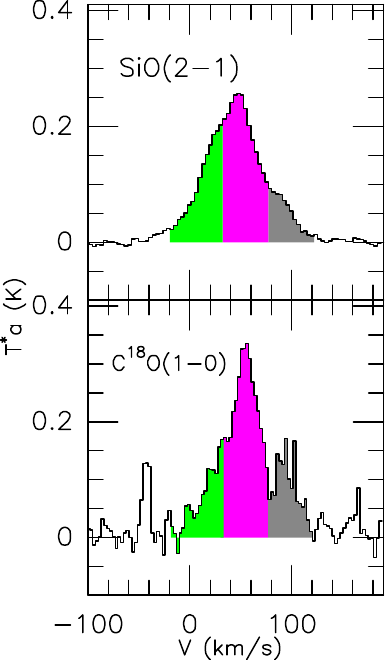}
    \caption{Spectrum of SiO J=2-1 (top panel) and C$^{18}$O J=1-0 (bottom panel), which were averaged over the entire mapped region of Sgr B2 shown in Figure \ref{fig:SiO_maps}. Areas delimited within the velocity ranges of [-20,+30], [+30,+75], and [+75,+120] km s$^{-1}$ are highlighted with different colors. These spectra were used to derive abundances of SiO in Section \ref{SiOabundances}.}
    \label{fig:SiOC18O}
\end{figure}

We study the average fractional abundance of SiO towards the mapped region of Sgr B2. The average spectrum of SiO J=2-1 is shown in the top panel of Figure~\ref{fig:SiOC18O}, whereas the bottom panel of this figure presents an average spectrum of C$^{18}$O J=1-0. A brief study of the fractional abundance of SiO towards seven selected positions of Sgr B2 is presented in Appendix \ref{AppendixA1}. These seven positions are indicated in Figure~\ref{fig:SiO_maps} and Table \ref{tab:Column_density}. As seen in Figure \ref{fig:SiO_maps}, SiO absorption is only detected towards the Sgr B2M and Sgr B2N cores, representing small regions over the whole mapped region. Therefore, the contribution of the absorption over the average spectrum of SiO J=2-1 is negligible.

We use the C$^{18}$O molecule to derive hydrogen column densities required to infer the fractional abundances of SiO, which is derived for three velocity ranges indicated in the SiO J=2-1 and C$^{18}$O J=1-0 lines in Figure~\ref{fig:SiOC18O}. These ranges are representative of different velocity components of the SiO gas as shown in Figures~\ref{fig:SiO_maps} and \ref{Histograms}. We have derived areas of the SiO J=2-1 and C$^{18}$O J=1-0 lines within the three velocity ranges by using the MADCUBA software. Then, the column densities of SiO and C$^{18}$O were derived with MADCUBA under Local Thermodynamic Equilibrium (LTE) conditions, considering line optical depth effects and extended emission \citep{Martin19}. In our analysis, the excitation temperatures (T$_{\rm ex}$) are assumed to be equal to 10 K and 7 K for C$^{18}$O and SiO, respectively. The T$_{\rm ex}$ of 10 K is an average value inferred from the J=2-1 and 1-0 transitions of C$^{18}$O towards several positions across Sgr B2 \citep{Martin08}, whereas the T$_{\rm ex}$ of \mbox{7 K} is derived for position 4 (shown in Figure~\ref{fig:SiO_maps}) by \cite{Rivilla18} (their G+0.693-0.03 position) using the J=2-1, 3-2 and 4-3 transitions of $^{29}$SiO.
Using MADCUBA we find optical depths of $<$0.08 for the SiO J=2-1 and C$^{18}$O J=1-0 lines shown in Figure ~\ref{fig:SiOC18O}.
The velocity ranges, the column densities of C$^{18}$O (N$_{\rm C^{18}O}$) and SiO (N$_{\rm SiO}$), as well as the integrated area of the SiO J=2-1 line (A$_{\rm SiO}$), are listed in Table \ref{Col_density}. The last column of this table gives the SiO relative abundance calculated as the N$_{\rm SiO}$/N$_{\rm H_2}$ ratio, where N$_{\rm H_2}$ is estimated from N$_{\rm C^{18}O}$ using the $^{16}$O/$^{18}$O isotopic ratio of 250 \citep{Wilson94} and the relative abundance of CO to H$_2$ of 10$^{-4}$ \citep{Frerking1982}. Table \ref{Col_density} shows that the average SiO abundance is $\sim$10$^{-9}$, which agrees with the average value derived from the SiO abundances found for the seven positions of Sgr B2 given in Table \ref{tab:Column_density}.


\begin{table}
    \caption{Parameters of C$^{18}$O and SiO derived for Sgr B2}\label{Col_density}
    \begin{threeparttable}
    \begin{tabular}{crcrc}
    \hline
    \multicolumn{2}{c}{C$^{18}$O} & \multicolumn{3}{c}{SiO}\\
    \hline
V$_{\rm LSR}$\tnote{(a)} & N$_{\rm C^{18}O}$ & A$_{\rm SiO}$ & N$_{\rm SiO}$ & X\tnote{(b)}\\
(km s$^{-1}$) & ($\times$10$^{15}$ cm$^{-2}$) & (K km s$^{-1}$) & ($\times$10$^{12}$ cm$^{-2}$) & ($\times$10$^{-9}$)\\
	\hline
    $[-20,+30]$ & 2.9$\pm$0.2 & 4.9$\pm$0.02 & 10.7$\pm$0.1 & 1.5$\pm$0.1\\
    $[+30,+75]$ & 10.5$\pm$0.2 & 9.1$\pm$0.02 & 19.9$\pm$0.1& 0.8$\pm$0.02\\
    $[+75,+120]$ & 4.1$\pm$0.2 & 2.5$\pm$0.02 &
    5.5$\pm$0.1 & 0.5$\pm$0.03\\
    \hline
    \end{tabular}
    \begin{tablenotes}
	\item[(a)] Velocity ranges tracing the main velocity components shown by the SiO gas in Figure \ref{fig:SiO_maps}.
    \item[(b)] The SiO abundance is calculated as the N$_{\rm SiO}$/N$_{\rm H_2}$ ratio, where N$_{\rm H_2}$ is derived from N$_{\rm C^{18}O}$ (see Section \ref{SiOabundances}).
    \end{tablenotes}
    \end{threeparttable}
\end{table}


\subsection{SiO abundance in the GC and other Galactic sources}
\label{sec:shockvel}
As seen in Table \ref{Col_density}, the relative abundances of SiO derived for Sgr B2 are within (0.5-1.5)$\times$10$^{-9}$. These SiO abundances agree with those estimated for several GC clouds \citep{Pintado97,Amo2011}. It has been proposed that large-scale shocks are responsible for the large SiO abundances found in GC regions \citep{MartinP1992,Pintado97}. The shocks in Sgr B2 may be produced by the cloud-cloud collision scenario discussed in Section \ref{Maps_sio}.

\begin{figure}
	\includegraphics[trim={0.6cm 1.3cm 0 1cm},clip,width=9cm]{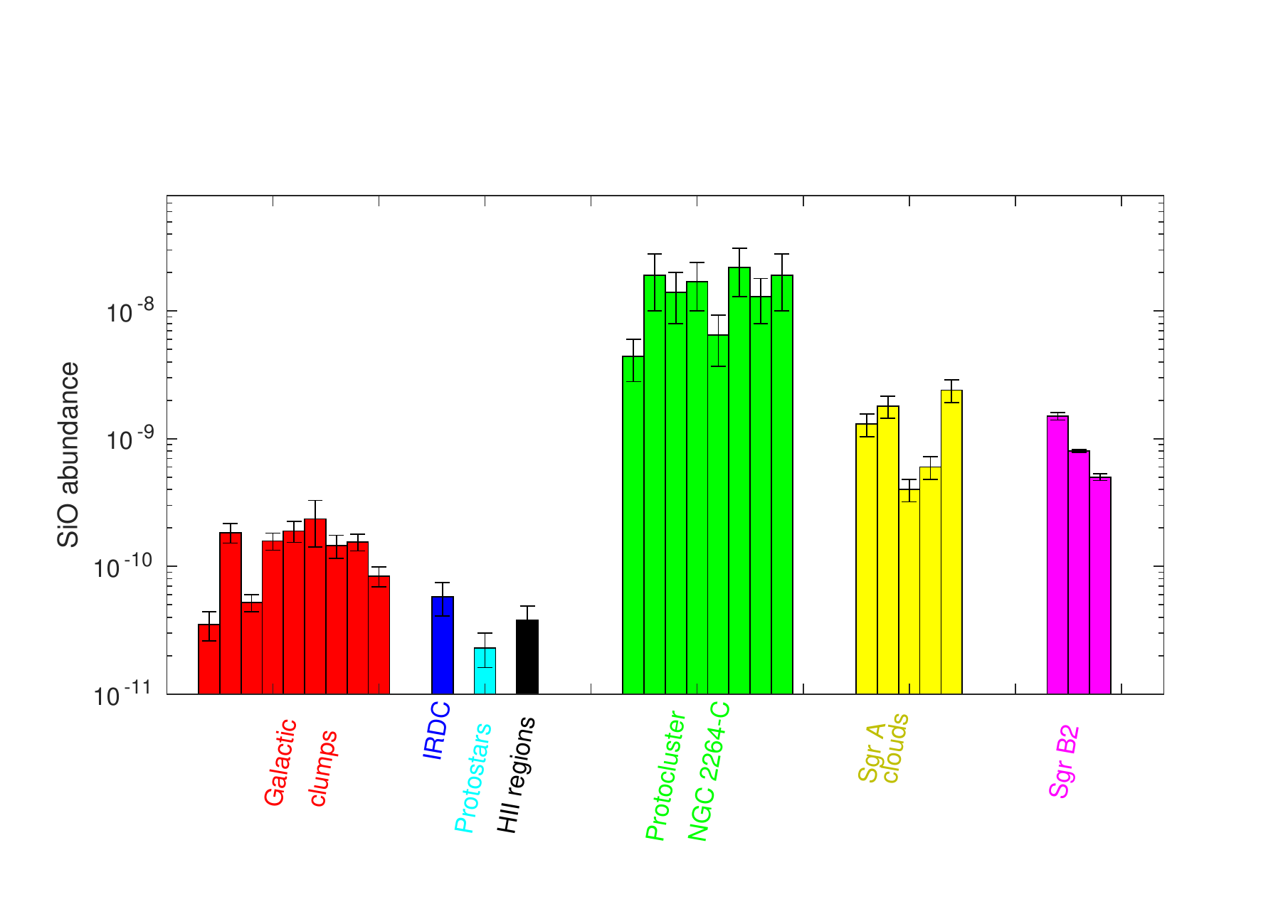}
    \caption{SiO abundances of different Galactic sources.
    The SiO abundances of the sources except those of Sgr B2 for different gas components are taken from the literature (see text).}\label{SiO_abunda}
\end{figure}

Figure~\ref{SiO_abunda} displays the comparison of the SiO abundances for the different velocity components of Sgr B2 with those estimated for other Galactic sources. In this comparison, we include SiO abundances found for Galactic clumps by \cite{Yu_2018}, average SiO abundances estimated for infrared dark clouds (IRDCs), protostars and young HII regions by \cite{Shanghuo19}, SiO abundances found for the high-mass protocluster NGC 2264-C by \cite{Lopez16}, as well as SiO abundances estimated for Sgr A clouds by \cite{Amo2011}.
The SiO abundances estimated for Sgr B2 are at least a factor of $\sim$2 higher than those of the Galactic clumps, while the SiO abundances found for Sgr B2 are at least a factor of $\sim$9, 22, and 13 higher than those towards IRDCs, protostars and young HII regions, respectively. \cite{Shanghuo19} argued that their estimates of the SiO column density, and therefore of the SiO relative abundance, can be considered as lower limits because the beam-filling factor can be smaller than 1 for 31\% of the studied sources. In this case, the differences in the SiO abundances between Sgr B2 and the sources studied by \cite{Shanghuo19} quoted above would be lower. On the other hand, the SiO abundances in the protocluster NGC 2264-C are at least a factor of 3 higher than the values of the SiO abundance in Sgr B2. Many of the Galactic clumps included in our comparison show signs of star formation \citep{Yu_2018} and it is likely that in these clumps SiO is mainly produced by shocks from outflows driven by protostars. The SiO abundances lower than $\sim$6$\times$10$^{-11}$ derived for the IRDCs, protostars and young HII regions by \cite{Shanghuo19} can also be mainly due to shocks affecting these sources with different evolutionary stages. The higher SiO abundances found in Sgr B2 than in the Galactic clumps, IRDC, protostars and young HII regions included in our comparison may be explained if Si is less depleted in Sgr B2 than in the other sources \citep{Pintado97}.
It is likely that more than one episode of star formation has taken place in NGC 2264-C \citep{Lopez16}, whose outflow activity would have injected turbulence into the medium \citep{Lopez16}. This could explain the high SiO abundances of $\sim$10$^{-8}$ found towards NGC 2264-C. As seen in Figure~~\ref{SiO_abunda}, on average the SiO abundances derived for Sgr B2 are similar to those of Sgr A clouds. It is thought that several of the Sgr A clouds have been impacted by the expanding shell of the Sgr A East supernova remnant \citep{Maeda02,Herrnstein02,Ferriere12}. The Sgr A clouds may also be affected by tidal perturbations during their close passage to the orbital pericentre in the GC \citep{2015MNRAS.447.1059K}. These mechanisms could be responsible for the turbulence and SiO formation through sputtering of grains by shocks \citep{MartinP1992}.\par

\subsection{SiO abundance predicted by chemical models}
\label{SiO_abundance}

C-type shock waves are known to produce SiO emission in molecular clouds via grain sputtering. Shock velocities below $\sim 25\,\rm km\,s^{-1}$ can only sputter SiO from the ice mantles of dust grains (e.g., see \citealt{1997A&A...321..293S,2013ApJ...775...88N,2016A&A...595A.122L}), while higher shock speeds $\gtrsim 25\,\rm km\,s^{-1}$ are needed to sputter Si-bearing material from grain cores (e.g., see \citealt{1997A&A...321..293S,Gusdorf08b}). Chemical models that mimic shocks with velocities of $\leq$25 km s$^{-1}$ show integrated SiO J=2-1 line intensities of $\sim$1-10 K km s$^{-1}$ \citep{1997A&A...321..293S,2016A&A...595A.122L}, which are in agreement with those observed in Sgr B2 (see Table \ref{Col_density} and Figure \ref{Histograms}), thus suggesting that the gas in this GC source is being mainly affected by very-low and low-velocity shocks, with some contributions from moderate-velocity shocks. Indeed, higher shock velocities >25 km s$^{-1}$ (which cover some of our moderate- and all our high-velocity shocks) produce integrated SiO J=2-1 line intensities $>$40 K km s$^{-1}$ for $n_{\rm H}=10^4-10^5\,\rm cm^{-3}$ \citep{1997A&A...321..293S}, which are more typical of interstellar regions with jets and outflows \citep{Gusdorf08b}.\par

In addition, chemical models of grain sputtering by C-type shocks show that the SiO abundances of $\sim 10^{-9}$ that we find for Sgr B2 can be readily obtained under a wide variety of conditions. For instance, SiO abundances of $\sim 10^{-9}$ only need shock speeds $v_{\rm shock}\gtrsim 4\,\rm km\,s^{-1}$ if the fraction of free silicon is 1-10\% in the pre-shock gas, or $v_{\rm shock}\gtrsim 7\,\rm km\,s^{-1}$ if the fraction of free SiO is 1-10\% in the grain mantles (e.g., see \citealt{2016A&A...595A.122L}). Similarly, \cite{Gusdorf08a} showed that shock velocities $\geq$ 25 km s$^{-1}$ can effectively eject all of the SiO in the mantles into the gas phase. Subsequently the SiO abundance decreases due to SiO freeze-out on grains in the cold post-shock gas. Their models show gas temperatures of \mbox{$\sim$20 K} for SiO abundances of $\sim$10$^{-9}$, which are reached on timescales $\gtrsim 5\times$10$^4$ years for gas densities of 10$^4$ cm$^{-3}$ and $\gtrsim 4\times$10$^3$ years for densities of 10$^5$ cm$^{-3}$ \citep{Gusdorf08b}.\par


Including the effects of shattering and vaporisation via grain-grain processing can also be important for gas with densities $\geq 10^5\,\rm cm^{-3}$ as seen in the work by \cite{Anderl2013}. They used a C-type shock model with a shock velocity of 20 km s$^{-1}$ to predict an integrated intensity of \mbox{$\sim$4 K km s$^{-1}$} for the J=2-1 rotational transition of SiO, which also agrees with the average value derived from those given in Table \ref{Col_density}. In general, chemical models of C-type shocks that include both grain core and grain mantle sputtering explain the observed SiO abundances and line widths in Sgr B2. 

Given the extreme environment in the GC, the best comparison with chemical model outputs can be done with results for another system, also in this region. \cite{Harada15} studied the chemistry of the circumnuclear disk surrounding the GC. This region has an estimated radiation field factor $G_0=10^4-10^5$, only slightly larger than Sgr B2's $G_0=10^3-10^4$ (\citealt{2004ApJ...600..214G}). \cite{Harada15} used a gas-grain chemical model that mimics grain sputtering by shocks. The average SiO abundance of $\sim$10$^{-9}$ found towards Sgr B2 agrees with that produced by the models by \cite{Harada15}, which consider a shock velocity of 10-20 km s$^{-1}$, a hydrogen density of 10$^5$ cm$^{-3}$, a cosmic-ray ionization rate $\zeta$ in the range of 10$^{-16}$-10$^{-15}$ s$^{-1}$ and timescales of around 10$^4$-10$^5$ years after the shock. Values of $\zeta$ within this range are estimated towards Sgr B2 \citep{vanderTak06,Bonfand19}. A SiO abundance of $\sim$10$^{-9}$ can also be explained by models with a $\zeta$ of 10$^{-16}$ s$^{-1}$, a density of 10$^5$ cm$^{-3}$, and shock velocities of 30-40 km s$^{-1}$. However, in this case the timescales to reach that abundance are 10$^2$-10$^3$ years after the shock \citep{Harada15}, which can be more restrictive for the age of Sgr B2. Figure 8 by \cite{Harada15} also shows that the time-averaged SiO abundance depends on the shock velocity as chemical models show higher SiO relative abundances at higher shock velocities for a given timescale, owing to the higher degree of Si sputtering from grain cores.\par

Finally, we exclude CJ- and J-type shocks for Sgr B2 as they would imply very short timescales, $\sim 10^2\,\rm years$ and much higher gas temperatures than those observed in this region. \cite{Gusdorf08a} found that CJ-type shock models, which consider both C- and J-type characteristics, predict gas temperatures higher than \mbox{10$^3$ K} for SiO abundances of 10$^{-9}$, which are at least a factor of 27 higher than those we found for Sgr B2. Similarly, chemical models that follow the production of SiO in dissociative J-type shocks of 25 and 35 km s$^{-1}$ show SiO abundances of 10$^{-9}$ at gas temperatures higher than 100 K \citep{Guillet09}. These temperatures are at least a factor of $\sim$3 higher than those of 30-37 K for Sgr B2, thus suggesting that J-type shocks may not be involved in the formation of SiO in this GC source.

\section{Excitation temperature, hydrogen density and kinetic temperature}\label{Excitation_T}
\begin{figure}
	\includegraphics[width=6cm]{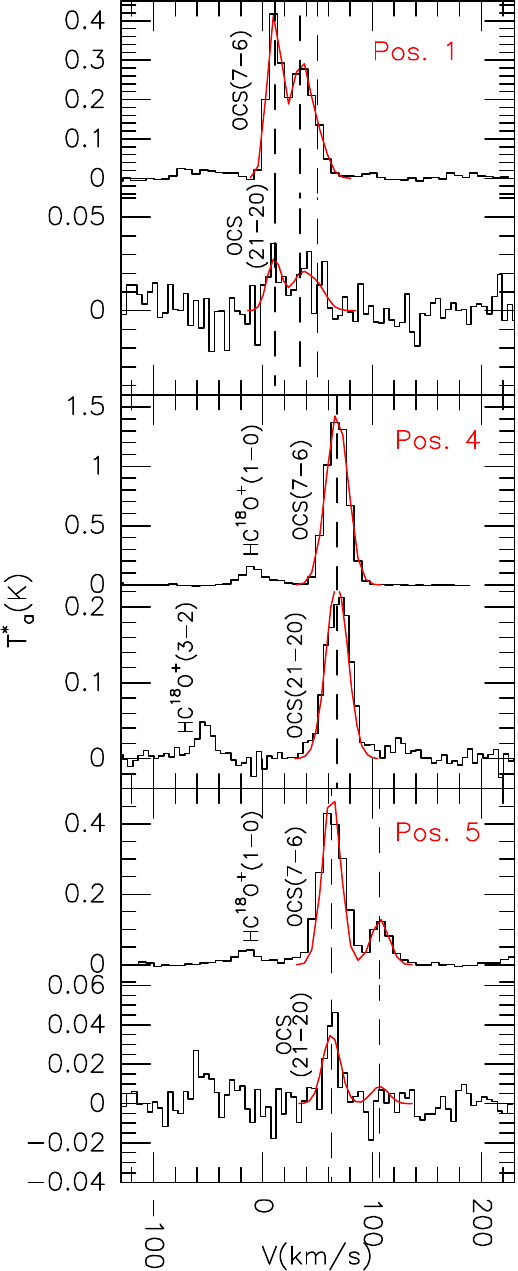}
    \caption{OCS J=7-6 and 21-20 lines observed towards positions 1, 4 and 5 of Sgr B2. 
    The dashed lines show velocities identified in the Gaussian fits of the OCS lines (see text). Lines of HC$^{18}$O$^+$ are also observed towards several positions. The best-fit LTE model lines are overlaid on top of the observed OCS lines.}
    \label{fig:OCS}
\end{figure}

Figure~\ref{fig:OCS} shows OCS lines of the J=7-6 and 21-20 transitions towards positions 1, 4 and 5 shown in Figure~\ref{fig:SiO_maps}.
Using the AUTOFIT tool of the MADCUBA software we have fitted synthetic lines to the observed OCS J=7-6 and 21-20 lines considering LTE conditions, thus deriving the local standard of rest (LSR) velocity, the full width half maximum (FWHM), the intensities (I) of the OCS J=7-6 and 21-20 lines, the OCS column density (N$_{\rm OCS}$), and the excitation temperature (T$_{\rm ex}$).
These derived parameters for position 1, 4, and 5 are listed in Table \ref{tab:Table_OCS}, and the LTE synthetic lines of OCS J=7-6 and 21-20 are shown with red lines in Figure~\ref{fig:OCS}. 
Multiple velocity components are needed to fit the observed OCS J=7-6 and 21-20 lines with MADCUBA. The FWHM and/or the T$_{\rm ex}$ were considered as fixed parameters when the AUTOFIT tool did not converge. We find T$_{\rm ex}$ within 24-29 K for different velocity components and the three positions selected in Sgr B2.

We have modeled the intensities of the OCS J=7-6 and 21-20 lines given in Table \ref{tab:Table_OCS} using the RADEX LVG code \citep{vandertak2007}.
Using this non-LTE code, we find that kinetic temperatures (T$_{\rm kin}$) of 30-37 K, H$_2$ densities (n$_{\rm H_2}$) of (3.0-4.0)$\times$10$^5$ cm$^{-3}$ and OCS column densities of (3.0-42.0)$\times$10$^{14}$ cm$^{-2}$ provide the best fits to both the derived LTE excitation temperature and the intensities of OCS J=7-6 and 21-20 lines listed in Table \ref{tab:Table_OCS}.
In the LVG modeling, the T$_{\rm kin}$, n$_{\rm H_2}$ and N$_{\rm OCS}$ were considered as free parameters, while the linewidth of 20 km s$^{-1}$ (the typical value found in GC clouds \citep{Amo2011,Armijos2015} and the FWHM peak in Figure \ref{Histograms}) was considered as a fixed parameter. The non-LTE column densities of OCS are consistent within a factor of 2 with those calculated using the LTE approach with MADCUBA. The T$_{\rm kin}$ of 30-37 K are slightly higher than the low kinetic temperature of $\sim$25 K found for GC clouds by \cite{Huttemeister93}.


\begin{table*}
	\caption{Parameters derived from OCS lines.}\label{tab:Table_OCS}
	\begin{threeparttable}
	\begin{tabular}{cccccccccc}
	\hline
Pos.&V$_{\rm LSR}$\tnote{(a)}& FWHM\tnote{(b)} & I (J=7-6) & I (J=21-20) & N$_{\rm OCS}^{\rm LTE}$ & T$_{\rm ex}$\tnote{(b)} &n$_{\rm H_2}$\tnote{(c)} & T$_{\rm kin}$\tnote{(c)}& N$_{\rm OCS}^{\rm non-LTE}$\,\,\tnote{(c)}\\
&(km s$^{-1}$) & (km s$^{-1}$) & (mK) & (mK) & ($\times$10$^{14}$ cm$^{-2}$) & (K) & ($\times$10$^5$ cm$^{-3}$) & (K) & ($\times$10$^{14}$ cm$^{-2}$)\\
\hline
1 & 10.6$\pm$0.2 & 15.6$\pm$0.4  & 414$\pm$16 & 28$\pm$6& 6.6$\pm$0.2 & 23.9$\pm$1.1 &3.5$^{+4.0}_{-1.7}$ & 30$^{+1}_{-2}$ & 11.0$^{+0.5}_{-1.0}$\\
  & 34.3$\pm$0.3  & 20 & 247$\pm$16 &  14$\pm$5 & 5.7$\pm$0.3 & 23.7$\pm$1.4 & 3.2$^{+5.1}_{-1.8}$ & 30$^{+1}_{-4}$ &6.5$^{+0.5}_{-0.5}$\\
  & 50.4$\pm$0.8 & 20 & 141$\pm$18 & 16$\pm$6 & 2.0$\pm$0.3 & 27.6$\pm$2.8 & 3.0$^{+12.0}_{-2.0}$ & 35$^{+4}_{-5}$ & 4.0$^{+0.5}_{-1.0}$\\
4  & 67.9$\pm$0.1 & 23.4$\pm$0.3  & 1447$\pm$27& 221$\pm$12 & 37.0$\pm$0.5 & 28.9$\pm$0.4 & 3.0$^{+1.0}_{-0.3}$ & 37$^{+1}_{-1}$ & 42.0$^{+1.0}_{-1.0}$\\
5 & 63.0$\pm$0.5 & 20 & 470$\pm$25  & 35$\pm$8 & 10.0$\pm$0.5 &24.0$\pm$2.4 & 4.0$^{^{+16.0}}_{-2.2}$ & 30$^{+1}_{-2}$& 13.0$^{+0.5}_{-2.0}$\\
  & 107.1$\pm$1.3 & 20 & 129$\pm$23 & 9$\pm$2 & 2.6$\pm$0.3 & 24 & 3.0$^{+7.0}_{-1.3}$ & 30$^{+3}_{-1}$& 3.0$^{+1.0}_{-0.2}$\\
	\hline
	\end{tabular}
	\begin{minipage}{\linewidth}\small
	$^{\rm (a)}$ Towards positions 1 and 5 several velocity components are identified in the LTE line modelling.\\
	$^{\rm (b)}$ The FWHM or T$_{\rm ex}$ is fixed in the LTE modelling when the parameter is listed without an error estimate.\\
	$^{\rm (c)}$ The H$_2$ density, kinetic temperature and column density of OCS give the best fits to both the line intensities of the OCS J=7-6 and 21-20 lines and excitation temperature in the non-LTE modelling with the RADEX code (see text).\\
    \end{minipage}
	\end{threeparttable}
\end{table*}

\section{A CLOUD-CLOUD COLLISION IN SGR B2}
\label{sec:cloud-cloud}

\subsection{Stellar feedback bubbles or a cloud-cloud collision?}
\label{BubblesOrCollisions}
As previously mentioned, the observed SiO maps of Sgr B2 show evidence of shocked gas in this molecular cloud complex. Is the shocked gas the result of stellar-wind bubbles sweeping their surrounding dense gas and/or the result of a large-scale cloud-cloud collision? Stellar winds are responsible for the formation of bubbles, bow shocks, wind-swept filaments in the interstellar medium (e.g., see \citealt{2016A&A...586A.114M,2017MNRAS.466.1857G}). Although the presence of $>200$ radio continuum sources in Sgr B2, including young stellar objects and $>70$ HII regions (e.g., see \citealt{1995ApJ...449..663G,Ginsburg2018}), indicates that winds and UV radiation fields from young stars are present, their compactness suggest they can be dynamically important on spatial scales ranging from sub-parsecs to a few parsecs. Thus, stellar wind bubbles and their associated HII regions could be responsible for shaping density sub-structures on those scales, but they are unlikely to dominate the large-scale dynamics of the Sgr B2 complex.\par

In a multi-bubble superposition scenario, several large-scale feedback bubbles would be needed to create a multi-shock structure spanning a 36$\times$36 pc$^2$ region, and the ensuing velocity field would show clear shock edges at the boundaries between ionisation fronts and molecular gas (e.g., see \citealt{2015A&A...573A..10M}). Such clear bubble signatures are not observed in our SiO emission maps in which the arc, shells, and cavities discussed in Section \ref{Maps_sio} do not have clearly delineated morphologies, and the cavities do not spatially coincide with the observed edges of the much smaller compact HII regions. In general, our SiO maps show that all these features display fractal sub-structures, which can be naturally explained by turbulence (e.g., see \citealt{2009ApJ...692..364F,2013MNRAS.436.1245F}). In this context, \cite{1986ApJ...303..667S} also found that the segmentary morphology found in dense gas in their $^{13}$CO maps of the Orion cloud can be attributed to the intrinsic inhomogeneous distribution of molecular gas rather than to the expansion of nearby HII regions.\par

In addition, the expansion velocities of compact and ultra-compact HII regions, such as those found in Sgr B2, are typically of the order of $\sim 10-40\,\rm km\,s^{-1}$ (see \citealt{2012MNRAS.419L..39R,2018MNRAS.479.2016W}), and the bubble expansion generally stops after $\sim 10^5$ years when the shock weakens and pressure equilibrium is reached. These shock speeds could in principle produce SiO via grain sputtering, but the emission would be confined to regions with sizes ranging from sub-parsecs to a few parsecs.\par



Therefore, the small size of the observed compact HII regions, the lack of spatial correspondence with regions of broad-line SiO emission, and the lack of spatial overlap between them and the cavities observed in our SiO emission maps suggest they are dynamically important at small scales, but also argue against a large-scale multi-bubble superposition scenario for this region. Instead, several previous studies have found signatures of a cloud-cloud collision in this region. For instance, \cite{Hasegawa1994} identified a clump at $70-80\,\rm km\,s^{-1}$ that spatially overlapped with a cavity at $40-50\,\rm km\,s^{-1}$ in their $^{13}$CO maps of Sgr B2, thus suggesting a two-cloud collision. Later, \cite{Sato00} spatially associated the local HII regions and molecular masers at $65-80\,\rm km\,s^{-1}$ with dense (potentially-compressed) gas with similar kinematics. The authors suggested that star formation in these regions was likely triggered by shocks induced by a cloud-cloud collision.\par

The above results are also supported by earlier recombination-line observations (\citealt{Mehringer93}), which show that ionised gas is moving at $\sim 50-70\,\rm km\,s^{-1}$, i.e., at speeds in between the velocities of the potentially-colliding clouds (see also \citealt{1995ApJ...451..284D,2011AJ....142..177D}). Interestingly, we also find that some of the radio continuum sources in the region overlap with shocked regions at $[40,85]\,\rm km\,s^{-1}$ as traced by our SiO maps (see Figure \ref{fig:SiO_maps}). Although the spatial coincidence is not perfect, the overlap may indicate a recent connection to stellar feedback at small scales. Furthermore, the non-uniform spatial distribution of shocked gas in the Sgr B2 region and its complex kinematics revealed by our maps strongly suggest that dense-gas collisions and turbulence emerging at larger scales play more important roles. Can the collision of two clouds reproduce such a structure? If yes, how does the initial density structure of the colliding clouds affect the resulting distribution of shocked gas?

\subsection{Numerical simulations of cloud-cloud collisions}
\label{subsec:Numerical}
In order to understand the kinematics of shocked gas in Sgr B2 and test the cloud-cloud collision hypothesis, we investigate the properties of shocked gas in idealised numerical simulations of cloud-cloud interactions. We study two cloud-cloud collision scenarios in 3D: one in which two uniform clouds collide (model UU with clouds U1 and U2), and one in which two fractal clouds collide (model FF with clouds F1 and F2), i.e. clouds with log-normal density distributions. In both cases the computational domain consists of a Cartesian prism with a spatial range $-40\,\rm pc\leq X_1\leq40\,pc$, $-60\,\rm pc\leq X_2\leq220\,pc$, and $-40\,{\rm pc}\leq X_3\leq40\,{\rm pc}$. We impose outflow boundary conditions on all sides of the simulation domain.\par

\begin{figure*}
\begin{center}
  \begin{tabular}{c c c c c}
    $t=0\,\rm Myr$ & $t=0.2\,\rm Myr$ & $t=0.5\,\rm Myr$ & $t=0.7\,\rm Myr$\\
    \hspace{-0.2cm}\resizebox{36mm}{!}{\includegraphics{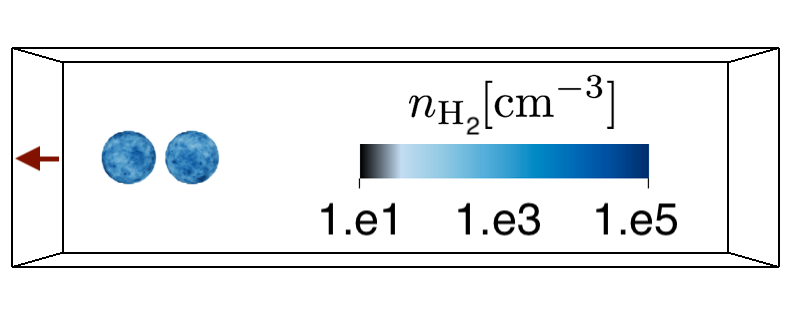}} & \hspace{-0.2cm}\resizebox{36mm}{!}{\includegraphics{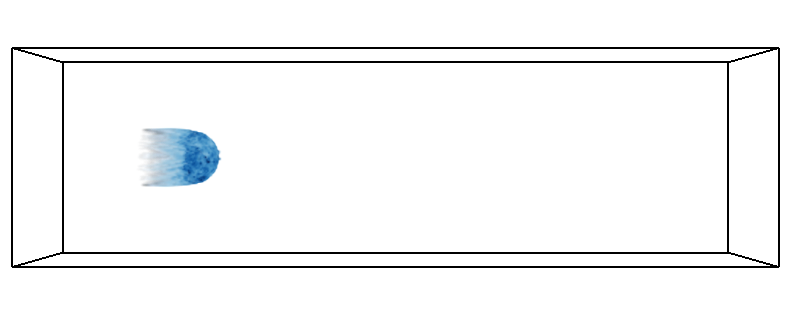}} & \hspace{-0.2cm}\resizebox{36mm}{!}{\includegraphics{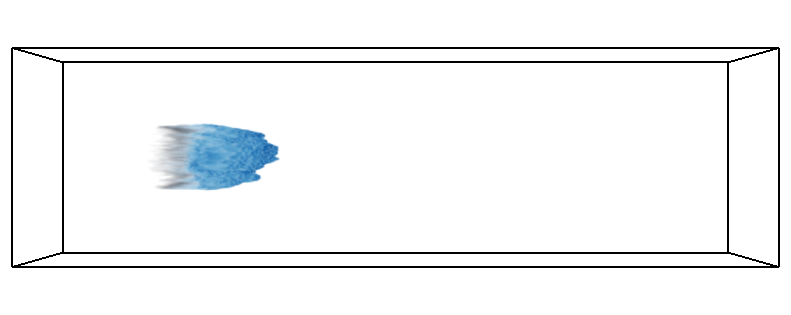}} & \hspace{-0.2cm}\resizebox{36mm}{!}{\includegraphics{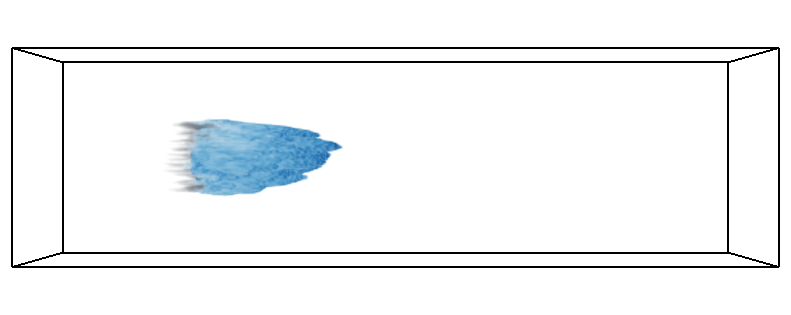}}\vspace{-0.5cm}\\    
    \hspace{-0.48cm}\resizebox{42mm}{!}{\includegraphics{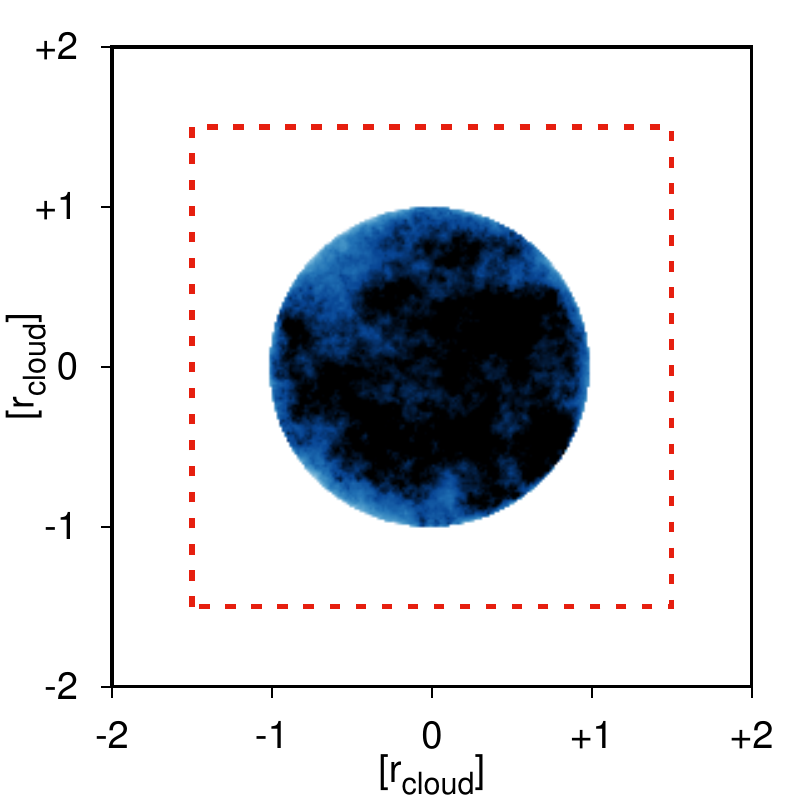}} & \hspace{-0.48cm}\resizebox{42mm}{!}{\includegraphics{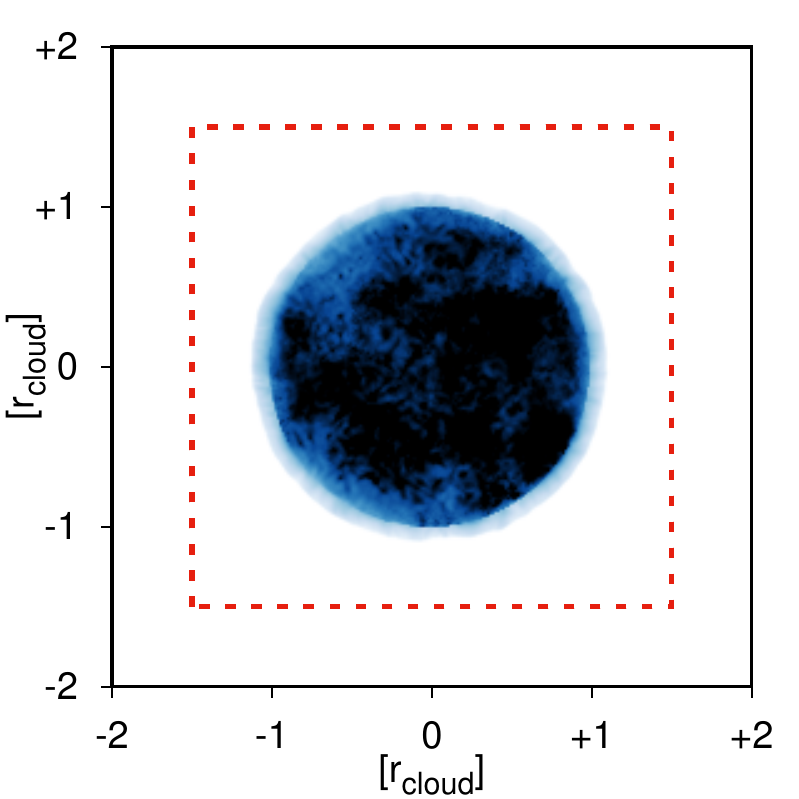}} & \hspace{-0.48cm}\resizebox{42mm}{!}{\includegraphics{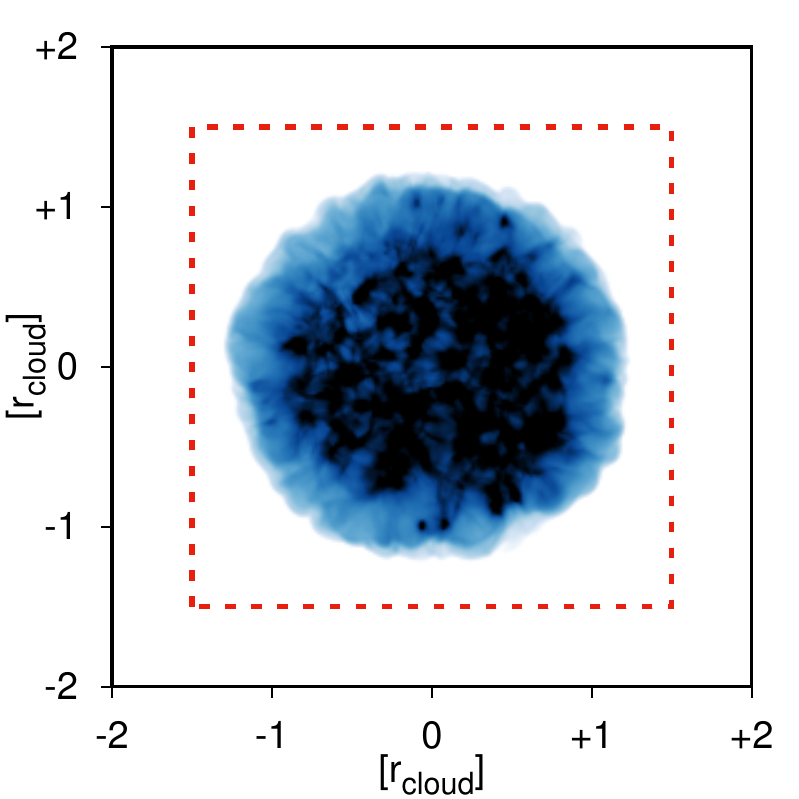}} & \hspace{-0.48cm}\resizebox{42mm}{!}{\includegraphics{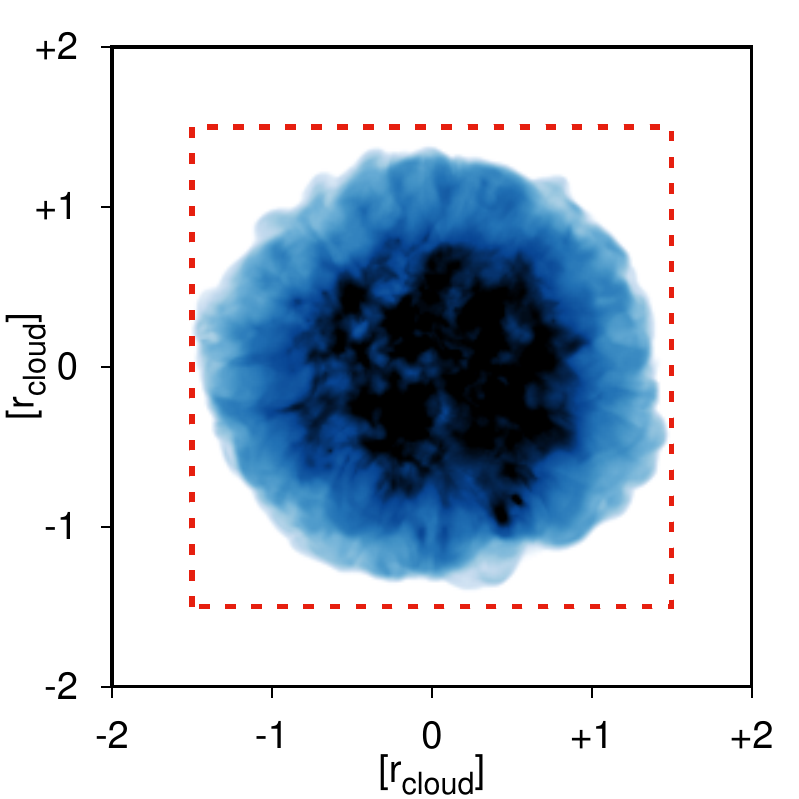}} &
    \hspace{-0.4cm}\resizebox{10mm}{!}{\includegraphics{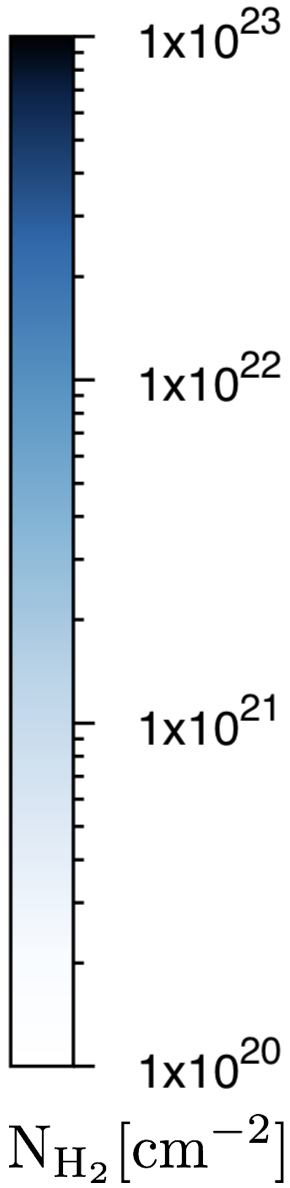}}\\
     $t=1.0\,\rm Myr$ & $t=1.2\,\rm Myr$ & $t=1.5\,\rm Myr$ & $t=1.7\,\rm Myr$\\
    \hspace{-0.2cm}\resizebox{36mm}{!}{\includegraphics{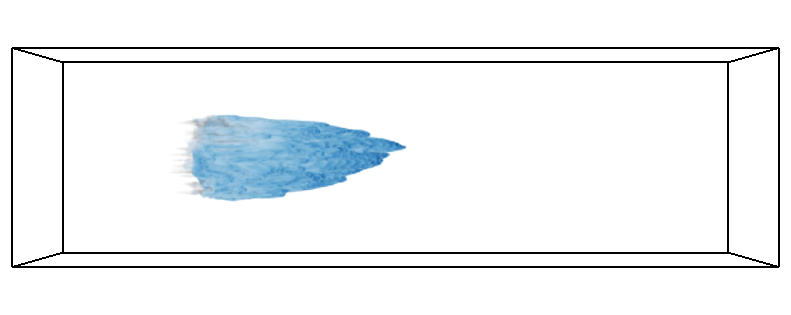}} & \hspace{-0.2cm}\resizebox{36mm}{!}{\includegraphics{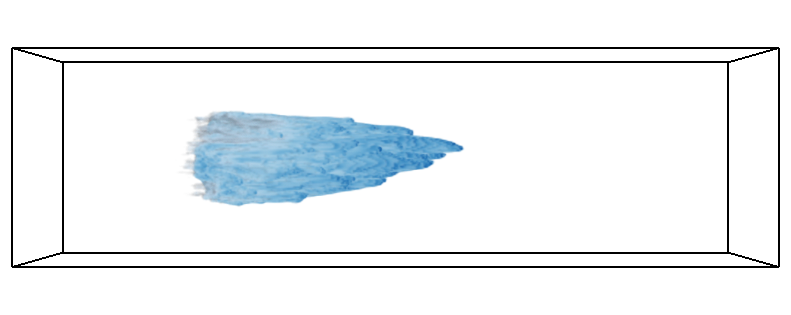}} & \hspace{-0.2cm}\resizebox{36mm}{!}{\includegraphics{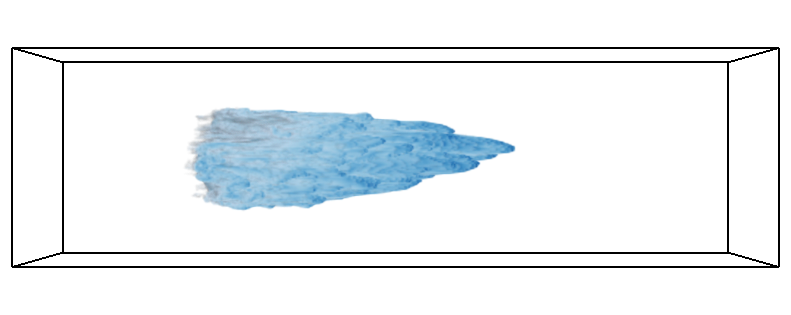}} & \hspace{-0.2cm}\resizebox{36mm}{!}{\includegraphics{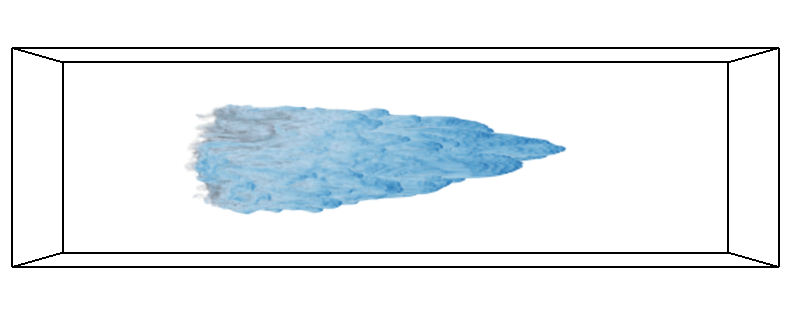}}\vspace{-0.5cm}\\   
    \hspace{-0.48cm}\resizebox{42mm}{!}{\includegraphics{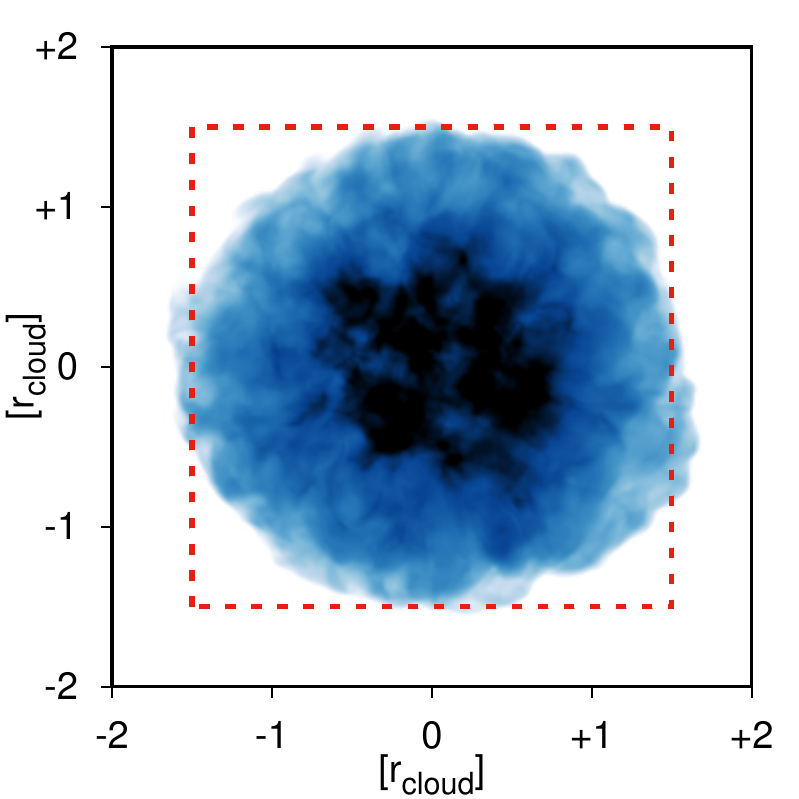}} & \hspace{-0.48cm}\resizebox{42mm}{!}{\includegraphics{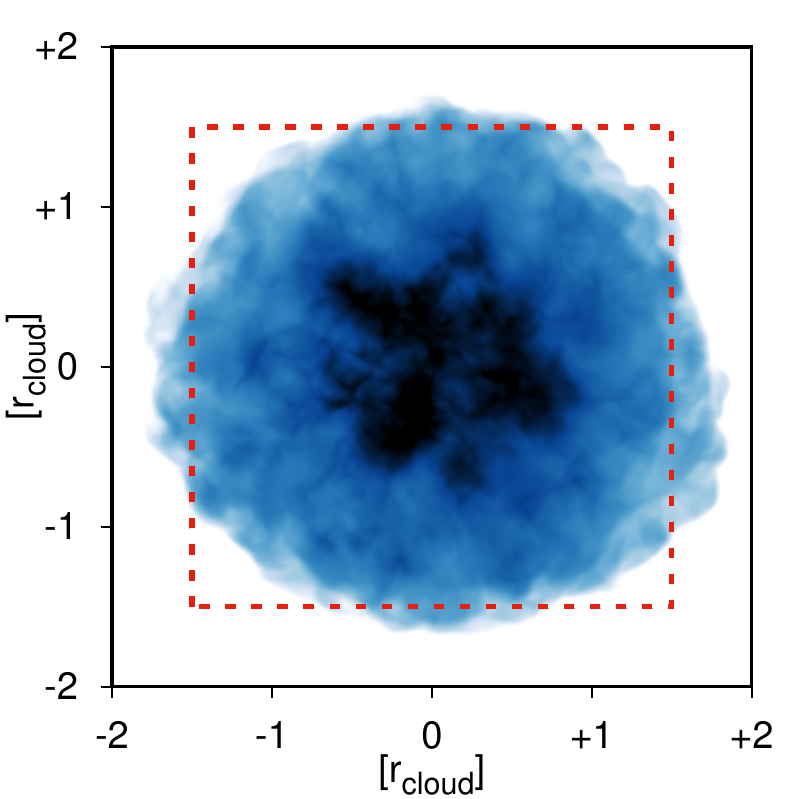}} & \hspace{-0.48cm}\resizebox{42mm}{!}{\includegraphics{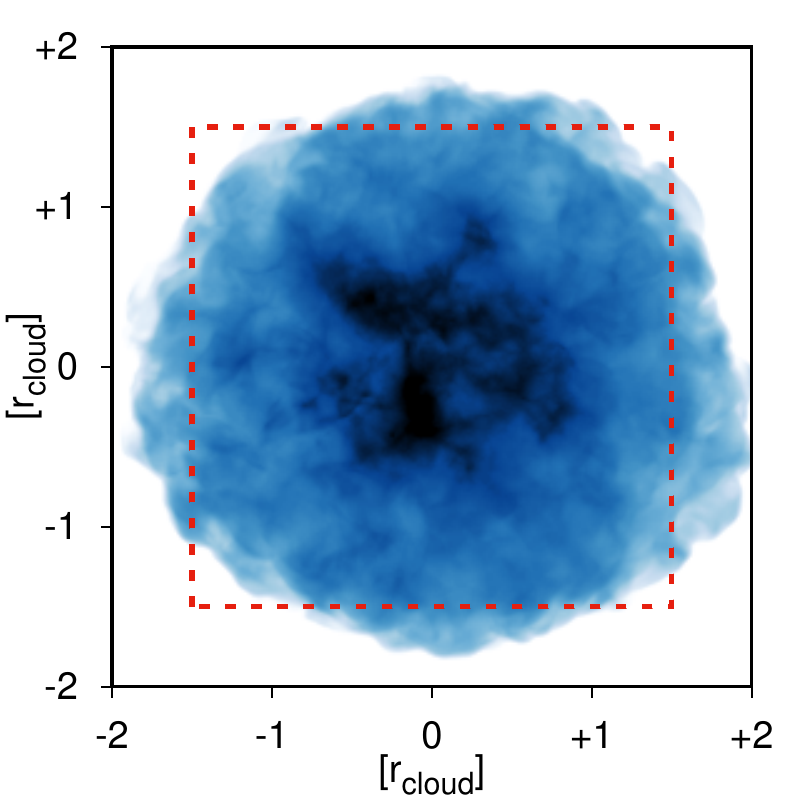}} & \hspace{-0.48cm}\resizebox{42mm}{!}{\includegraphics{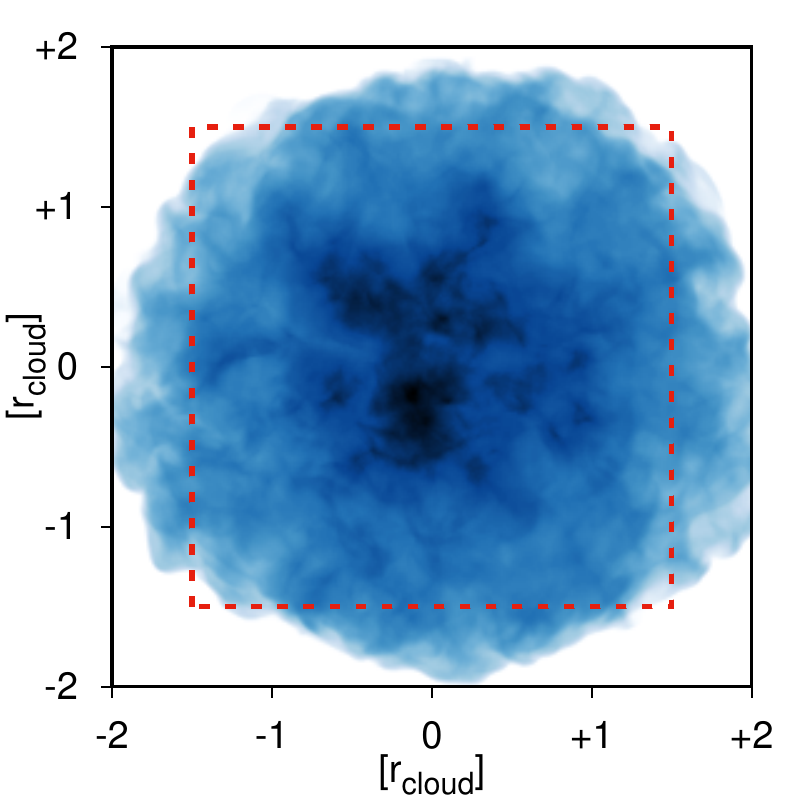}} &
    \hspace{-0.4cm}\resizebox{10mm}{!}{\includegraphics{barra.png}}\\
     \multicolumn{2}{c}{Optically-thick column density map of Sgr B2} & \multicolumn{2}{c}{Optically-thin column density map of Sgr B2}\\   
    \multicolumn{2}{l}{\resizebox{75mm}{!}{\includegraphics{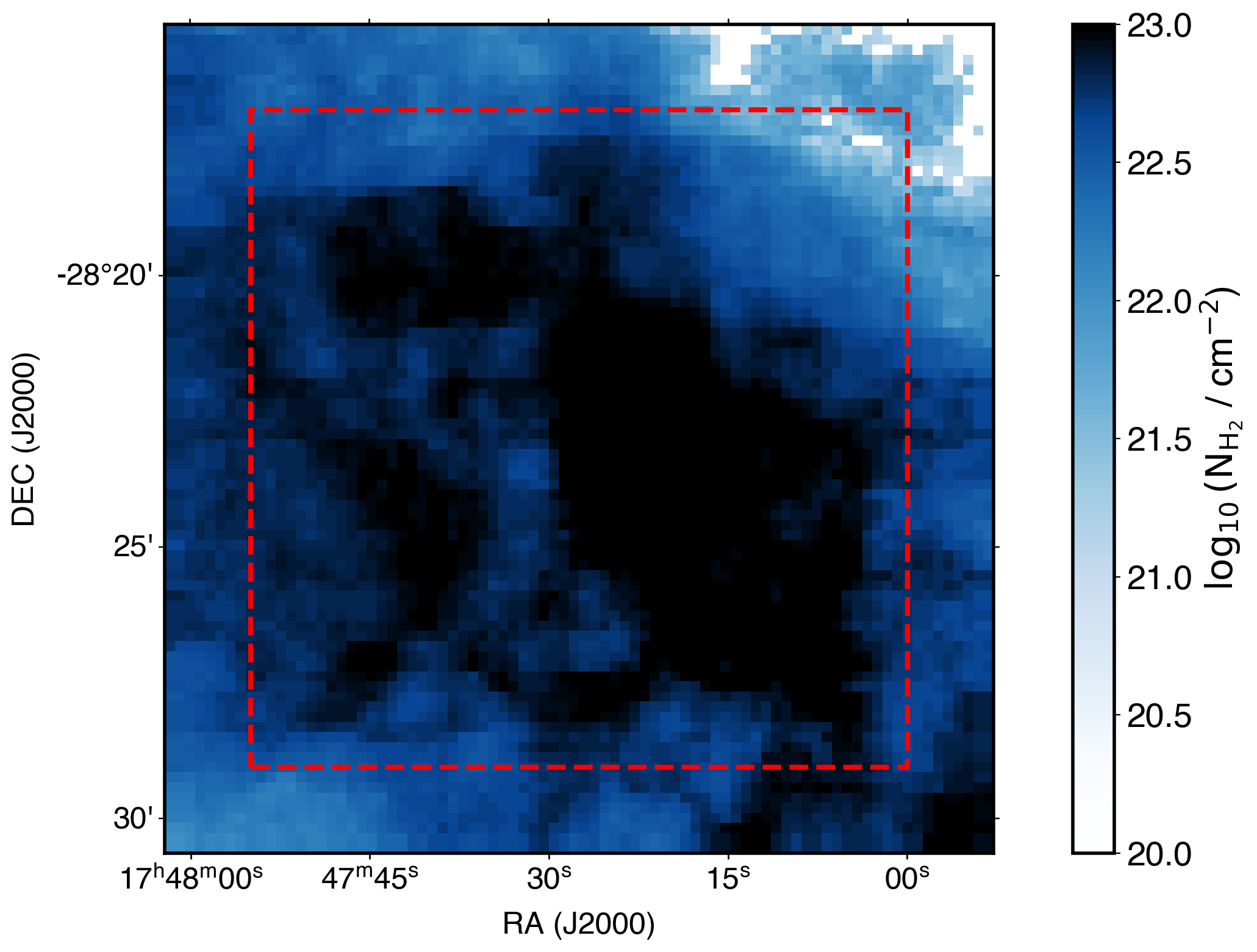}}} & \multicolumn{2}{l}{\resizebox{75mm}{!}{\includegraphics{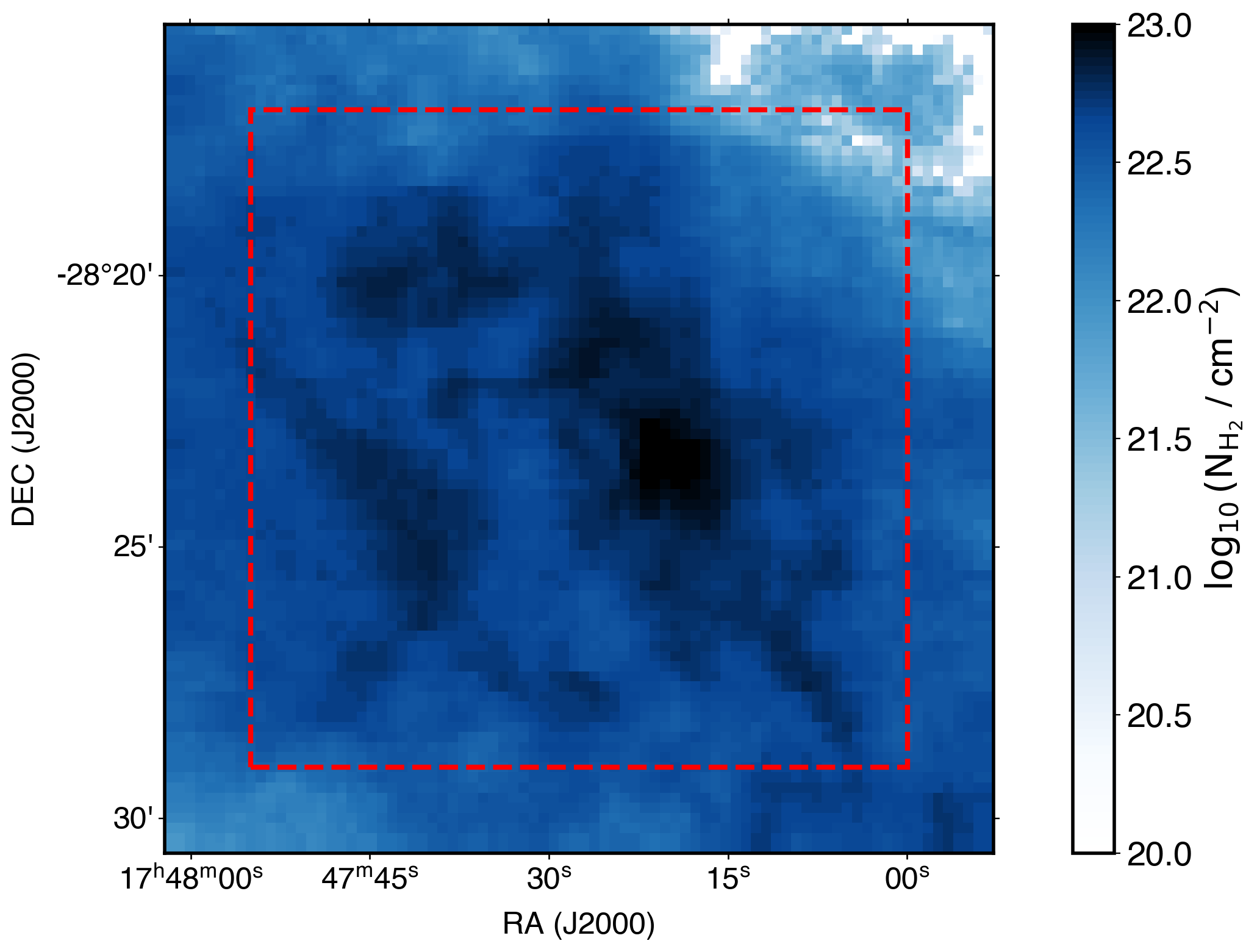}}}\\
  \end{tabular}
  \caption{Top panels: Maps of the number density (3D renderings) and column number density of hydrogen (2D projections along the observer's LOS, whose direction is indicated by a red arrow) in model FF, at increasing times in the range $0<t<1.7\,\rm Myr$. Bottom panels: observed maps of the column number density of hydrogen as inferred from our $^{13}$CO data cube assuming optically thick (bottom left panel) and optically thin regimes (bottom right panel). The red, dashed square indicates regions of the same size, $30^2\,\rm pc^2$, in both simulations and observations. Our collision model FF reproduces the turbulent structure observed in Sgr B2. The full-time evolution can be viewed in movies included in the electronic version of this paper.} 
  \label{Figure5}
\end{center}
\end{figure*}

The simulation grid is uniform and has a resolution of $(N_{\rm X_{1}}\times N_{\rm X_{2}}\times N_{\rm X_{3}})=(512\times1792\times512)$, and the pair of clouds in each model are centred on $(0,-15\,\rm pc,0)$ (cloud U1/F1) and on the origin $(0,0,0)$ (cloud U2/F2; see the top left panel of Figure \ref{Figure5}). The initial radius of both clouds is $10\,\rm pc$ and is covered by $64$ grid cells, so the resolution in physical units is $\approx 0.16\,\rm pc$ per grid cell. This resolution ensures that dynamical instabilities arising at the interfaces between the cloud and inter-cloud medium are well captured throughout the evolution (see \citealt{2016MNRAS.457.4470P,2018MNRAS.473.3454B,2019MNRAS.486.4526B} for additional details).\par

For these simulations we solve the system of conservation laws of ideal hydrodynamics using the {\sevensize PLUTO v4.3} code (see \citealt{2007ApJS..170..228M}) in a 3D $(X_1,X_2,X_3)$ Cartesian coordinate system. We employ the \verb#HLLC# approximate Riemann solver of \cite*{Toro:1994}, and a Courant-Friedrichs-Lewy (CFL) number of $C_{\rm a}=0.3$. We close the system of conservation laws by introducing a quasi-isothermal equation of state with a ratio of calorific capacities at constant pressure and volume of $\gamma=1.01$. This choice of polytropic index is adequate to model molecular gas (see \citealt{2000ApJ...531..350M,2005MNRAS.359..211L}). We note that while our numerical setup implies that our models are scale-free, we choose to report the initial conditions and all the results in physical units that are constrained by the observations discussed so far.\par

In the first scenario, model UU, we initialise two spherical clouds with uniform density fields, and in the second scenario, model FF, we initialise two fractal clouds with density profiles described by log-normal distributions (see Table \ref{tab:simulations}). We generate these fractal density fields using the pyFC library\footnote{Publicly available at: \url{https://bitbucket.org/pandante/pyfc}}, which relies on a Fourier method developed by \cite{Lewis:2002ug} to generate random, log-normal density fields with user-defined power-law spectra. For this set, we generate two solenoidal clouds (i.e., clouds with density distributions that arise from divergence-free driving of supersonic turbulence; see \citealt*{2008ApJ...688L..79F}) with different seeds and the same initial mean density $\bar{\rho}_{\rm cloud}=1.53\times 10^{-20}\,\rm g\,cm^{-3}$, and initial standard deviation of $\sigma_{\rho_{\rm cloud}}=1.8\,\bar{\rho}_{\rm cloud}$. Before interpolating the clouds into the computational domain, we mask regions in the fractal cloud domain outside a radius $r_{\rm cloud}=10\,\rm pc$ and scale the average density of each cloud to $\bar{\rho}_{\rm cloud}=1.53\times 10^{-20}\,\rm g\,cm^{-3}$ (i.e., to a mean hydrogen number density of $\bar{n}_{\rm H}=4\times 10^{3}\,\rm cm^{-3}$) so that both clouds start with the same mass and initial mean density. Our choice of initial density implies a total mass of $2\times 10^6\,\rm M_{\odot}$. These densities and the resulting mass are in agreement with those inferred by previous authors for Sgr B2 (e.g. see \citealt{2004ApJ...600..214G,2016A&A...588A.143S}).\par

\begin{table}
\caption{Initial conditions for the 3D cloud-cloud collision models. Column 1 indicates the medium (U=uniform, F=fractal). Column 2 shows the line-of-sight (LOS) initial velocity in the rest frame of the observer. Column 3 shows the initial hydrogen number density. Columns 4 and 5 show the initial mean temperature and sound speed, respectively. Column 6 shows the initial thermal pressure equilibrium ($k_B$=Boltzmann constant).}
\begin{adjustbox}{max width=8cm}
\begin{tabular}{c c c c c c}
\hline
\textbf{(1)} & \textbf{(2)} & \textbf{(3)} & \textbf{(4)} & \textbf{(5)} & \textbf{(6)}\\
\textbf{Medium} & $v_{\rm LOS}$ & $\bar{n}_{\rm H}$ & $\bar{T}$ & $\bar{c}_{s}$ & $P/k_{B}$ \\
 & $[\rm km\,s^{-1}]$ & $[\rm cm^{-3}]$ & $[\rm K]$ & $[\rm km\,s^{-1}]$ & $\rm [K\,cm^{-3}]$ \\ \hline
Ambient & $+20$ & $4\times10^2$ & $390$ & $1.20$ & $2\times 10^5$ \\
Cloud U1/F1 & $+140$ & $4\times10^3$ & $39$ & $0.37$ & $2\times 10^5$\\
Cloud U2/F2 & $+20$ & $4\times10^3$ & $39$ & $0.37$ & $2\times 10^5$\\\hline
\end{tabular}
\end{adjustbox}
\label{tab:simulations}
\end{table} 

Next, we assume for simplicity that the collision occurs along the observers' line of sight (LOS). Thus, in each scenario cloud U1/F1 has an initial LOS velocity of $+140\,\rm km\,s^{-1}$, while cloud U2/F2 has a LOS velocity of $+20\,\rm km\,s^{-1}$. Our choice of initial velocities for the simulations is constrained by our SiO observations. In particular, our observations show that a large fraction of the SiO emission is in the velocity channels $[10,40]\,\rm km\,s^{-1}$, which restricts the rest-frame velocity of the slower cloud to that range. In addition, our observations also show that there exists very fast shocked gas at velocities $[70,120]\,\rm km\,s^{-1}$, which in turn restricts the initial speed of the faster cloud to that range. These velocities are also consistent with current dynamical models of the CMZ. For instance, \cite{Molinari2011} found that a constant orbital velocity of $80\,\rm km\,s^{-1}$ fits their twisted ring model, and more recently \cite{2015MNRAS.447.1059K} proposed a model with multiple gas streams moving at even higher velocities $100-200\,\rm km\,s^{-1}$. In the latter model, Sgr B2 has an orbital velocity of $\sim 130\,\rm km\,s^{-1}$. \cite{Hasegawa1994} and \cite{Sato00} considered a lower relative velocity of $30\,\rm km\,s^{-1}$ for the colliding clouds, but we find that such low velocity is insufficient to reproduce the vast range of gas velocities seen in our SiO emission maps. Thus, motivated by the kinematics of shocked gas in Sgr B2 and our simulations, we propose a higher relative velocity of $\sim 120\,\rm km\,s^{-1}$ for the cloud-cloud collision.\par

After the cloud density fields are interpolated into the computational domain, we use the aforementioned relative velocity and initialise the simulations in the rest frame of the slower cloud U2/F2, with both clouds (U1/F1 and U2/F2) and the inter-cloud medium in thermal pressure equilibrium, $P/k_B=2\times 10^5\,\rm K\,cm^{-3}$, where $k_B$ is the Boltzmann constant. This pressure value is lower than the range of pressures predicted for the GC region from equilibrium considerations (see \citealt{1996ARA&A..34..645M,2012MNRAS.423.3512C} and references therein), but is still higher than the pressures inferred for clouds in the Galactic disk, $P/k_B\sim 10^4\,\rm K\,cm^{-3}$. Before presenting the results from our models, we first discuss the limitations of our current simulations in the context of previous cloud-cloud collision models.\par

\subsection{Previous cloud-cloud collision models and limitations of our numerical work}
\label{Limitations}
Cloud-cloud collisions are ubiquitous in the interstellar medium and play a key role in shaping the gas structure in the GC (e.g., see \citealt{Enokiya19} and references therein). Cloud-cloud collisions contribute to the formation of molecular gas (\citealt{2017MNRAS.469..383J,2017MNRAS.472.2496K}) and they can also trigger star formation by inducing gravitational collapse in the shocked gas (\citealt{2013ApJ...774L..31I,2014ApJ...792...63T,2017ApJ...835..137W,2018PASJ...70S..54S}). Self-gravity and sink particles are key ingredients in star formation studies, but they would be important on much smaller scales ($<0.05\,\rm pc$; see e.g., \citealt{2012ApJ...761..156F}) than what we can resolve with our current simulations. Moreover, as we discussed in Section \ref{BubblesOrCollisions} stellar feedback does not appear to be playing a major role in the large-scale dynamics of shocked gas in Sgr B2. Therefore, for the purposes of this study we exclude such recipes in the numerical models.\par

In addition, our idealised, single-fluid hydrodynamic simulations do not allow us to self-consistently study two-fluid shocks (e.g. J-, CJ-, or C-type shocks; see \citealt{1993ARA&A..31..373D}), which would be needed to produce synthetic maps of the intensity of SiO emission. Our models lack magnetic fields, a chemical network, and radiative cooling and heating, which are all needed to study two-fluid shocks in detail (e.g., see \citealt{1997A&A...326..801S,2016MNRAS.455.2066L}). Despite this, our single-fluid shocks can be proxies of two-fluid shocks as a first approximation as their estimated thicknesses, $L_{\rm shock}$, for the gas conditions in Sgr B2 are $\sim 3-100$ times smaller than the cell size of our simulations ($\Delta x_i=0.16\,\rm pc$). Assuming molecular hydrogen number densities of $10^3\,\rm cm^{-3}-10^5\,\rm cm^{-3}$ with ionisation fractions $\chi_{\rm e}\sim 10^{-7}- 10^{-8}$ and magnetic field strengths $B/\mu G=b\,(n_{\rm H}/\rm cm^{-3})^{1/2}$ (\citealt{2010ApJ...725..466C}), where $b$ is the scaling parameter, we follow \cite{1980ApJ...241.1021D,1990MNRAS.246...98W} and estimate C-type shock thicknesses of the order of $0.010-0.050\,\rm pc$ for $b=1$ and $0.003-0.100\,\rm pc$ for $b=2$. These values imply magnetic precursor length scales consistent with those found in models with SiO \citep{Gusdorf08b,Serra2008,2012A&A...544A.150R}. Similarly, the inferred magnetic field strengths are $0.1-1.2\,\rm mG$ (see also \citealt{1996ApJ...462L..79C}), which correspond to magnetosonic speeds $v_{\rm ms}>5\times 10^3\,\rm km\, s^{-1}$ (see \citealt{1980ApJ...241.1021D}). This implies that all shocks for these conditions are C-type shocks as shock speeds of $5-50\,\rm km\,s^{-1}$ are characteristic in cloud-cloud collisions (see Section \ref{Shocks}).\par

The above calculations signify that: (i) two-fluid shocks in our simulations are unresolved and therefore indistinguishable from single-fluid shocks, and (ii) magnetohydrodynamical (MHD) cloud-cloud collision models would require numerical resolutions $200-2000$ cells per cloud radius to capture two-fluid shocks in conditions relevant for Sgr B2. Such requirement makes self-consistent 3D MHD simulations computationally prohibitive at this stage of our modelling, but future work along this line is warranted as previous studies suggest magnetic fields and radiative processes can also have dynamical effects not captured by our current models. For instance, magnetic fields can change both the dynamics of clouds moving through a more diffuse medium (e.g., see \citealt{2016MNRAS.455.1309B,2020ApJ...892...59C}) and the radiative signatures of shocks (\citealt{2016MNRAS.455.2066L}). If strong, they are also expected to affect the dynamics of a collision between clouds and the resulting star formation efficiency (e.g., see \citealt{2017ApJ...835..137W,2020ApJ...891..168W}). Similarly, the balance between heating driven by ion-neutral collisions and radiative cooling by molecular-line emission can alter the chemistry of the gas (\citealt*{1983ApJ...264..485D}) and have an impact on the energy dissipation and temperatures at shock fronts (\citealt*{2016MNRAS.463.1026L}), which our nearly isothermal equation of state does not capture.\par

Despite the above caveats, our idealised simulations do allow us to study: (i) the overall density structure that results from the collision between two clouds, (ii) the range of shock velocities (and Mach numbers) expected for such a region, both as a function of initial cloud geometry and time, and (iii) the structure of shocked gas and how this compares to the morphology of the shocked gas revealed by our molecular line observations. As we explain in the following sections, these aspects can be investigated by studying the hydrodynamical interaction of two colliding clouds, and they are important to build our understanding of Sgr B2 as well as inform future self-consistent models of SiO emission in cloud-cloud collisions.\par



\subsection{Evolution of cloud-cloud systems and gas kinematics}
\label{subsec:Num-results}
Having reviewed some numerical aspects of cloud-cloud systems, we now discuss the evolution of our collision models and how they compare to observations. The top panels of Figure \ref{Figure5} show the evolution of the number density (3D renderings) and the column number density (2D projections along the LOS) of cloud material in model FF, at different times, $0\leq t \leq 1.7\,\rm Myr$. The region of interest, which corresponds to the physical size of the observed region, is demarcated by a dashed, red square.\par

For comparison, in the bottom panels of Figure \ref{Figure5} we show the column number density maps of molecular hydrogen, calculated from the $^{13}$CO J=1-0 data cube (see Section \ref{observation}), assuming that the gas is optically thick (bottom left panel) and optically thin (bottom right panel). To obtain the H$_2$ column density map we first calculate a $^{13}$CO column density map following equations 12 and 13 in \cite{Heiderman10}. We assume a beam-filling factor of unity and a constant excitation temperature of 10 K, based on the average T$_{\rm ex}$ of 10 K derived by \cite{Martin08} for Sgr B2 from C$^{18}$O data. We calculate the optical depth for each pixel of the data cube, and then following equations 13 in \cite{Heiderman10} we calculate the $^{13}$CO column density. For the column density calculations we integrate the $^{13}$CO J=1-0 line intensity over the range of [-20,120] km s$^{-1}$. We obtain the optically-thick molecular hydrogen column density map from the $^{13}$CO column density map by assuming an H$_2$/$^{13}$CO ratio of 4$\times$10$^5$ \citep{Heiderman10}. For the optically-thin molecular hydrogen column density map we follow the same procedure, but we assume that the $^{13}$CO(1-0) emission is optically thin.\par

These panels show that the cloud-cloud collision consists of three global stages, in both models UU and FF. In the first stage, the initial contact triggers shocks in both clouds. Most shocks move forward, but some move backwards as gas is reflected from high-density cores. In the second stage, these shocks travel through the clouds heating them up and contributing to their lateral expansion. The softer polytropic index included in our models limits the expansion rate, thus mimicking the dynamical effects of radiative cooling. At the same time internal clumps continue colliding and mixing as vorticity is deposited at cloud-cloud and cloud-intercloud interfaces.\par

The collision between high-density gas in both clouds contributes to momentum transfer from the faster cloud (cloud U1/F1) to the slower cloud (cloud U2/F2). Thus, some of the gas in cloud U2/F2 starts accelerating, while gas in cloud U1/F1 decelerates. In the third stage the fastest transmitted shocks reach the rear side of the slower cloud and low-density, mixed gas is driven downstream. The exiting, fast-moving gas provides a signature of the initial velocity field of the faster cloud, while the slower gas that stays behind probes gas moving at speeds close to the initial rest-frame velocity of the slower cloud. The backflow of reflected shocks can explain the presence of shocked gas with negative radial velocities in the observations.\par

\subsection{The importance of turbulence}
\label{subsec:Turbulence}
Although the above stages are common to both models there are differences in the density structure and shocked gas morphology between models UU and FF. In model UU the collision is symmetric as the colliding clouds are spheres, while in model FF the collision produces a non-isotropic, asymmetric, turbulent structure. In model UU the high-column-density gas is located on the middle point, along the collision axis (i.e., along the LOS), while in model FF there are high-density cores anisotropically distributed across the collision area.\par

\begin{figure}
\begin{center}
  \begin{tabular}{c c c}
   \multicolumn{2}{c}{Model UU vs. model FF at $t=0.5\,\rm Myr$}\\
    \hspace{-0.3cm}{\resizebox{38mm}{!}{\includegraphics{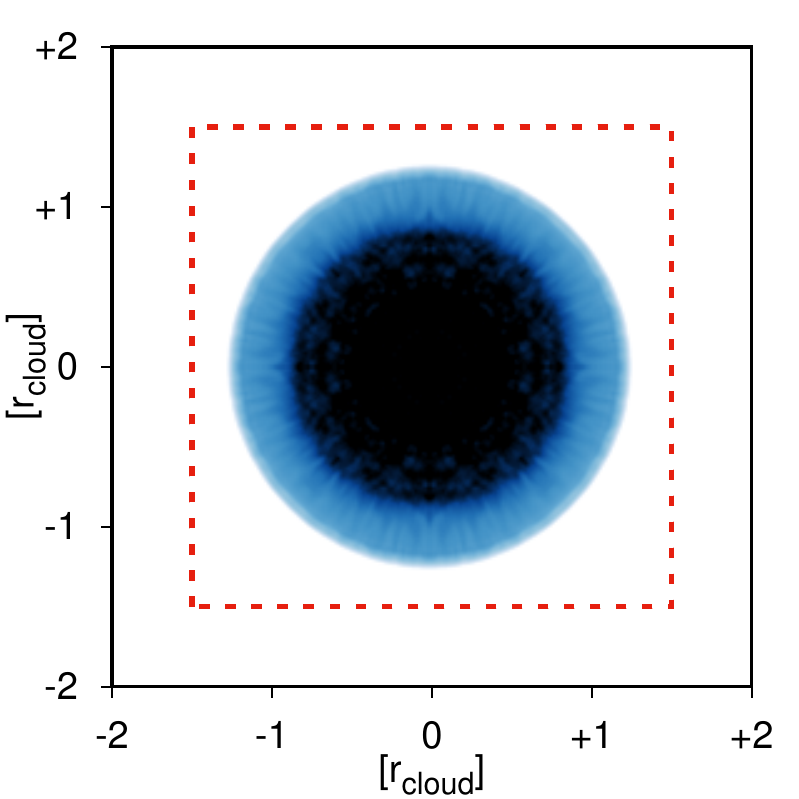}}} & \hspace{-0.5cm}{\resizebox{38mm}{!}{\includegraphics{nh006-eps-converted-to.pdf}}} & \hspace{-0.45cm}{\resizebox{9.5mm}{!}{\includegraphics{barra.png}}}\\
  \end{tabular}
  \caption{Density structure in model UU (left) and model FF (right) at $t=0.5\,\rm Myr$. Taking into account the turbulent nature of clouds is essential to reproduce the observational density maps.}
  \label{FigureA1}
\end{center}
\end{figure}

The panels in Figures \ref{Figure5} and \ref{FigureA1} show that the shocked and mixed gas in model FF precludes through the collision region more easily than in the model UU owing to the intrinsic porosity of the colliding clouds. Thus, when projected along the observer's LOS, model FF provides a better match to observations. In model FF, clouds F1 and F2 have differently-seeded log-normal density fields, so fast-moving gas hollows out of the slower cloud anisotropically. The collision stirs the gas in both clouds generating turbulence and shocks, which create regions of strong compression and also cavities, in a similar fashion to what has been reported in previous cloud-cloud collision models (e.g. see \citealt{2010MNRAS.405.1431A,2018ApJ...859..166F}) and to what we see in the observations of Sgr B2. Thus, we find that the collision between two uniform clouds (model UU) does not match the observations, and that considering the turbulent nature of clouds (as in the initial conditions of model FF) in collision models is crucial to understand the morphology and kinematics of the molecular gas in Sgr B2.\par

\subsection{Density structure in cloud-cloud collisions and in Sgr B2}
\label{subsec:PDFs}
\begin{figure}
\begin{center}
  \begin{tabular}{c}
    Column number density PDFs\vspace{-0.4cm}\\  
    \resizebox{80mm}{!}{\includegraphics{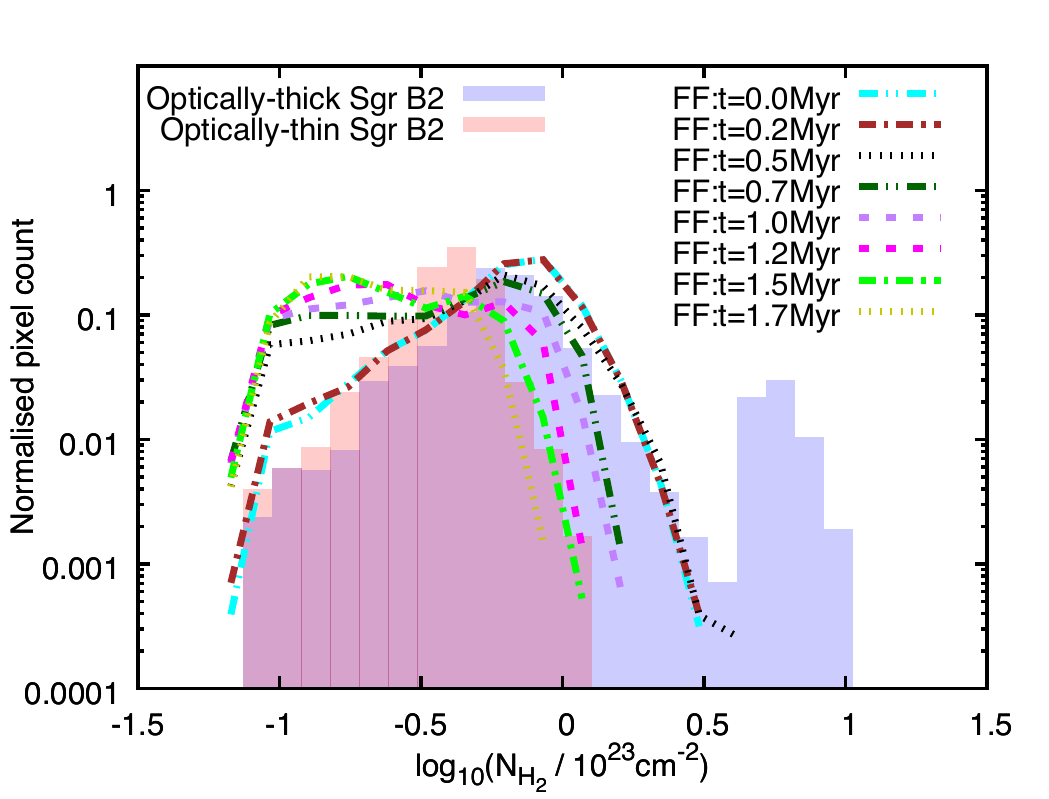}}\\
  \end{tabular}
  \caption{PDFs of the column number density in observations and simulations calculated for gas inside the red, dashed squares shown in Figure \ref{Figure5}. The optically-thick PDF (represented by blue boxes) is bi-modal, while the optically-thin PDF (represented by red boxes) is log-normal. The simulated PDFs are depicted at different times by different lines. The shape and peak show the closest resemblance to the optically-thick PDF for $t\leq 0.5\,\rm Myr$.} 
  \label{Fig_PDF}
\end{center}
\end{figure}

We also compare the PDFs of the column number density in observations and simulations (see Figure \ref{Fig_PDF}). Assuming the gas is optically thick, the PDF is bi-modal and has a mean column number density $\bar{N}_{\rm H_2}\sim 10^{23}\,\rm cm^{-2}$. In contrast, assuming the emission is optically thin, the PDF is log-normal and has a mean column number density of $\bar{N}_{\rm H_2}\approx 5\times 10^{22}\,\rm cm^{-2}$. In general, we find good agreement between the PDFs obtained from model FF and the observational PDFs. The shape and peak of the simulated PDFs keep close resemblance to the optically-thick PDF for times $t\leq 0.5\,\rm Myr$. At later times, the PDFs in the simulation develop a flatter plateau and even a secondary peak at low densities as mixed, diffuse gas becomes more abundant than high-density gas.\par

While the exact column density numbers depend on the modelling of the $^{13}$CO data and on the normalisation in our simulations, the density structure we observe in Sgr B2 is consistent with a scenario in which two clouds have collided. Additionally, the time evolution of the simulated PDFs also suggests that if a cloud-cloud collision triggers star formation, this should take place during the early stages of the interaction as high-density cores become less prominent at late times due to gas mixing, dynamical instabilities, and turbulence decay. 

\subsection{Shocked gas dynamics in cloud-cloud collisions}
\label{Shocks}
\begin{figure*}
\begin{center}
  \begin{tabular}{c c c c c}
    $[-5,10]\,\rm km\,s^{-1}$ & $[10,25]\,\rm km\,s^{-1}$ & $[25,40]\,\rm km\,s^{-1}$ & $[40,55]\,\rm km\,s^{-1}$\\
    \hspace{-0.48cm}\resizebox{42mm}{!}{\includegraphics{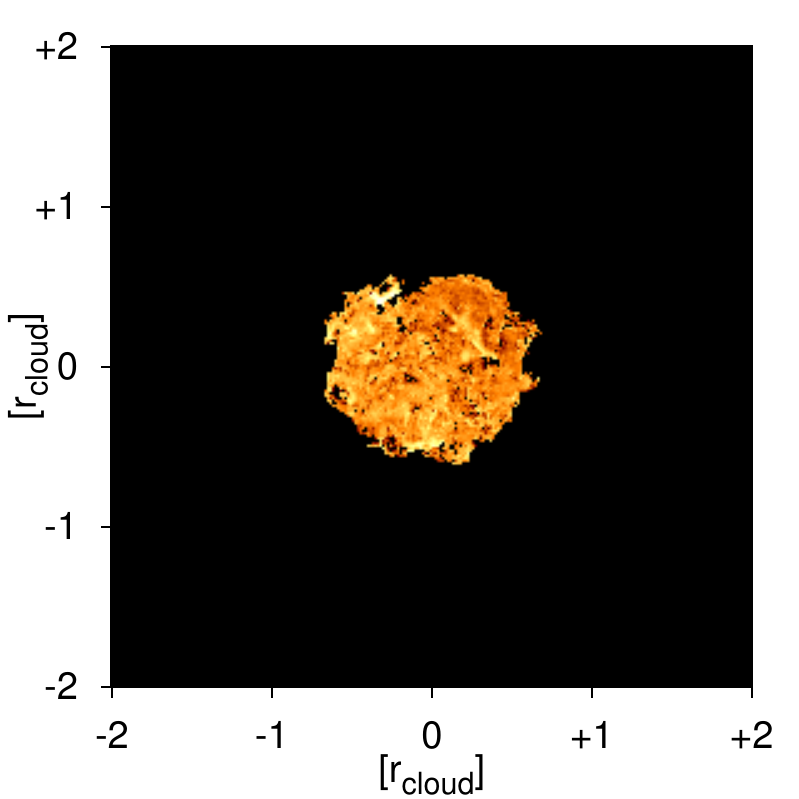}} & \hspace{-0.48cm}\resizebox{42mm}{!}{\includegraphics{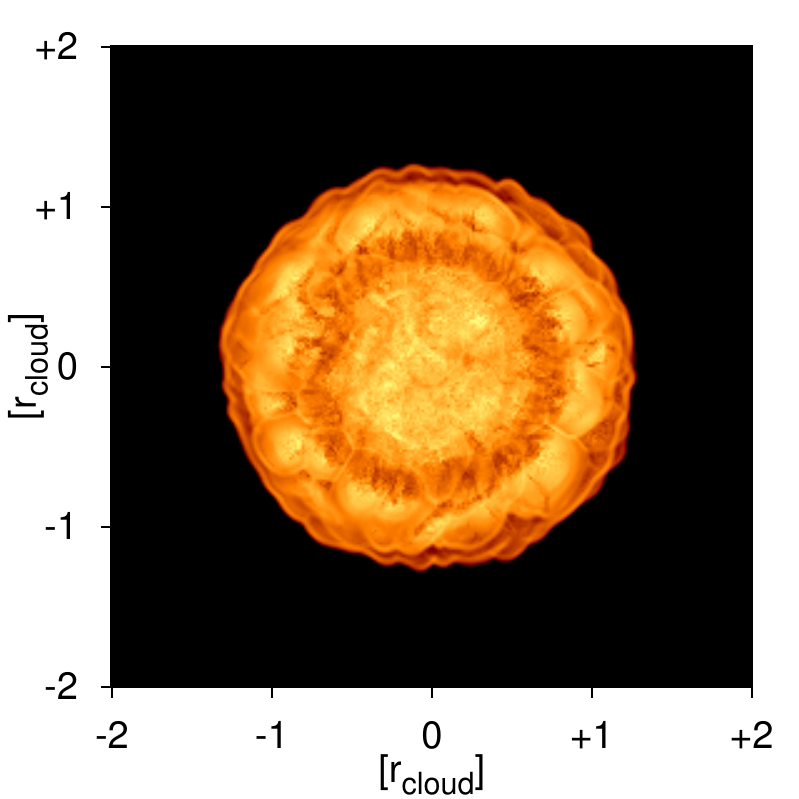}} & \hspace{-0.48cm}\resizebox{42mm}{!}{\includegraphics{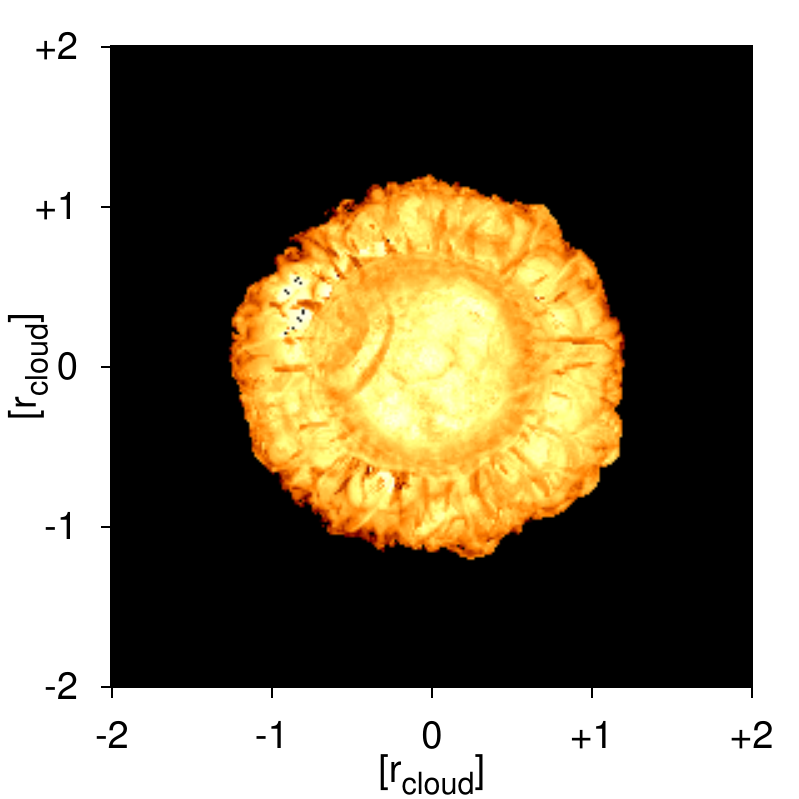}} & \hspace{-0.48cm}\resizebox{42mm}{!}{\includegraphics{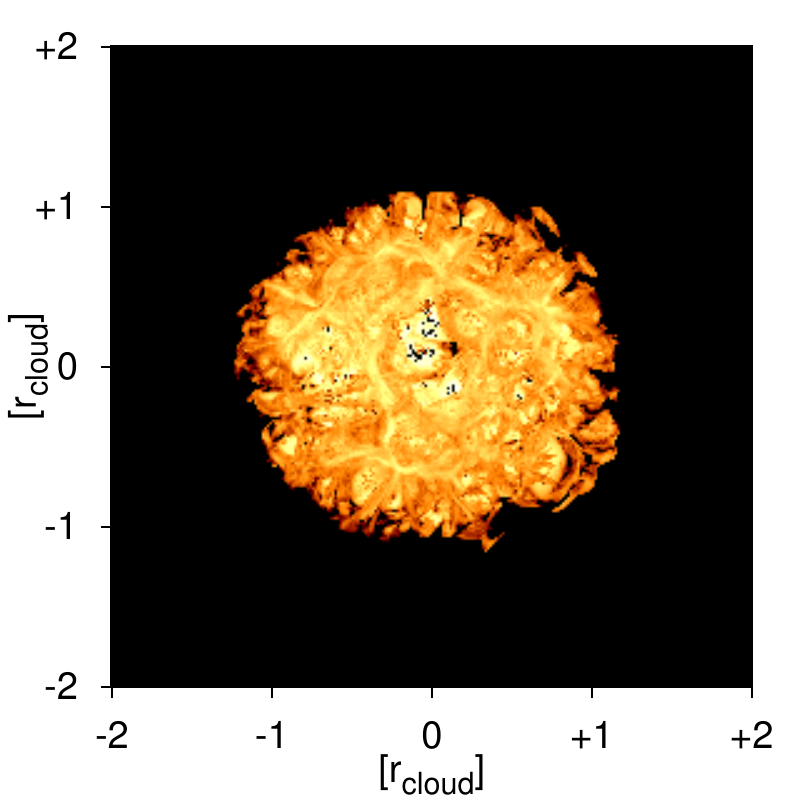}} & \hspace{-0.2cm}\resizebox{8mm}{!}{\includegraphics{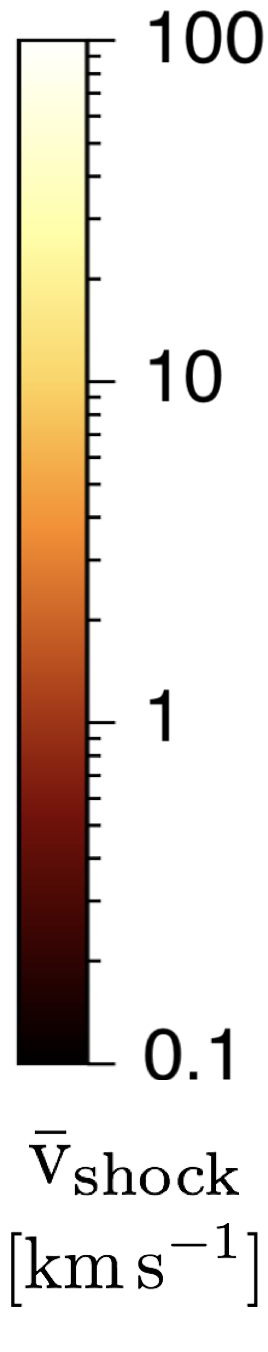}}\\
    $[55,70]\,\rm km\,s^{-1}$ & $[70,85]\,\rm km\,s^{-1}$ & $[85,100]\,\rm km\,s^{-1}$ & $[100,115]\,\rm km\,s^{-1}$\\
    \hspace{-0.48cm}\resizebox{42mm}{!}{\includegraphics{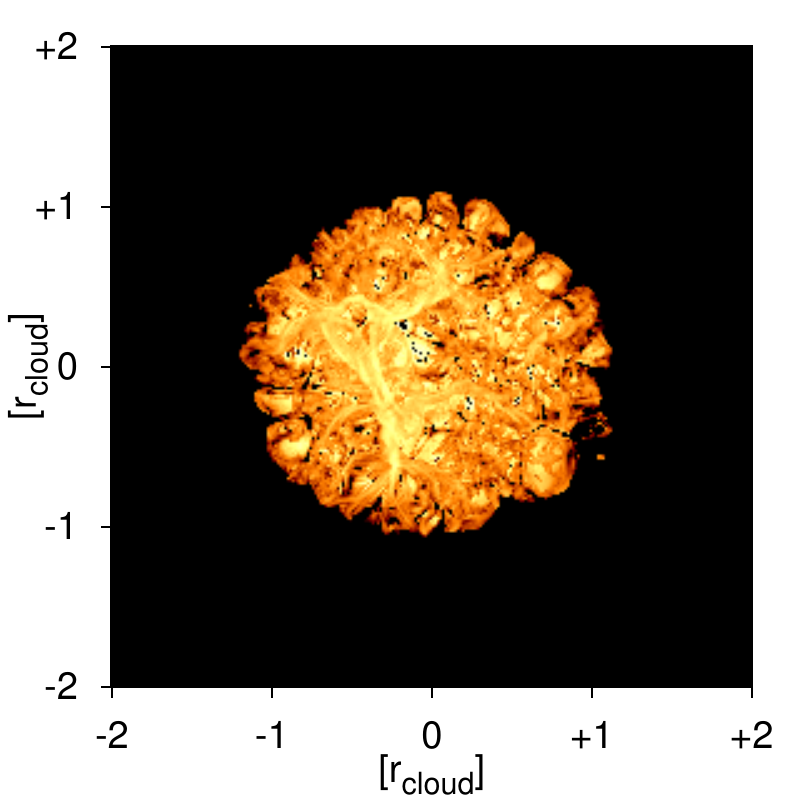}} & \hspace{-0.48cm}\resizebox{42mm}{!}{\includegraphics{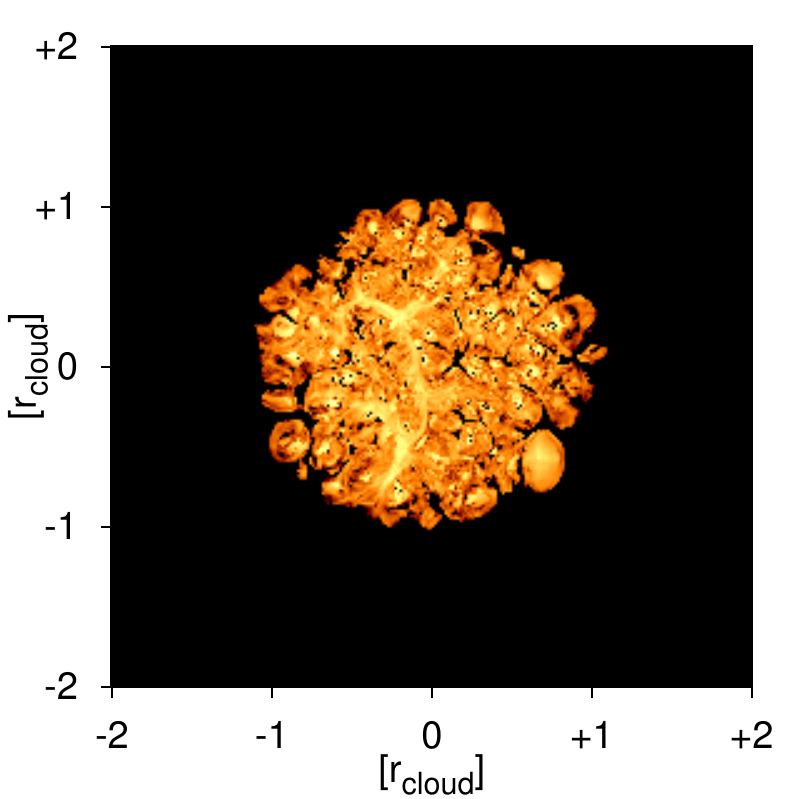}} & \hspace{-0.48cm}\resizebox{42mm}{!}{\includegraphics{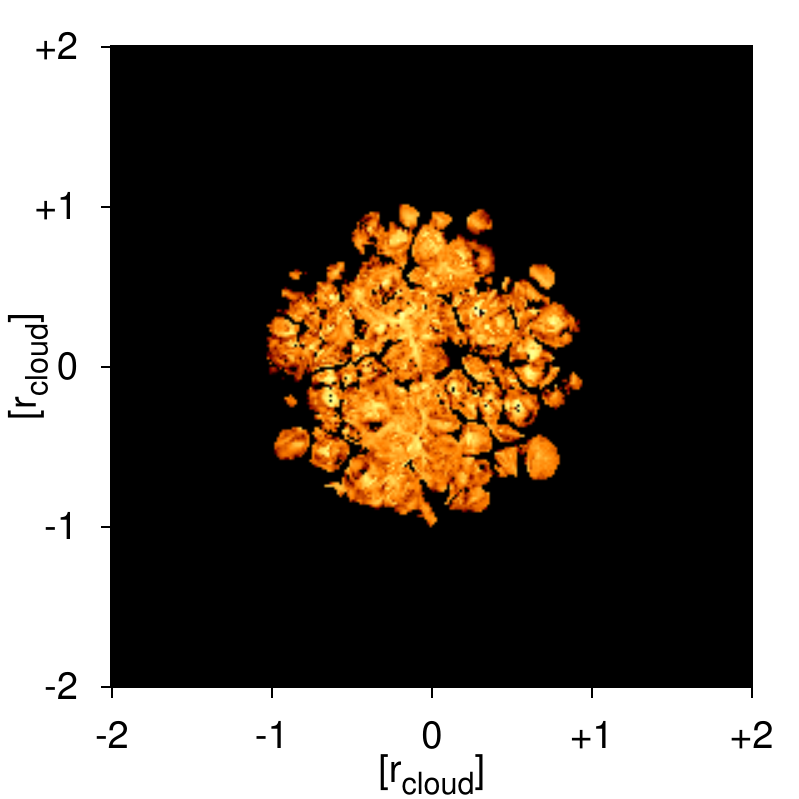}} & \hspace{-0.48cm}\resizebox{42mm}{!}{\includegraphics{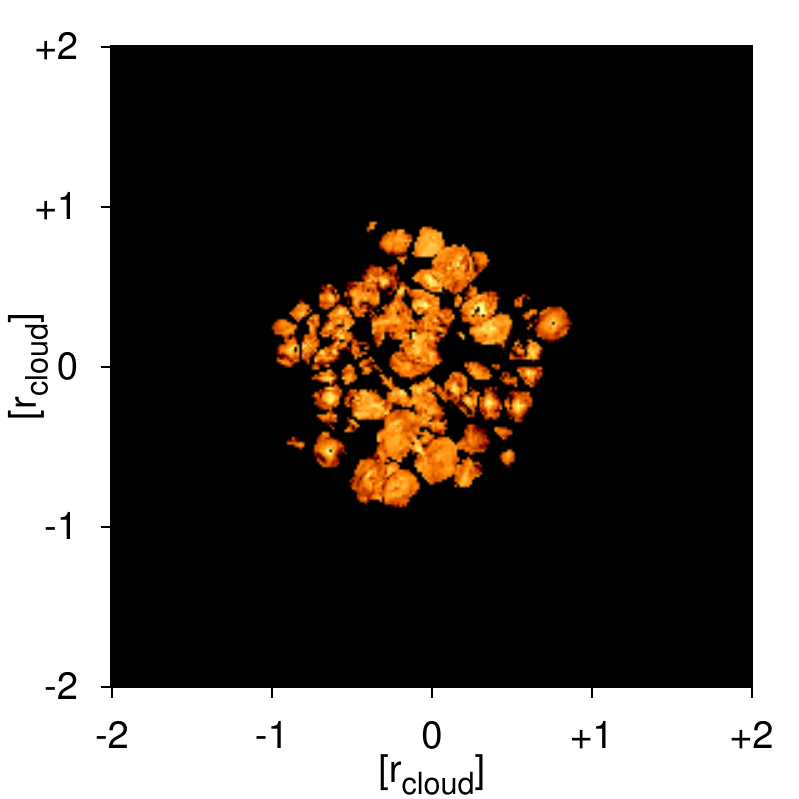}} &  \hspace{-0.2cm}\resizebox{8mm}{!}{\includegraphics{barshock.png}}\\
    \multicolumn{2}{l}{\hspace{+2.3cm} Shock velocity histogram at $t=0.5\,\rm Myr$} & \multicolumn{2}{c}{\hspace{-0.8cm} Shock velocity histogram at $t=1.7\,\rm Myr$}\vspace{-0.1cm}\\
    \multicolumn{2}{l}{\resizebox{80mm}{!}{\includegraphics{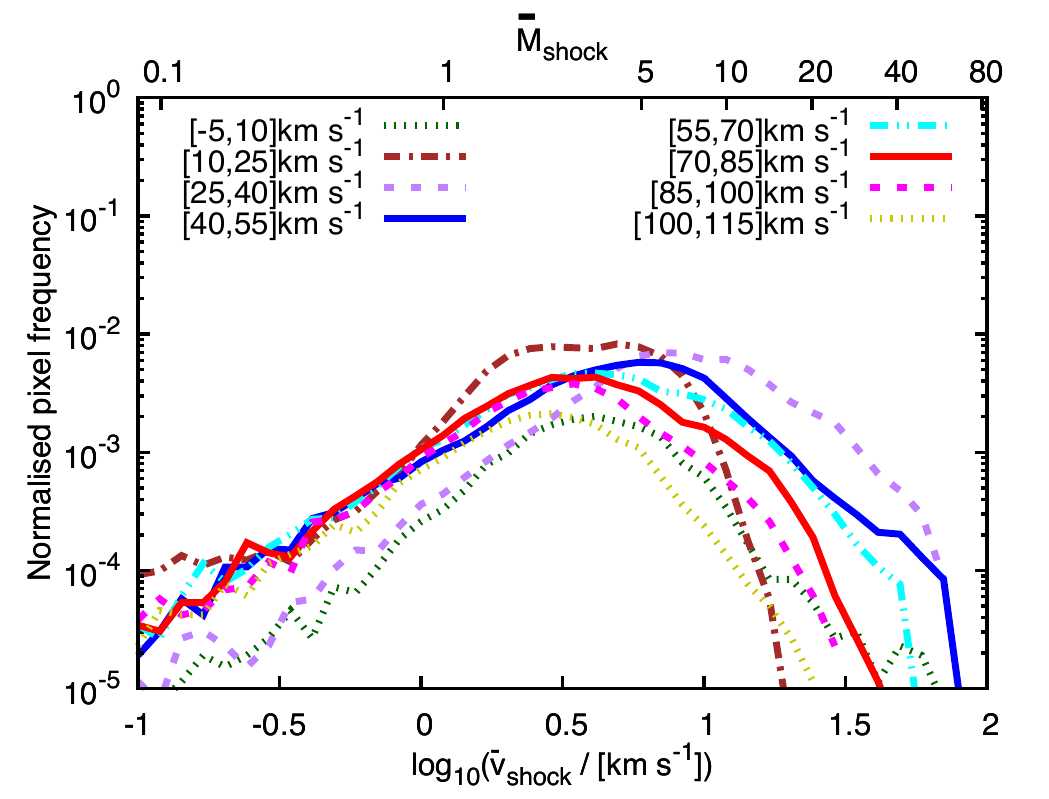}}} & \multicolumn{2}{l}{\resizebox{80mm}{!}{\includegraphics{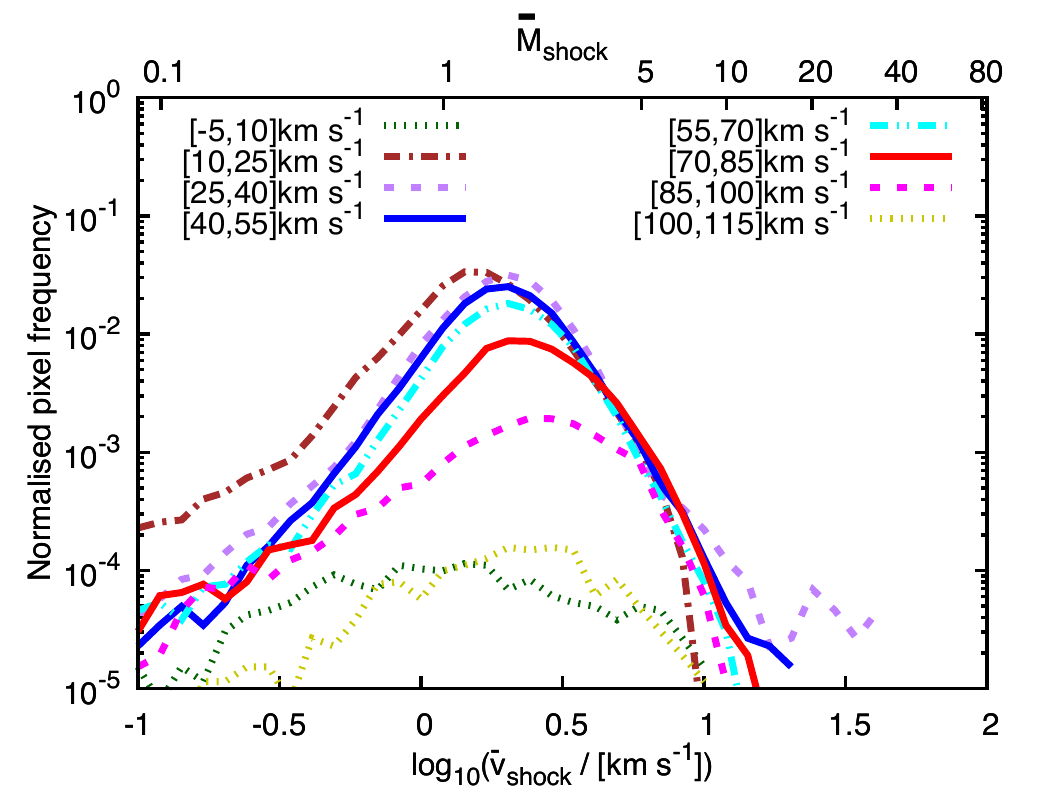}}}\vspace{-0.2cm}\\
  \end{tabular}
  \caption{Top panels: 2D maps of the structure of shocked gas in model FF at $t=0.5\rm \,Myr$. We show the average shock velocity in different velocity channels (selected to match those on the SiO maps in Figure \ref{fig:SiO_maps}). The shock velocities are in the range from $2\,\rm km\,s^{-1}$ (dark regions) to $100\,\rm km\,s^{-1}$ (bright regions), with typical values between $5\,\rm km\,s^{-1}$ and $40\,\rm km\,s^{-1}$. SiO molecules are expected to form on timescales of the order of $10^4$-$10^5\,\rm yr$ for these speeds. Most of the SiO emission is expected at radial velocities of $[25,55]\,\rm km\,s^{-1}$, which corresponds to gas accelerated from cloud F2 and decelerated from cloud F1. At higher speeds, less SiO is expected to form as gas from cloud F1 loses momentum while travelling across cloud F2. Bottom panels: shock velocity (Mach number) PDF obtained from model FF at $t=0.5\rm \,Myr$. Note that Mach numbers < 1 and their respective speeds correspond to waves, rather than shocks. The entire simulation can be viewed in an animation included in the electronic version of this paper.}
  \label{Figure6}
\end{center}
\end{figure*}

In this section we discuss the morphology and properties of shocked gas in model FF and how they compare with the SiO observations. While a direct comparison to the SiO emission intensity is not possible as we do not track the chemistry of the gas in our idealised computational models, an indirect comparison can be done as we do track the properties of shocked gas with a scalar tracer. In order to study the dynamics of shocked gas, Figure \ref{Figure6} shows 2D maps of the average internal shock velocities in model FF for some of the same velocity channels studied in our SiO observations.\par

To construct the Mach number/shock velocity maps shown in Figure \ref{Figure6} we first use a discontinuity finder routine (based on \citealt{2011MNRAS.418..960V} and \citealt{2016MNRAS.463.1026L}) to detect shocked cells in the computational domain. For this we search for cells containing cloud material, and then for convergent flows, i.e. ${\bm{\nabla \cdot v}}<0$, associated with large pressure gradients. Once a cloud cell has been flagged as potentially shocked, we calculate the associated Mach number from the velocity differences across the cell. In some cases we obtain $M\leq 1$, so such cells represent waves rather than shocks. To obtain the Mach numbers and shock/wave velocities we use a central difference method, in which we first calculate the derivatives in each direction and then sum the components to obtain the global Mach numbers. As a final step we average the Mach numbers and shock/wave velocities along the LOS and project them onto the maps shown in Figure \ref{Figure6}. Thus, these maps represent the average strength of shocks at time $t=0.5\,\rm Myr$, which corresponds to a time at which the column density maps and the shocked gas distribution are analogous in simulations and observations.\par

The bottom panels of Figure \ref{Figure6} show the histograms of shock/wave velocities and Mach numbers corresponding to this time, $t=0.5\,\rm Myr$, and for comparison to the latest time of the simulation, $t=1.7\,\rm Myr$. The full-time evolution of shocked gas can be viewed in the movies accompanying this paper. At $t=0.5\,\rm Myr$ shocks have velocities in the range from $\sim 2\,\rm km\,s^{-1}$ (dark regions) to $\sim 70\,\rm km\,s^{-1}$ (bright regions), with typical values between $5\,\rm km\,s^{-1}$ and $50\,\rm km\,s^{-1}$. These shock speeds correspond to Mach numbers between $4-42$. In addition, most of the shocks (and their associated SiO emission) are expected between $[25,70]\,\rm km\,s^{-1}$, where the lower limit roughly corresponds to the rest frame of the slower cloud F2, the in-between to mixed gas from both clouds, and the higher limit to gas decelerated from the faster cloud F1. Production of SiO at lower radial velocities is not significant as there is not much backflow, while at higher radial velocities, less SiO is expected to form as very little gas percolates through low-density gas in cloud F2.\par

\begin{figure}
\begin{center}
  \begin{tabular}{c}
    \hspace{-1.4cm}Number density versus shock speed\\  
    \resizebox{82mm}{!}{\hspace{-1.1cm}\includegraphics{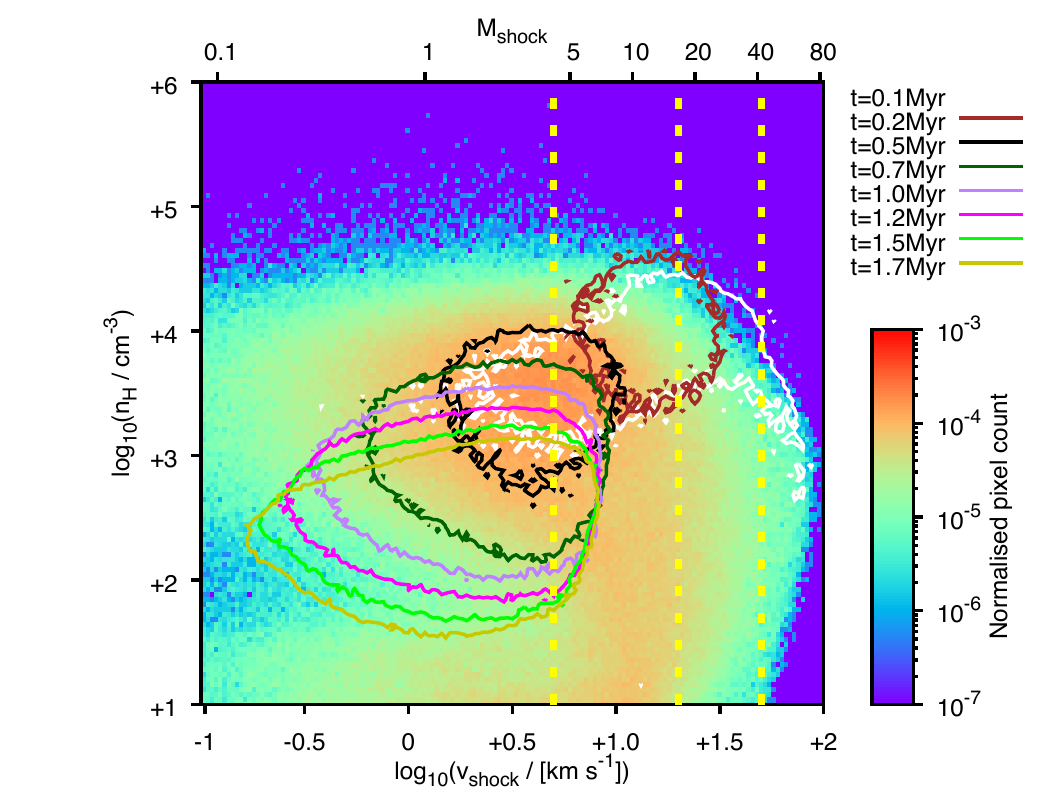}}\\
    \hspace{-1.4cm}Column number density versus shock speed\\  
    \resizebox{82mm}{!}{\hspace{-1.1cm}\includegraphics{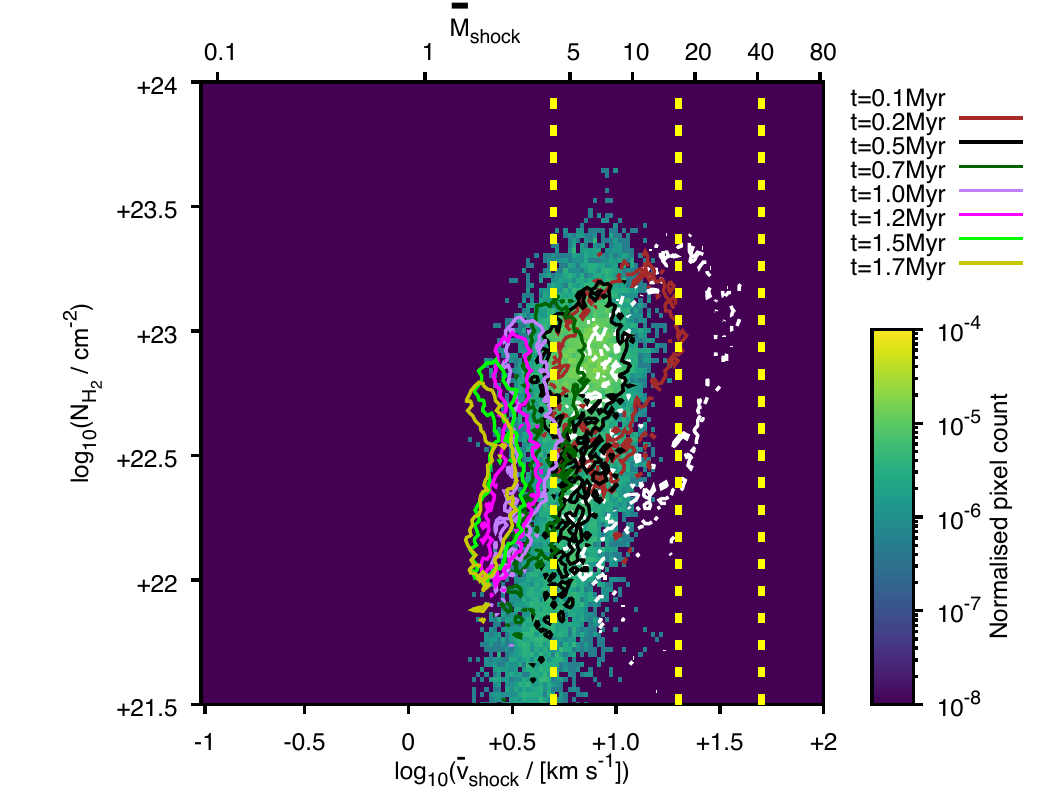}}\\
  \end{tabular}
  \caption{Top panel: 2D histogram of the gas number densities, $n_{\rm H}$, versus shock velocities, $v_{\rm shock}$, for $t=0.5\,\rm Myr$. Bottom panel: 2D histogram of the gas column number densities, $N_{\rm H_2}$, versus the average shock velocities, $\bar{v}_{\rm shock}$, along the LOS (see Figure \ref{Figure6}). The contours in both panels show the time evolution of the bulk of shock velocities for $t=0.1-1.7\,\rm Myr$, and the yellow lines correspond to $v_{\rm shock}=5$, $20$, and $50\,\rm km\,s^{-1}$, i.e., to the velocity ranges where C-type shocks are dominant and can produce SiO efficiently. A low number of high-velocity shocks are produced early on in the collision, while a high number of moderate- to low-velocity and very-low-velocity shocks become prevalent for $t>0.2\,\rm Myr$ and $t>0.5\,\rm Myr$, respectively. The entire sequence can be viewed in movies included in the electronic version of this paper.} 
  \label{NH2vsVshock}
\end{center}
\end{figure}

In contrast, at $t=1.7\,\rm Myr$ the PDFs are dominated by very-low-velocity shocks with typical values between $2\,\rm km\,s^{-1}$ and $5\,\rm km\,s^{-1}$, which correspond to Mach numbers between $1.5-4$. At this time, very little shocked gas remains in the velocity channels $[-5,10]\,\rm km\,s^{-1}$ and $[100,115]\,\rm km\,s^{-1}$, and the kinematics of shocked gas is restricted to velocities closer to the post-collision rest frame of the cloud-cloud system, i.e. to the $[25,70]\,\rm km\,s^{-1}$ range. To visualise the global correlation between gas densities and shock speeds, we can analyse 2D histograms of these variables at different times. Figure \ref{NH2vsVshock} reports the 2D histograms of $n_{\rm H}$ (top panel) and $N_{\rm H_{2}}$ (bottom panel) versus $v_{\rm shock}$ and its LOS average, respectively, for $t=0.5\,\rm Myr$, jointly with $>10^{-4}$-level and $>5\times10^{-6}$-level contours, respectively, for times between $0.1-1.7\,\rm Myr$.\par

These panels show that the amount of shocks goes up rapidly after the impact and that the shock distribution is dominated by high-velocity shocks ($v_{\rm shock}>50\,\rm km\,s^{-1}$) only in the very early stages of the cloud-cloud interaction at $t<0.2\,\rm Myr$. This is when the relative velocity between gas in both clouds is the highest and when gas from both clouds start colliding and producing regions of strong compression and rarefaction, which are characteristic of supersonic turbulence. As time progresses, the shock distribution moves towards lower shock velocities (and Mach numbers) as the collision-induced turbulence steadily dissipates. Between $t=0.2-0.5\,\rm Myr$ the distribution is dominated by moderate- and low-velocity shocks ($v_{\rm shock}=5-50\,\rm km\,s^{-1}$), and at late times for $t>0.5\,\rm Myr$ by very-low-velocity shocks ($v_{\rm shock}=2-5\,\rm km\,s^{-1}$) and subsonic waves.\par
This analysis constrains the time scales for the production of SiO to the availability of shocks with the appropriate velocities and pre-shock densities to trigger either grain core or grain mantle sputtering. On the one hand, high-velocity shocks ($v_{\rm shock}>50\,\rm km\,s^{-1}$) only need $10^2-10^3\,\rm yr$ to produce enough SiO to explain the abundances estimated for Sgr B2 as they can very efficiently produce SiO via core sputtering (which generally needs $v_{\rm shock}>20\,\rm km\,s^{-1}$). Thus, they could favour a scenario where Sgr B2 is very young, $<10^4\,\rm yr$. However, our simulation indicates that such high-velocity shocks are only available at the very beginning of the cloud-cloud interaction. Thus, while possible, we think it would be difficult for shocks of such speeds to explain all the wide-spread SiO emission seen in all channels. On the other hand, moderate- and low-velocity shocks ($v_{\rm shock}=5-50\,\rm km\,s^{-1}$) need longer timescales $10^3-10^5\,\rm yr$ to produce the observed SiO abundance, but they are much more common, wide-spread, and long-lived in our simulation than high-velocity shocks. Thus, we think that these moderate- and low-velocity shocks are the most likely sources for the ubiquity of SiO emission in this region.\par

Moderate- and low-velocity shocks, associated with pre-shock densities between $10^3-10^4\,\rm cm^{-3}$, are common during the first $\approx 0.5\,\rm Myr$ of the cloud-cloud collision, so they favour a scenario where Sgr B2 is the site where a collision has been taking place for $\lesssim 0.5\,\rm Myr$ and SiO has been continuously replenished by shocks with the appropriate speeds to trigger grain mantle sputtering, which only needs shocks speeds $>7\, \rm km\, s^{-1}$ (if the medium is pre-enriched with Si-bearing material) and can produce broad SiO lines (\citealt{Gusdorf08a}). The colliding clouds could have been moving along the same orbit/stream at different speeds, in neighbouring orbits/streams, or along bar-driven inflows onto the CMZ (e.g., see \citealt{Sormani18}). Thus, our results are consistent with the standard dynamical models for the GC's CMZ by \citealt{1991MNRAS.252..210B,Molinari2011} and by \citealt{2015MNRAS.447.1059K,2019MNRAS.484.5734K}. However, in the latter comparison, SiO emission and our analysis in this paper do point to a local origin for the extensive star formation in Sgr B2.\par



\subsection{Additional cloud-cloud collision signatures}
\label{Signatures}
Cloud-cloud collisions have distinct signatures when they occur in the interstellar medium (ISM), such as complementary spatial distributions between the colliding gas, complementary kinematics revealed by first-moment maps, and bridging `V-shaped' features visible in position-velocity diagrams (see \citealt{2015MNRAS.450...10H,2018ApJ...859..166F}). As highlighted by \cite{2018ApJ...859..166F} these features are short-lived ($<1\,\rm Myr$) during a collision and they would fade after triggered star formation and stellar feedback introduces additional stirring in the gas. Thus, if found in observations, these features can show that a recent cloud-cloud collision has occurred.\par

\begin{figure*}
\begin{center}
  \begin{tabular}{c c c c}
     \hspace{-1cm}P-V diagram at $t=0.4\,\rm Myr$ & \hspace{-1cm}P-V diagram at $t=0.5\,\rm Myr$ & \hspace{-1cm}P-V diagram at $t=1.7\,\rm Myr$ & \vspace{-0.3cm}\\
    \resizebox{48mm}{!}{\hspace{-1cm}\includegraphics{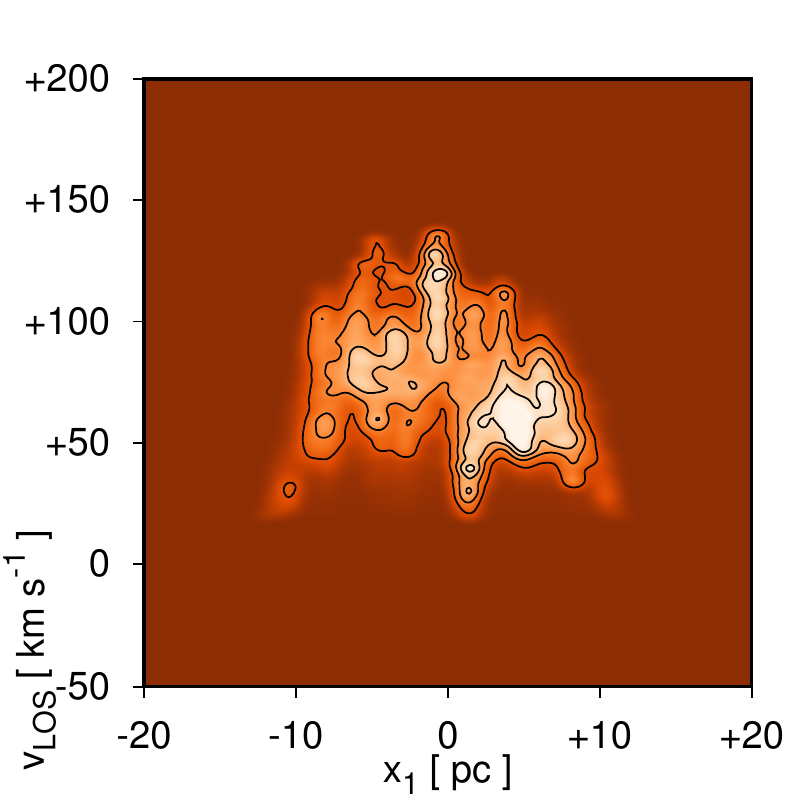}} & \resizebox{48mm}{!}{\hspace{-1cm}\includegraphics{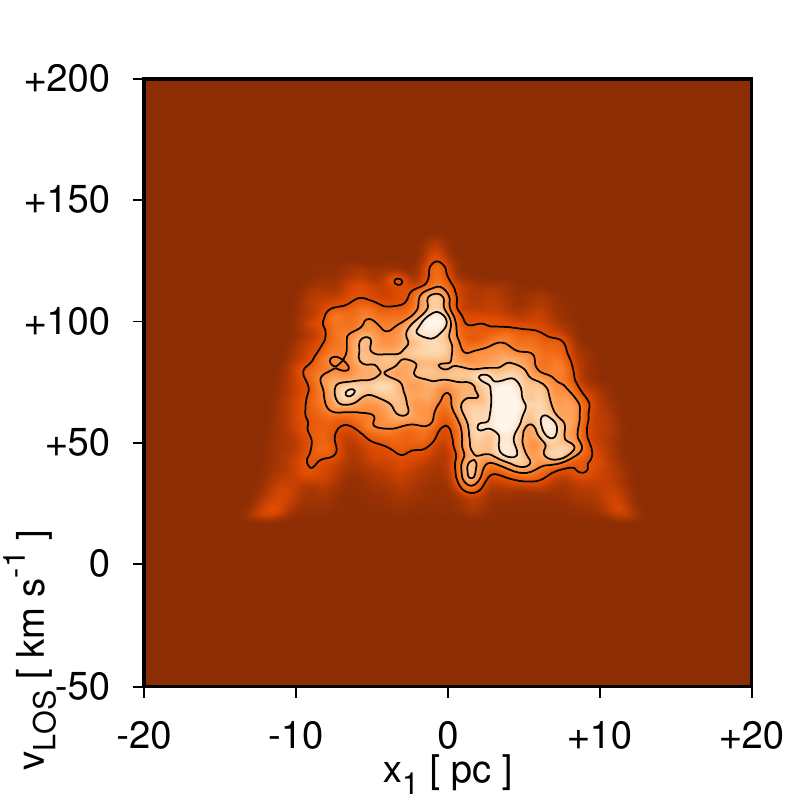}} & \resizebox{48mm}{!}{\hspace{-1cm}\includegraphics{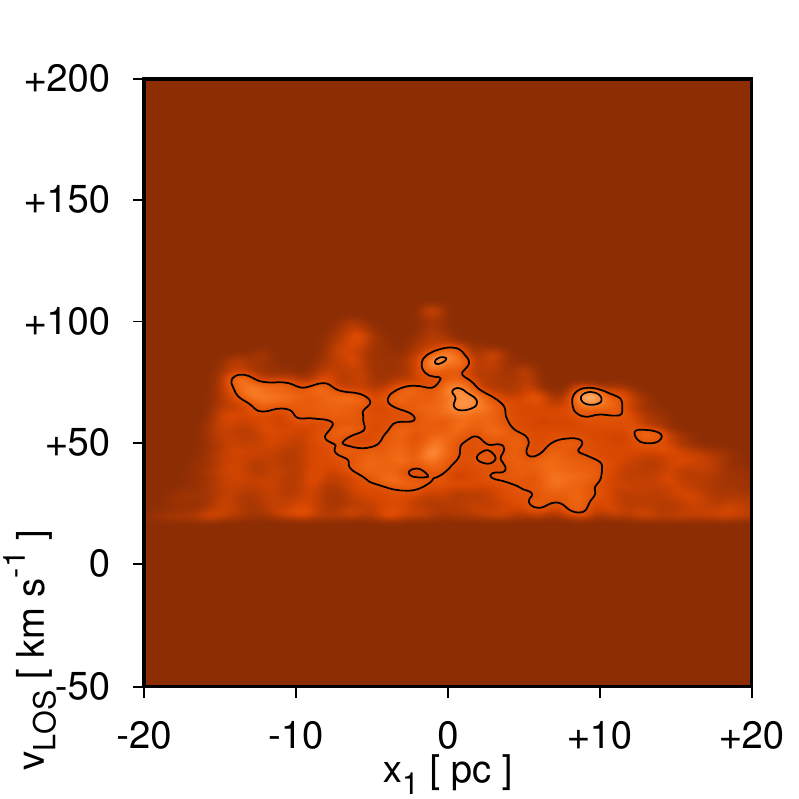}} & \resizebox{9.4mm}{!}{\hspace{-1.2cm}\includegraphics{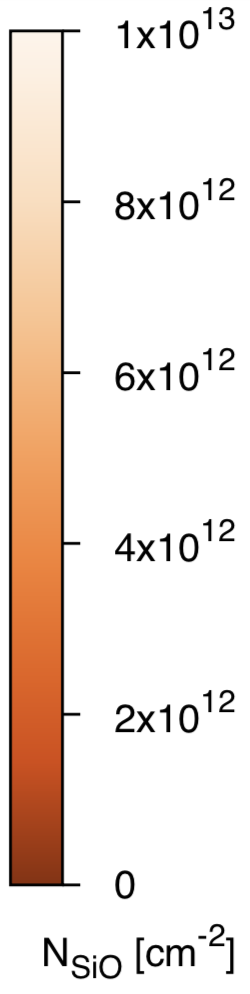}}\\
     \hspace{-1cm}P-V diagram at -28:22:03 & \hspace{-1cm} P-V diagram at Dec: -28:24:25 & \hspace{-1cm}Diagonal P-V diagram & \\  
    \resizebox{51mm}{!}{\hspace{-1.2cm}\includegraphics{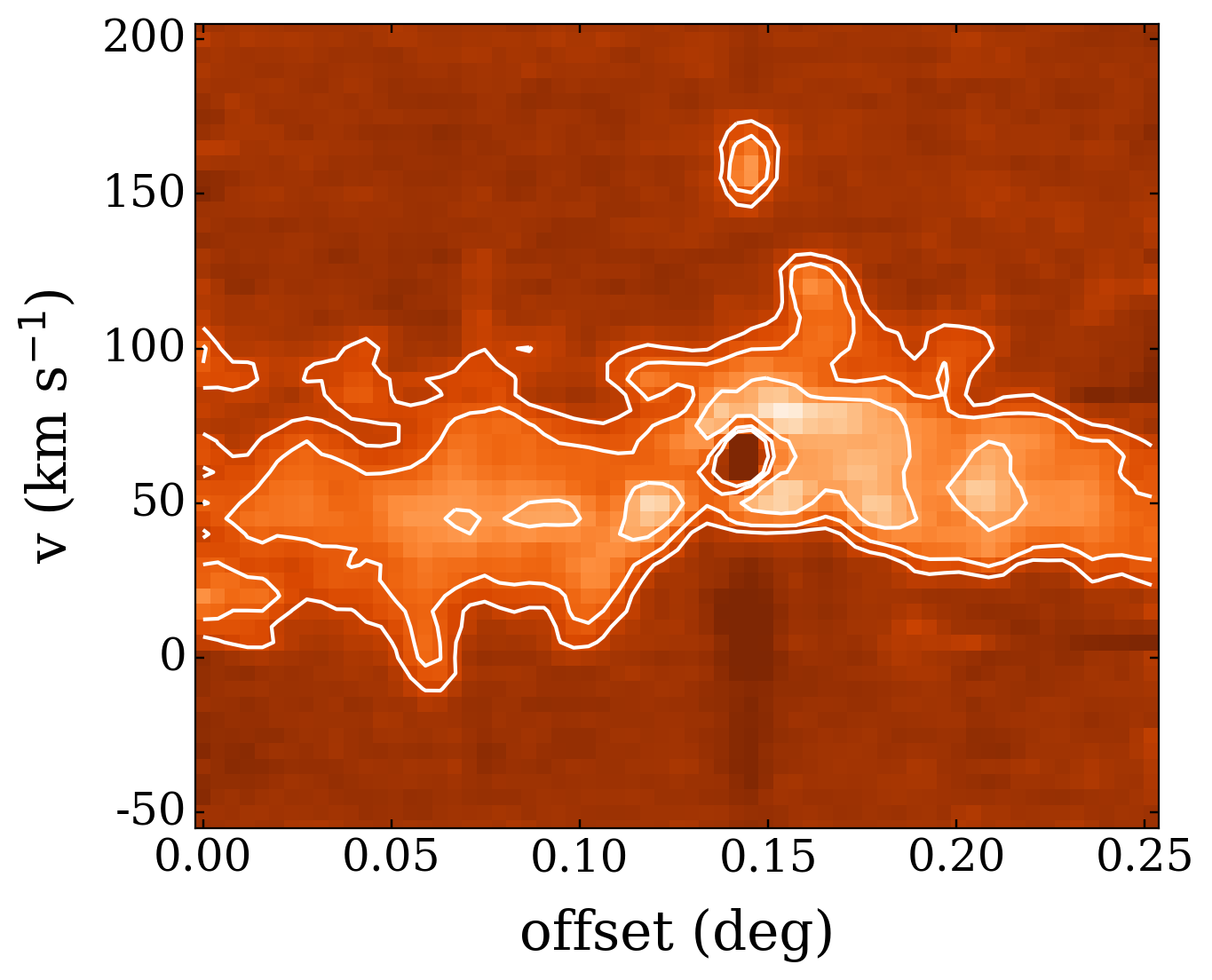}} & \resizebox{51mm}{!}{\hspace{-1.2cm}\includegraphics{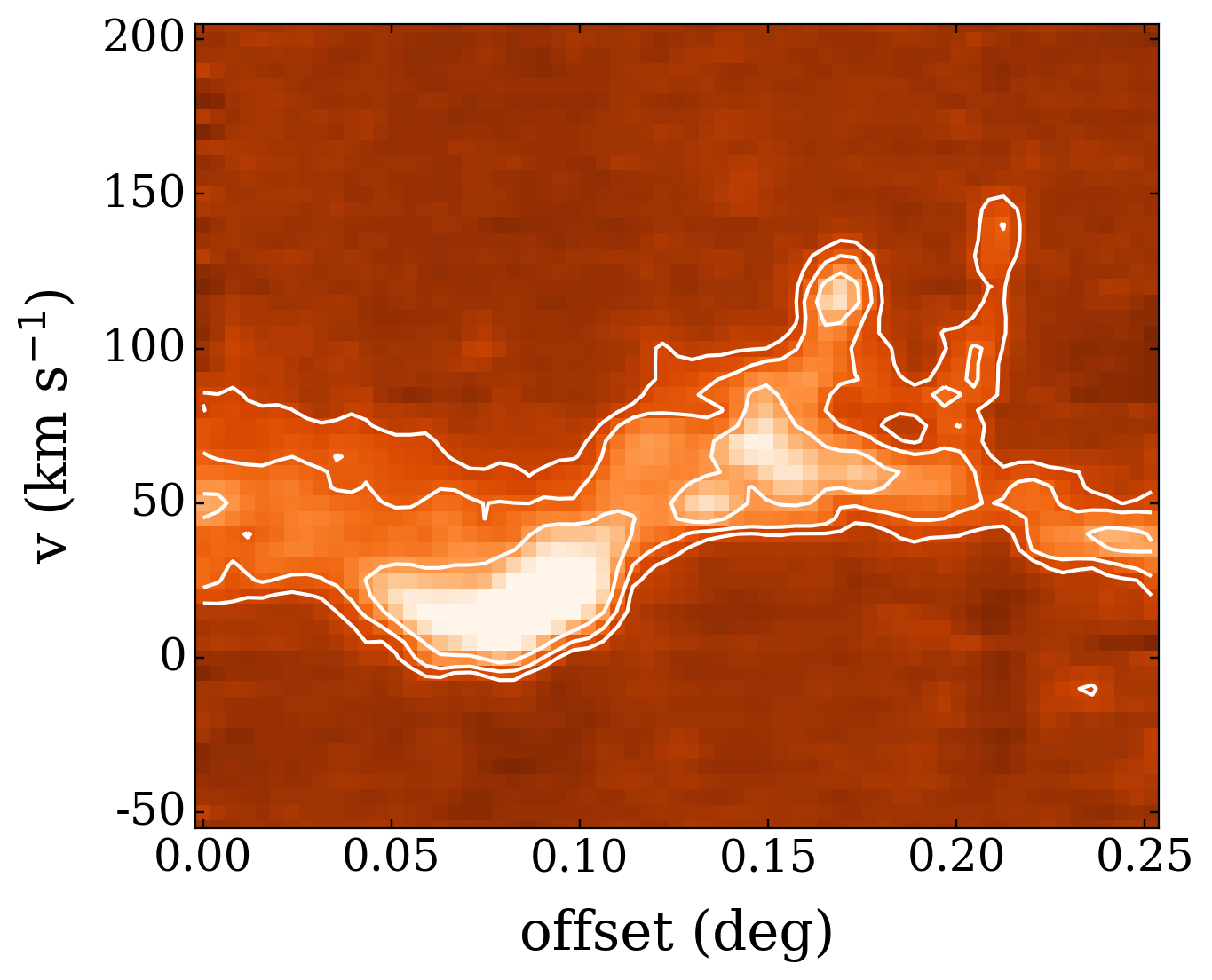}} & \resizebox{49.5mm}{!}{\hspace{-1.2cm}\includegraphics{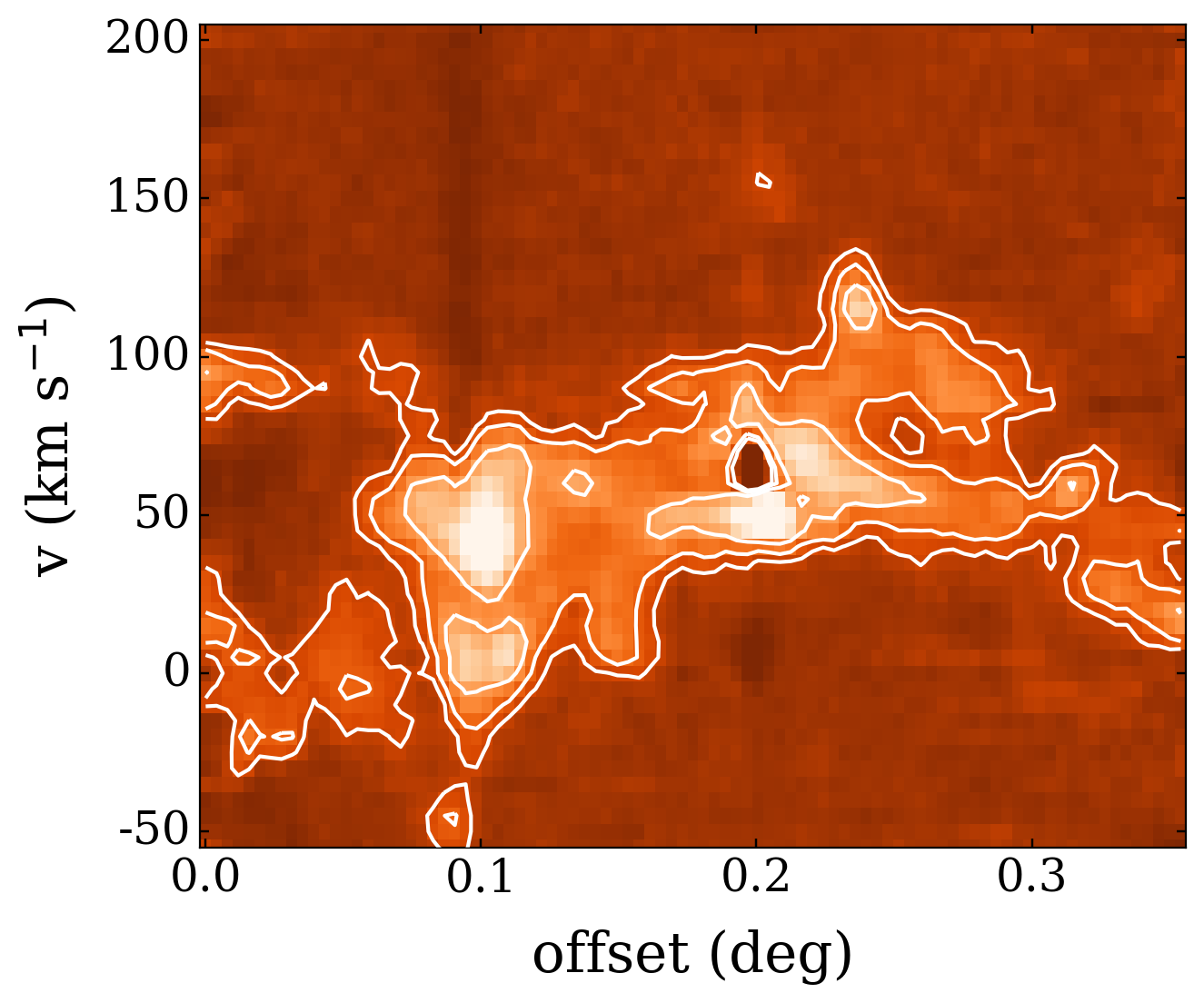}} & \resizebox{7.4mm}{!}{\hspace{-1.2cm}\includegraphics{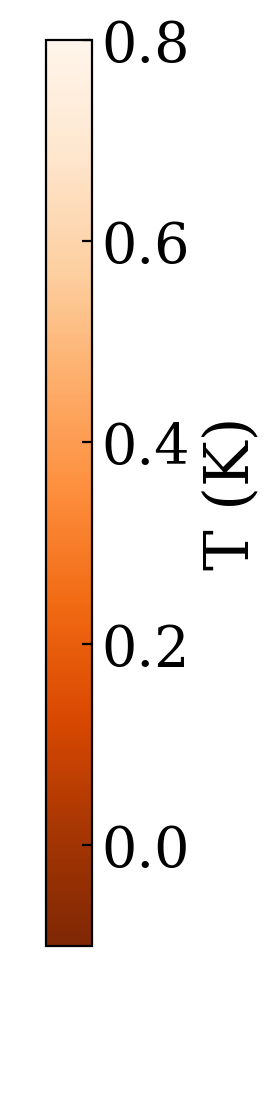}}\\
  \end{tabular}
  \caption{Top panels: Position-velocity (P-V) diagrams in our simulation FF (top panels) for three different times, $t=0.4\,\rm Myr$, $t=0.5\,\rm Myr$, and $1.7\,\rm Myr$, respectively. Bottom panels: P-V diagrams for three different declination slices across the SiO emission map. Bridging 'zig-zag' features with V-shaped ends at their turning points can be viewed during the early stages of the cloud-cloud collision in the simulation, and in all of the presented observational maps. The existence and shape of these features provide further evidence for a cloud-cloud collision in Sgr B2. Since these features vanish with time, Sgr B2 must be younger than $\approx 0.5\,\rm Myr$. The full-time sequence can be viewed in a movie included in the electronic version of this paper.} 
  \label{PPVplots}
\end{center}
\end{figure*}

Our analysis of both observations and simulations supports this view and favours a cloud-cloud collision scenario for Sgr B2. As discussed in Section \ref{Maps_sio}, we find a complementary distribution between shocked gas at $[10,25]\,\rm km\,s^{-1}$ and gas at $[70,85]\,\rm km\,s^{-1}$. In addition, we also find features akin to the bridging features reported by earlier studies in position-velocity (P-V) diagrams (\citealt{2017ApJ...835..142T}). Figure \ref{PPVplots} shows our P-V diagrams for three different times in our simulation FF, $t=0.4\,\rm Myr$, $0.5\,\rm Myr$, $1.7\,\rm Myr$ (top row) and for three different declination slices in our SiO observations (bottom row). The simulation maps show that `zig-zag' bridges (with 'V-shaped' features at their turning points) between colliding parcels of gas are characteristic during the early stages of the collision, while they vanish at late stages. Similar bridging features can also be seen in the observations, where high-velocity gas appears to be connected to medium-velocity gas and this to low-velocity gas via `zig-zag'-shaped bridges.\par

The multiplicity of bridging features on these diagrams suggests that several cloudlets are interacting in the collision as expected from turbulent clouds. `Zig-zag' shapes are more accentuated in the diagonal slice (bottom right panel), which was done along the Galactic plane. This may indicate that interacting shocked gas lies primarily on the Galactic plane. Similarly, the bridging features connecting high- and medium-velocity gas in the observations appear to be fainting at this stage of the interaction. In our simulations such features faint for times $>0.7\,\rm Myr$, which also adds an additional time constraint for shocked gas in this region. This is in agreement with our previous findings and with previous studies on other systems, which suggest that V-shaped features diffuse away after $1\,\rm Myr$ (e.g., \citealt{2018ApJ...859..166F}).\par

\subsection{Implication for star formation in Sgr B2}
\label{SF_SgrB2}
These results also have important implications for the star formation history in this region. The gas kinematics, its density structure, and multiple shock diagnostics are consistent with a scenario in which two clouds have been colliding for $t\approx0.5\,\rm Myr$. Our modelling suggests that star formation in Sgr B2 was triggered by high-velocity shocks in compressed gas during the initial stages of the collision between two clouds, in agreement with earlier results by \cite{Hasegawa1994,Sato00}. Our simulations show that during the first $0.2\,\rm Myr$ of the cloud-cloud interaction, the compression due to high-velocity shocks can readily produce gas with hydrogen number densities of $>10^{5}\,\rm cm^{-3}$. The free-fall time of such gas is only $0.1\,\rm Myr$, so assuming that the observed compact HII regions needed another $0.1\,\rm Myr$ to reach their sizes, our dynamical time scale is still consistent with an overall age of $\lesssim0.5\,\rm Myr$ for Sgr B2. Similarly, observations of stars and the inferred star-formation age of Sgr B2 are in agreement with this result, as most stars found in this region are relatively young O- and B-type stars, and the ultra-compact HII regions have estimated ages $10^3-10^5\,\rm yr$ (\citealt{1995ApJ...449..663G}).\par

Our estimated age for Sgr B2 ($t\approx0.5\,\rm Myr$) from our analysis on SiO emission is lower than the estimate by \citealt{2015MNRAS.447.1059K}, who proposed star formation in this region was ignited during the last pericentre passage of this cloud. Based on this dynamical model, \cite{Ginsburg2018} gives a lower estimate of $0.4\,\rm Myr$ taking a feature known as `the Brick' (G0.253+0.016, whose low star formation can be attributed to the solenoidal driving of turbulence; see \citealt{2016ApJ...832..143F}), rather than the pericentre passage, as a reference; and \citealt{2017MNRAS.469.2263B} provides a broader time scale of $0.5-0.9\,\rm Myr$, assuming that clouds remain quiescent for $0.3-0.5\,\rm Myr$ after star formation is triggered by tidal compression. Our analysis and modelling on SiO emission favour a local origin for star formation in Sgr B2.

\section{Conclusions}
We have presented a comprehensive study of the structure and kinematics of shocked gas in Sgr B2. By comparing molecular line observations of this region with empirically-constrained numerical simulations, we have found that a cloud-cloud collision scenario is consistent with the observed morphology and kinematics of shocked gas in this region. To summarise:

\begin{itemize}
    \item We show integrated intensity maps of SiO J=2-1 towards Sgr B2, which reveal shocked gas with a turbulent substructure featuring an arc labeled as A at velocities of [-20,10] km s$^{-1}$ and three cavities labeled as B, C and D at [10,40] km s$^{-1}$ (see Figure~\ref{fig:SiO_maps}). Cavities B, C and D are observed towards the western edge of a known shell in Sgr B2. The morphology of these features is consistent with the sub-structure of gas formed by turbulent stirring resulting from a large-scale cloud-cloud collision.
    \item The northern and eastern edges of the SiO gas at velocities of [70,85] km s$^{-1}$ lie along cavity B at velocities of [10,25] km s$^{-1}$ (see Figure~\ref{fig:SiO_maps}). The spatial anti-correlation of low- and high-velocity gas supports the idea that a large-scale collision between two clouds has happened in Sgr B2.
    \item Known (ultra-)compact HII regions of Sgr B2 are too small to justify a large-scale multi-bubble scenario for this region. In the context of shock production, their sizes suggest stellar feedback is dynamically important on scales ranging from sub-parsecs to parsecs in Sgr B2. HII regions spatially overlap with some sites of strong SiO emission at [40,85] km s$^{-1}$, and they are between or along the edges of the SiO gas features at [100,120] km s$^{-1}$. This supports the idea that the formation of the stars responsible for the ionization of the compact HII regions happened in gas compressed by colliding flows.
    \item We determine the velocity, integrated line intensity, and line width distribution of the SiO gas by decomposing all of the spatial pixels of the SiO data cube. We find 3 peaks in the velocity distribution of SiO at $\sim$ 25, 45 and 80 km s$^{-1}$, and an average velocity of 47 $\pm$ 2 km s$^{-1}$. For the integrated line intensity, we find a mean of $11 \pm 2\,\rm K\,km\,s^{-1}$; and for the line widths (FWHM), we find a mean of 31 $\pm$ 5 km s$^{-1}$ and a log-normal distribution with a peak at 21 $\rm km\,s^{-1}$. These values of integrated intensity and line widths indicate that moderate- and low-velocity shocks are dominant in this region.
    \item From analysing OCS emission, we find an H$_2$ density of $\sim$10$^{5}$ cm$^{-3}$ and kinetic temperatures around 30 K for dense gas in three positions of Sgr B2. From analysing $^{13}$CO emission we obtain $\bar{N}_{\rm H_2}\sim 10^{23}\,\rm cm^{-2}$ assuming the emission is optically thick, and $\bar{N}_{\rm H_2}\sim 5\times 10^{22}\,\rm cm^{-2}$ assuming the emission is optically thin.
    \item We find an average SiO abundance of $\sim$10$^{-9}$ for Sgr B2 by comparing C$^{18}$O with SiO emission. The abundance, the gas temperature of $\sim$30 K, and the integrated SiO J=2-1 line intensities with a mean of 11 K km s$^{-1}$ agree with the predictions of chemical models of grain sputtering by C-type shocks. In particular, our results agree well with models of a similar system in the GC, with a cosmic-ray ionization rate within 10$^{-16}$-10$^{-15}$ s$^{-1}$, shock velocities within 5-50 km s$^{-1}$, and production timescales of $>10^3$ years (for $v_{\rm shock}\geq25\,\rm km\,s^{-1}$) and $>$10$^5$ years (for $v_{\rm shock}<25\,\rm km\,s^{-1}$).
    \item Our numerical modelling reveals that a cloud-cloud collision can explain the structure and kinematics of shocked gas in Sgr B2. Our models show that most of the SiO is produced for $0.1\,\rm Myr\lesssim t\lesssim 0.5\,\rm Myr$ at all velocities, and predominantly in the $[25,40]\,\rm km\,s^{-1}$ and $[40,55]\,\rm km\,s^{-1}$ velocity channels. Less emission is expected at high velocities, $[85,100]\,\rm km\,s^{-1}$ and $[100,115]\,\rm km\,s^{-1}$, as gas slows down after the collision. The backflow caused by reflected shocks produces shocked gas at $[-20,10]\,\rm km\,s^{-1}$.
    \item In our model, high-velocity shocks ($v_{\rm shock}>50\,\rm km\,s^{-1}$) ignite star formation during the first $\sim 0.2\,\rm Myr$, while moderate- and low-velocity shocks ($v_{\rm shock}=5-50\,\rm km\,s^{-1}$) are responsible for the wide-spread SiO emission up until $\sim 0.5\,\rm Myr$. Internal Mach numbers of $4-42$ are typical in such a collision.  
    \item By simultaneously comparing gas densities and shock velocities in observations and simulations, we constrain the collision age to $\lesssim0.5\,\rm Myr$. Longer time scales are disfavoured as column densities and shock speeds decrease below the limits at which grain sputtering is efficient.
    \item The aforementioned conclusion is also supported by the presence of `zig-zag' features with `V-shaped' turning points in position-velocity diagrams of SiO emission. Our simulations show that such features are only detectable during the early stages ($<0.7\,\rm Myr$) of the collision.
\end{itemize}    

Sgr B2 is a star-forming region with a complex gas kinematics. Using SiO maps we have shown that turbulent stirring associated with a cloud-cloud collision provides a simple and viable solution to explain the kinematics of gas and the presence of shocks and star formation in this region. Exploring models where MHD simulations are coupled to sub-grid C-type shock models and with recipes for sink particles and self-gravity would be interesting to complement this view on Sgr B2 in the future.

\section*{Acknowledgements}
We gratefully acknowledge discussions with Andrew Lehmann and Fabien Louvet, and thank the anonymous referee for very helpful and constructive feedback. This work is based on observations carried out under project number 137-14 with the IRAM 30m telescope. IRAM is supported by INSU/CNRS (France), MPG (Germany) and IGN (Spain). WBB is supported by the Deutsche Forschungsgemeinschaft (DFG) via grant BR2026125, and by the National Secretariat of Higher Education, Science, Technology, and Innovation of Ecuador, SENESCYT. C.~F.~acknowledges funding provided by the Australian Research Council (Discovery Project DP170100603 and Future Fellowship FT180100495), and the Australia-Germany Joint Research Cooperation Scheme (UA-DAAD). The authors gratefully acknowledge the Gauss Centre for Supercomputing e.V. (\url{www.gauss-centre.eu}) for funding this project (pn34qu) by providing computing time on the GCS Supercomputer SuperMUC-NG at Leibniz Supercomputing Centre (\url{www.lrz.de}). Part of the numerical work presented here was conducted on the Hummel supercomputer at Universit\"at Hamburg. This work has made use of the pyFC package by A. Y. Wagner (available at \url{https://bitbucket.org/pandante/pyfc}) to generate log-normal, fractal clouds for the initial conditions in the simulations, the VisIt visualisation software \citep{HPV:VisIt}, the gnuplot program (\url{http://www.gnuplot.info}), {\sc matplotlib} \citep{Hunter2007}, {\sc NumPy} \citep{vanderWalt2011}, and {\sc Astropy}, a community-developed core {\sc Python} package for Astronomy (\citealt{Astropy, Astropy_2}; \url{http://www.astropy.org}).

\section*{Data availability}
The data underlying this article will be shared on reasonable request to the corresponding author.



\bibliographystyle{mnras}

\bibliography{mnras_template.bib} 




\appendix
\section{SiO abundances in selected positions}
\label{AppendixA1}
\begin{figure*}
	\includegraphics[width=18cm]{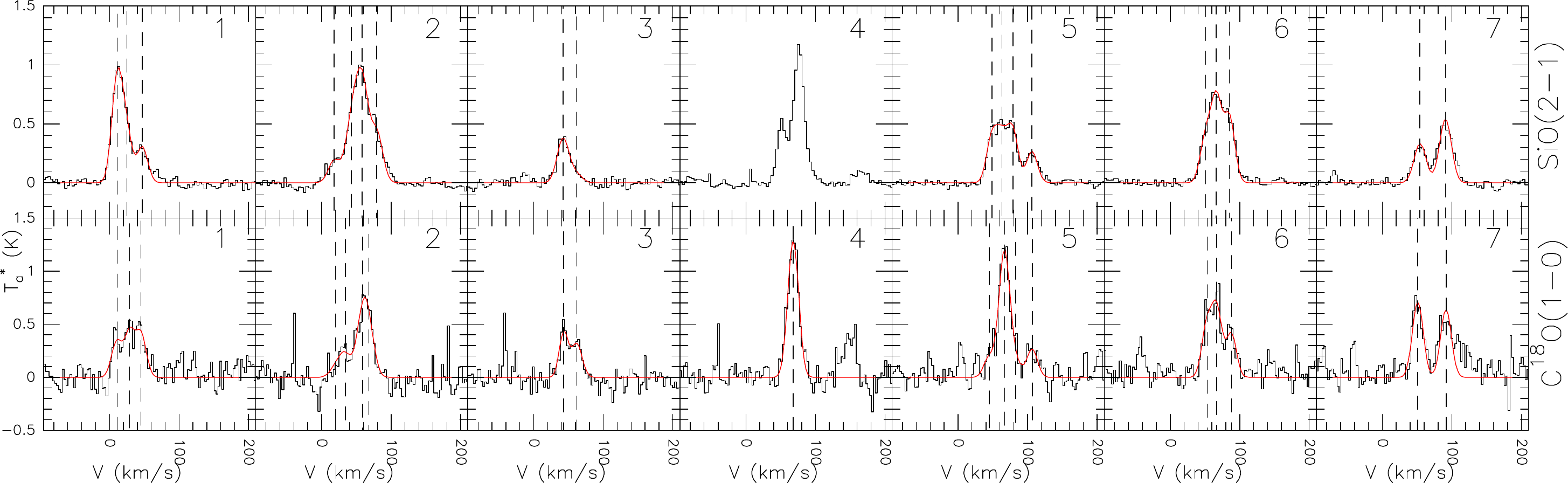}
    \caption{SiO J=2-1 (top panels) and C$^{18}$O J=1-0 (bottom panels) lines observed towards seven positions of Sgr B2 indicated in Figure~\ref{fig:SiO_maps}. The LTE best fits obtained with MADCUBA are shown with red lines. The SiO J=2-1 line is not fitted in position 4 as this line is self-absorbed. The dashed lines show the different velocity components identified in the LTE best fits.}
    \label{fig:SiO_C18O_spectra}
\end{figure*}

Figure~\ref{fig:SiO_C18O_spectra} shows SiO J=2-1 and C$^{18}$O J=1-0 spectra extracted over 28 arcsec regions (the HPBW of the SiO data cube) centered on the seven selected positions shown in Figure~\ref{fig:SiO_maps}. 
We observe multiple velocity components in the SiO J=2-1 and C$^{18}$O J=1-0 line profiles towards almost all positions. 
As seen in Figure~\ref{fig:SiO_C18O_spectra}, the SiO J=2-1 and C$^{18}$O J=1-0 lines in positions 1-3 reveal peak intensities within $\sim$10-80 km s$^{-1}$, while both lines in positions 5-7 show peak intensities within $\sim$50-110 km s$^{-1}$.
The SiO J=2-1 line profile towards position 4 appears to show two velocity components, but the C$^{18}$O J=1-0 line reveals only one velocity component, thus demonstrating that the SiO J=2-1 line is affected by self-absorption.
We use the C$^{18}$O molecule to derive hydrogen column densities required to infer the fractional abundances of SiO. Ground state lines of HCO$^+$, HCN, and HNC show also self-absorption effects towards the Sgr B2M and Sgr B2N cores \citep{Jones_2008}, which is due to the presence of a hot and diffuse envelope around both cores \citep{Vicente97,Martin1999}. It is expected that only the SiO J=2-1 line of position 4, between all the seven positions, is affected by self-absorption because this position is the closest to the hot cores Sgr B2M and Sgr B2N, where SiO J=2-1 absorption is detected (see Figure~\ref{fig:SiO_maps}).

We have derived SiO and C$^{18}$O column densities by using the AUTOFIT tool of the MADCUBA software, fitting synthetic spectra to the observed spectra under LTE conditions and considering line optical depth effects.
In our analysis, the linewidths of both molecules is assumed to be equal to 20 km s$^{-1}$, which is the typical linewidth observed in GC clouds
\citep{Amo2011,Armijos2015} and the FWHM peak in Figure \ref{Histograms}, while the T$_{\rm ex}$ are assumed to be equal to 10 K and 7 K for C$^{18}$O and SiO, respectively, as in Section \ref{SiOabundances}.
Multiple velocity components were required to fit the line profiles observed towards positions 1-3 and 5-7. Our best fits to the SiO J=2-1 and C$^{18}$O J=1-0 lines are shown in Figure~\ref{fig:SiO_C18O_spectra}. The SiO J=2-1 and C$^{18}$O J=1-0 lines were fitted separately with the AUTOFIT tool. 
We find optical depths of $\leq$0.3 for the C$^{18}$O J=1-0 and SiO J=2-1 lines towards the seven positions. The line optical depth of 0.3 for the SiO J=2-1 line in position 4 is biased as its real peak intensity is unknown. The column density of SiO is a lower limit for position 4 as the SiO J=2-1 line is affected by self-absorption. This value is derived over the velocity range of [40,100] km s$^{-1}$ and considering the same assumptions as above. The derived LSR velocities and column densities (N) are listed in Table \ref{tab:Column_density}. The last column of this table gives the SiO relative abundance calculated as the N$_{\rm SiO}$/N$_{\rm H_2}$ ratio, where N$_{\rm H_2}$ is estimated from N$_{\rm C^{18}O}$ using the $^{16}$O/$^{18}$O isotopic ratio of 250 \citep{Wilson94} and the relative abundance of CO to H$_2$ of 10$^{-4}$ \citep{Frerking1982}. We find SiO abundances within (0.2-3.2)$\times$10$^{-9}$ for the different velocity components and the seven positions of Sgr B2.

\begin{table*}
	\caption{Studied positions and parameters of C$^{18}$O and SiO derived for seven positions of Sgr B2.}\label{tab:Column_density}
	\begin{threeparttable}
	\begin{tabular}{cccccccc} 
	\hline
	Pos.& $\rmn{RA}(J2000)$ & $\rmn{Dec.}(J2000)$ &\multicolumn{2}{c}{C$^{18}$O} & \multicolumn{3}{c}{SiO}\\
	\cline{4-5}
	\cline{6-8}
&&  &V$_{\rm LSR}$ & N$_{\rm C^{18}O}$ & V$_{\rm LSR}$ & N$_{\rm SiO}$ & X\tnote{(a)} \\ 	
&&  &(km s$^{-1}$) & ($\times$10$^{15}$ cm$^{-2}$) & (km s$^{-1}$) & ($\times$10$^{13}$ cm$^{-2}$) & ($\times$10$^{-9}$) \\
		\hline
1 & $17^{\rmn{h}} 47^{\rmn{m}} 41\fs272$ & $-28\degr 25\arcmin 07\farcs 80$  & 10.8$\pm$2.0 & 6.8$\pm$1.1 & 10.9$\pm$0.5 & 2.3$\pm$0.2 & 1.3$\pm$0.2\\
		  &&& 28.7$\pm$2.3 & 7.1$\pm$1.2 & 24.6$\pm$1.6 & 5.7$\pm$0.2 &3.2$\pm$0.6\\
		  &&& 44.8$\pm$1.8 & 8.3$\pm$1.3 & 47.3$\pm$0.8 & 1.4$\pm$0.2 & 0.7$\pm$0.1\\
		  \hline
	2 & $17^{\rmn{h}} 47^{\rmn{m}} 21\fs839$ & $-28\degr 20\arcmin 36\farcs 52$  & 20\tnote{(b)} & 1.8$\pm$1.0 & 18.2$\pm$1.5 & 0.7$\pm$0.1 &1.7$\pm$0.9\\
		 & && 33.7$\pm$1.8 & 4.6$\pm$0.7 & 42.8$\pm$1.5 & 1.9$\pm$ 0.2 &1.7$\pm$0.3\\
		 & && 58.8$\pm$0.6 & 11.7$\pm$0.6 & 57.9$\pm$1.0 & 5.0$\pm$0.2&1.7$\pm$0.1\\
		 & && 68.1$\pm$1.2 & 7.2$\pm$0.7 & 79.3$\pm$0.8 & 2.3$\pm$0.1&1.3$\pm$0.1\\\hline
	3 & $17^{\rmn{h}} 46^{\rmn{m}} 57\fs697$ & $-28\degr 23\arcmin 58\farcs 11$ & 43.1$\pm$0.9 & 6.9$\pm$0.6 & 42.2$\pm$0.6 & 1.9$\pm$0.1&1.1$\pm$0.1\\
		 & && 61.9$\pm$1.3 & 6.2$\pm$0.7 & 61.5$\pm$3.5 & 0.3$\pm$0.1&0.2$\pm$0.1\\\hline
	4 & $17^{\rmn{h}} 47^{\rmn{m}} 21\fs862$ & $-28\degr 21\arcmin 26\farcs 29$ & 68.1$\pm$0.4 & 26.9$\pm$1.0 & [40,100]\tnote{(c)} & >6.8\tnote{(c)}&>1.0\tnote{(c)}\\\hline
	5 & $17^{\rmn{h}} 47^{\rmn{m}} 12\fs089$ & $-28\degr 22\arcmin 33\farcs 72$ & 45.0$\pm$3.5 & 3.8$\pm$1.1 & 48.9$\pm$1.1 & 2.3$\pm$0.1&2.4$\pm$0.7\\
		 & && 66.7$\pm$1.0 & 25.1$\pm$1.8 & 63.2$\pm$2.0 & 1.8$\pm$0.1&0.3$\pm$0.03\\
		 & && 83.2$\pm$7.9 & 2.6$\pm$1.6 & 78.6$\pm$0.9 & 1.2$\pm$0.1&1.9$\pm$1.2\\
		 & && 106.6$\pm$2.4 & 5.1$\pm$1.0 & 106.4$\pm$0.7 & 1.2$\pm$0.1&0.9$\pm$0.2\\\hline
	6 & $17^{\rmn{h}} 47^{\rmn{m}} 18\fs752$ & $-28\degr 24\arcmin 57\farcs 36$ & 52.8$\pm$1.8 & 9.1$\pm$1.7 & 50.9$\pm$0.9 & 1.5$\pm$0.2&0.6$\pm$0.2\\
		  &&& 67.2$\pm$1.6 & 12.0$\pm$1.6& 66.4$\pm$0.6 & 3.4$\pm$0.2&1.1$\pm$0.1\\
		&  && 88.0$\pm$1.4 & 7.6$\pm$1.0 & 85.2$\pm$0.5 & 2.6$\pm$0.1&1.4$\pm$0.2\\\hline  
	7 & $17^{\rmn{h}} 47^{\rmn{m}} 07\fs417$ & $-28\degr 26\arcmin 16\farcs 51$ & 51.0$\pm$1.0 & 14.1$\pm$1.4& 54.4$\pm$0.8 & 1.6$\pm$0.1&0.5$\pm$0.1\\
		 & && 91.6$\pm$1.1 & 12.9$\pm$1.4& 90.8$\pm$0.5 & 3.1$\pm$0.1 &1.0$\pm$0.1\\
		\hline
	\end{tabular}
	\begin{tablenotes}
	\item[(a)] The SiO abundance is calculated as the N$_{\rm SiO}$/N$_{\rm H_2}$ ratio, where N$_{\rm H_2}$ is estimated from N$_{\rm C^{18}O}$ using the $^{16}$O/$^{18}$O isotopic ratio of 250 \citep{Wilson94} and the relative abundance of CO to H$_2$ of 10$^{-4}$ \citep{Frerking1982}.
	\item[(b)] Parameter fixed in the MADCUBA analysis to obtain simultaneous fitting for the multiple velocity components.
    \item[(c)] The SiO line is self-absorbed in this position (see Figure~\ref{fig:SiO_C18O_spectra}), thus N$_{\rm SiO}$ and also the relative abundance of SiO are lower limits. N$_{\rm SiO}$ is calculated from the integrated intensity of the SiO J=2-1 line over the velocity range of [40,100] \mbox{km s$^{-1}$}.
	\end{tablenotes}
	\end{threeparttable}
\end{table*}

\section{Decomposition with GaussPy+}
\label{Sec:AppendixDecomposition}

\begin{figure*}
\begin{center}
  \begin{tabular}{c c c c }
    \hspace{-0.3cm}{\resizebox{45mm}{!}{\includegraphics{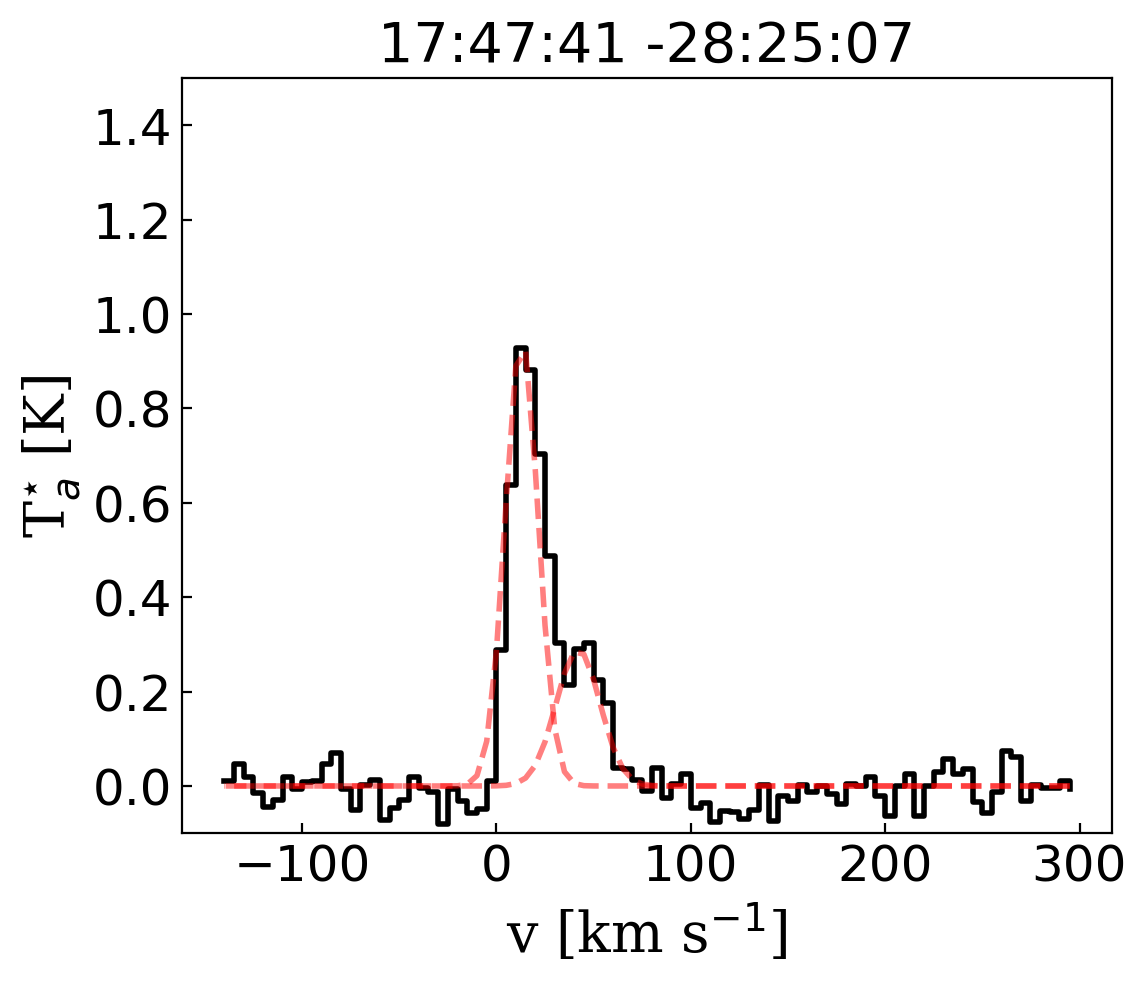}}} & \hspace{-0.3cm}{\resizebox{45mm}{!}{\includegraphics{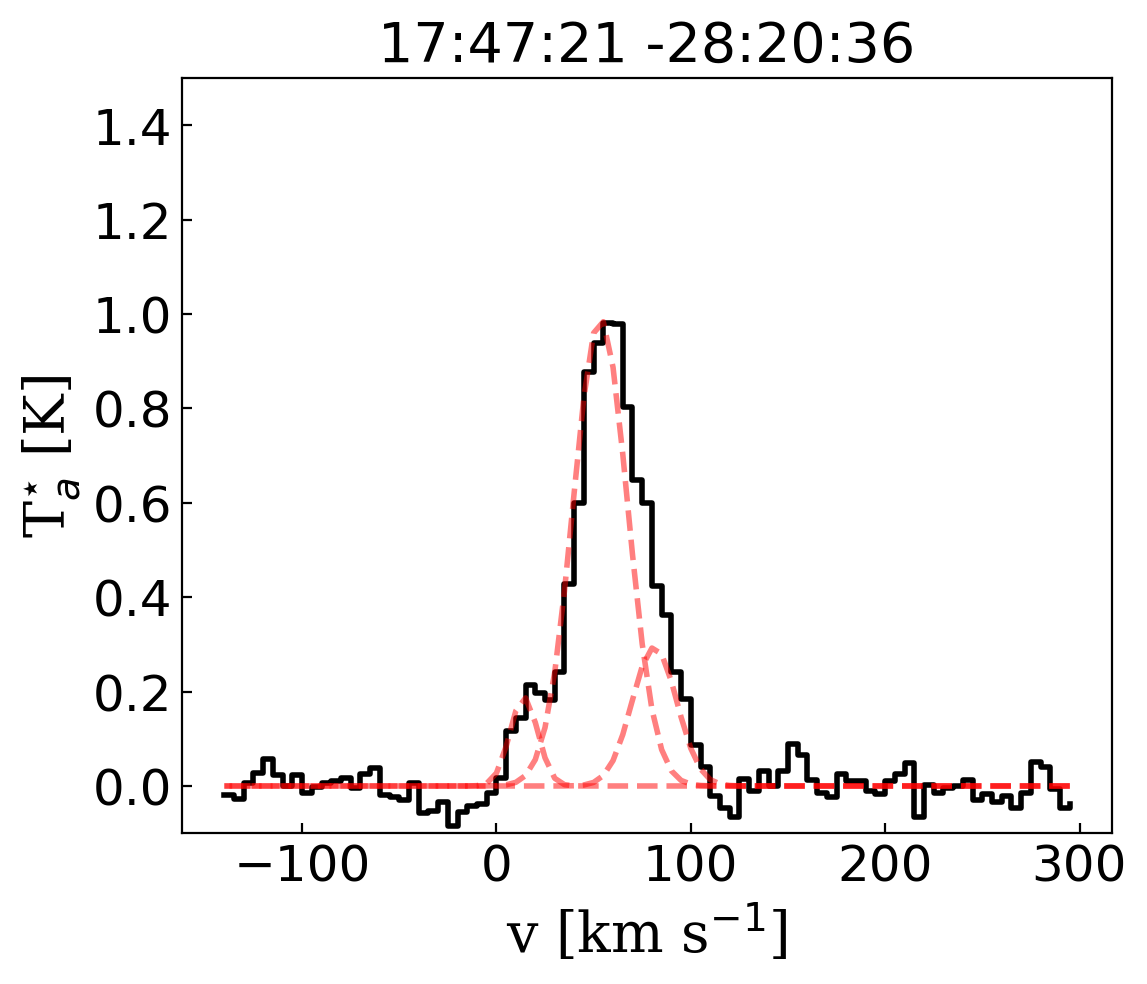}}} &
    \hspace{-0.3cm}{\resizebox{45mm}{!}{\includegraphics{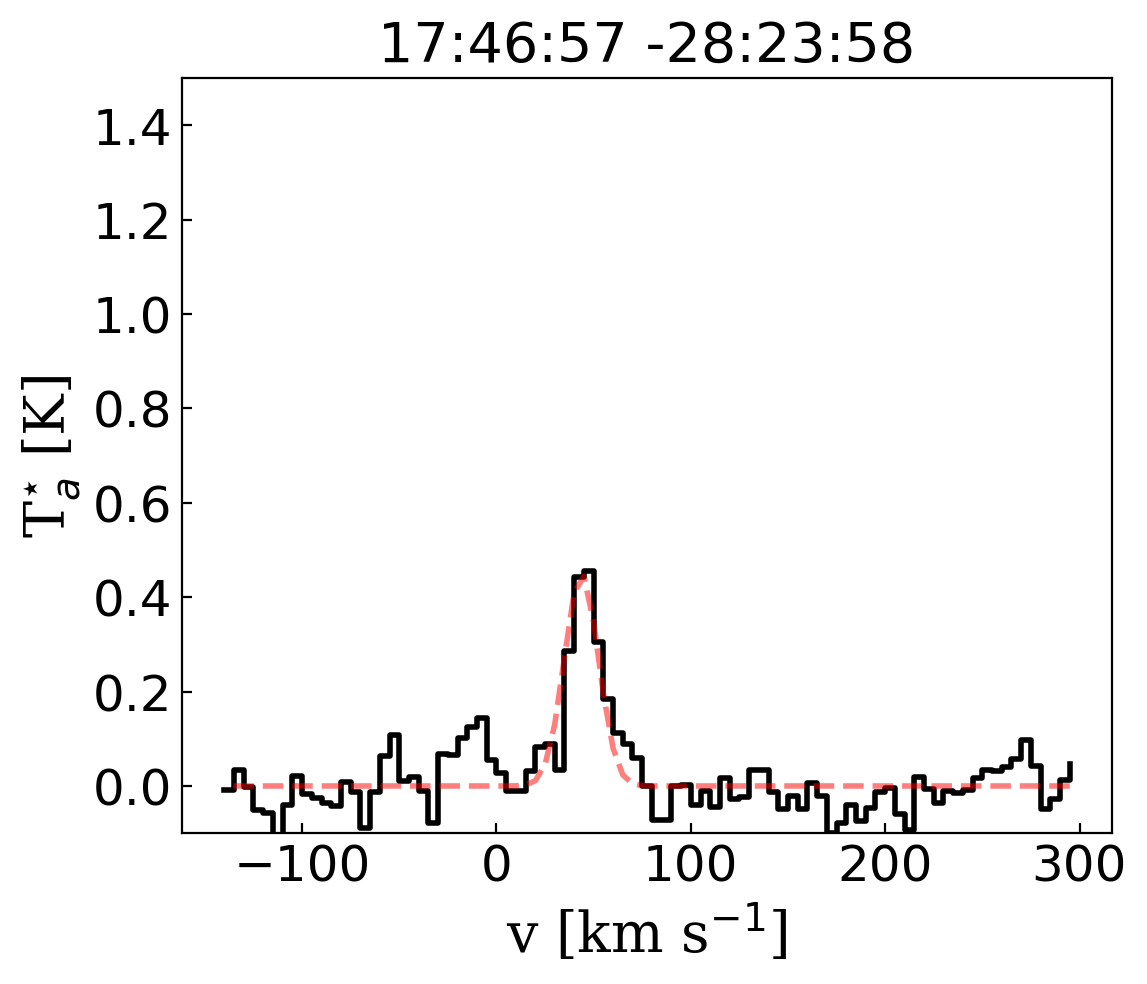}}} &
    \hspace{-0.3cm}{\resizebox{45mm}{!}{\includegraphics{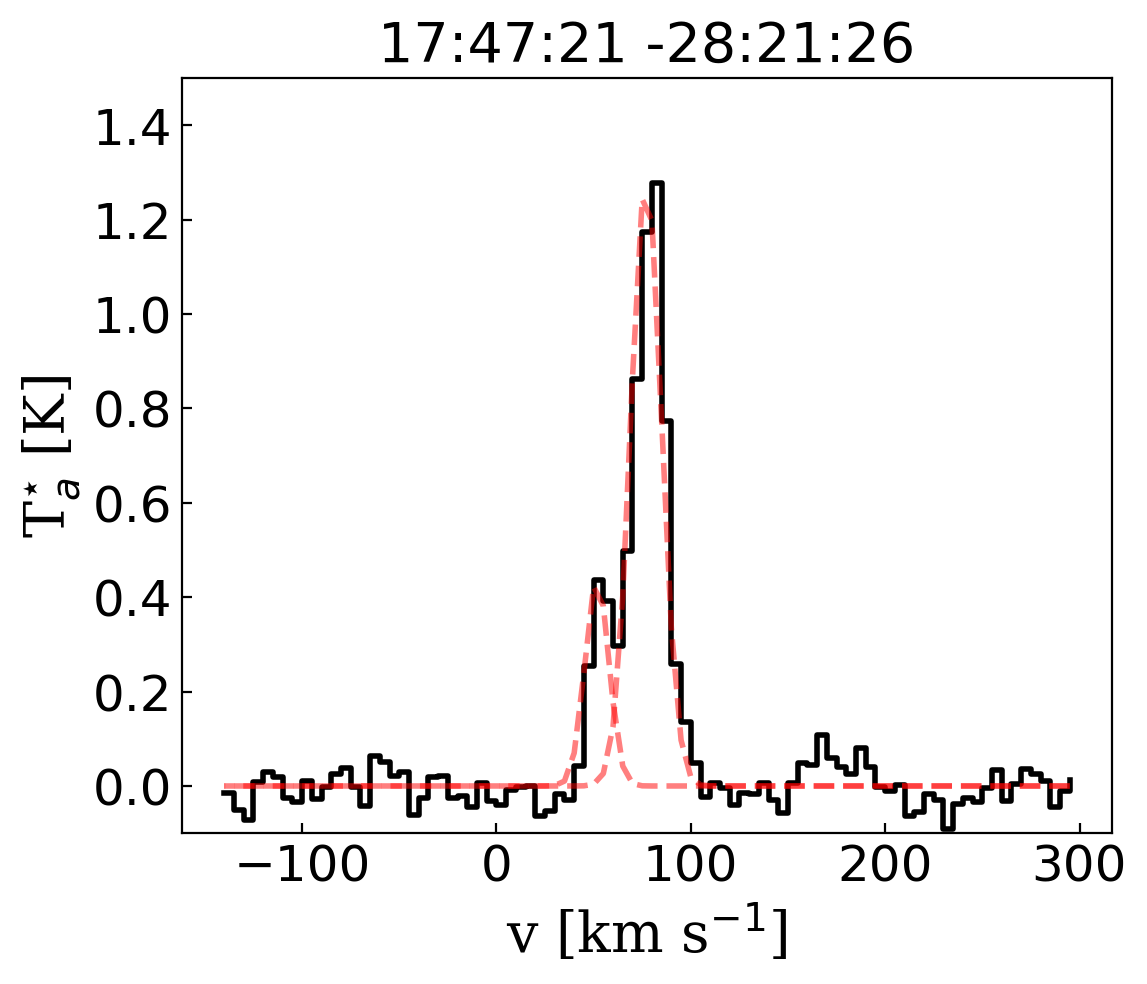}}} \\
    \hspace{-0.3cm}{\resizebox{45mm}{!}{\includegraphics{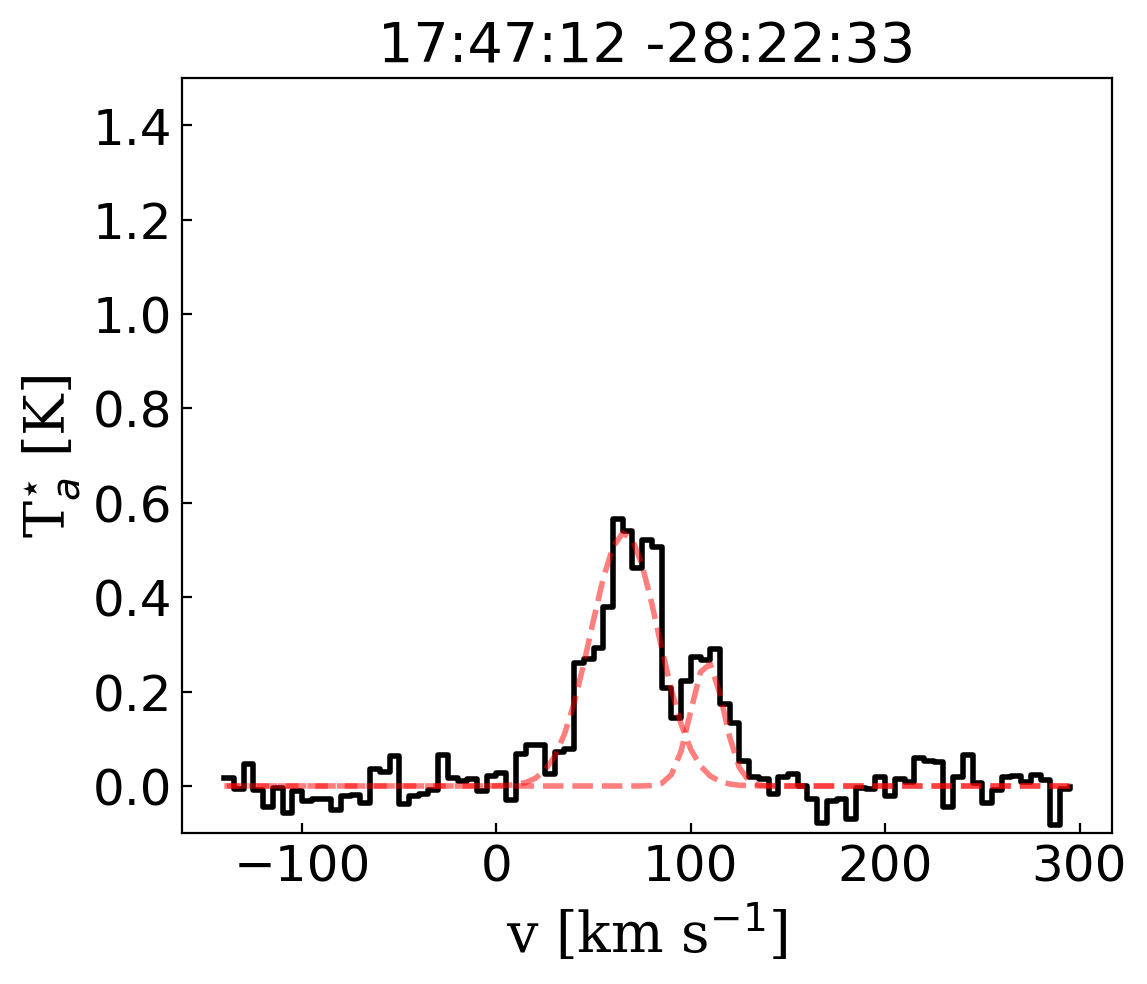}}} &
    \hspace{-0.3cm}{\resizebox{45mm}{!}{\includegraphics{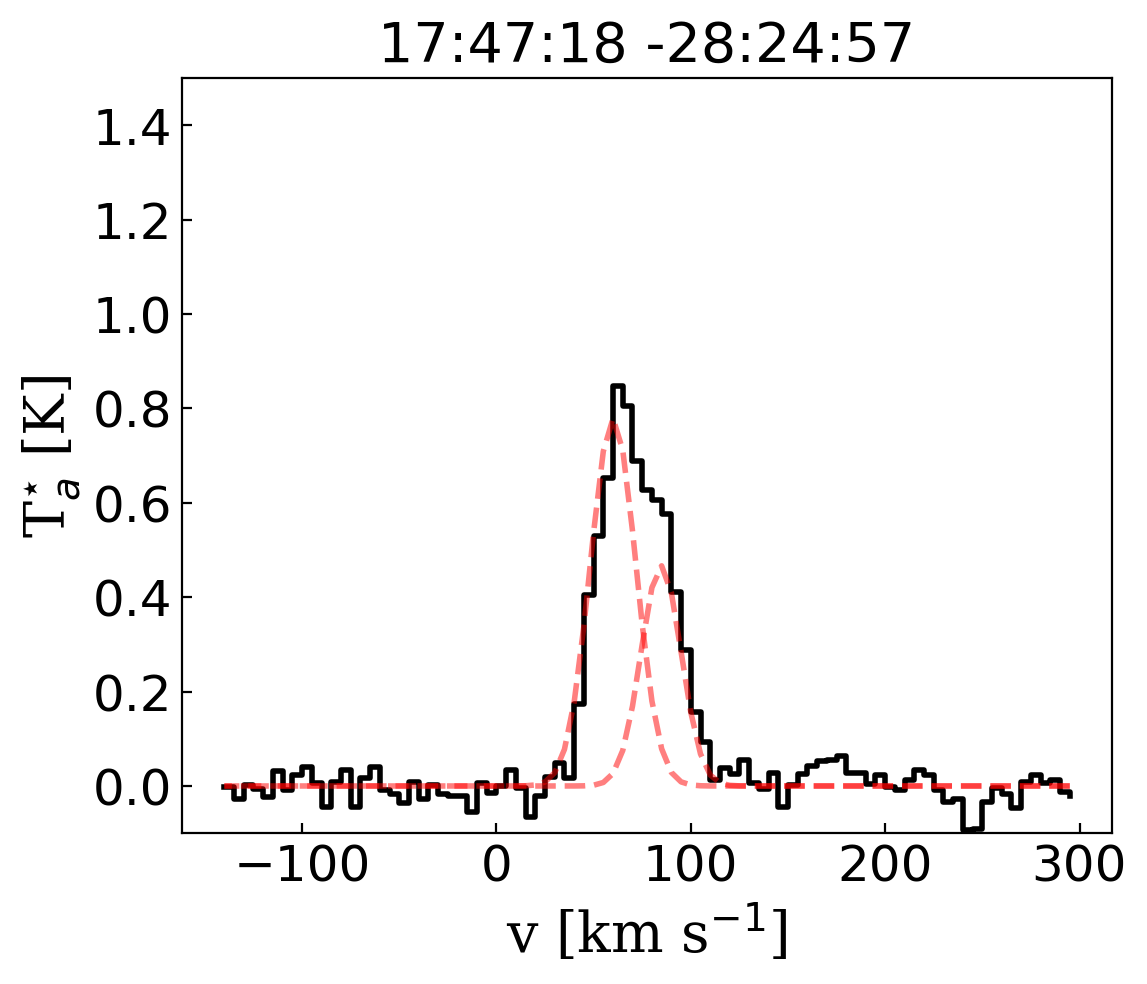}}} &
    \hspace{-0.3cm}{\resizebox{45mm}{!}{\includegraphics{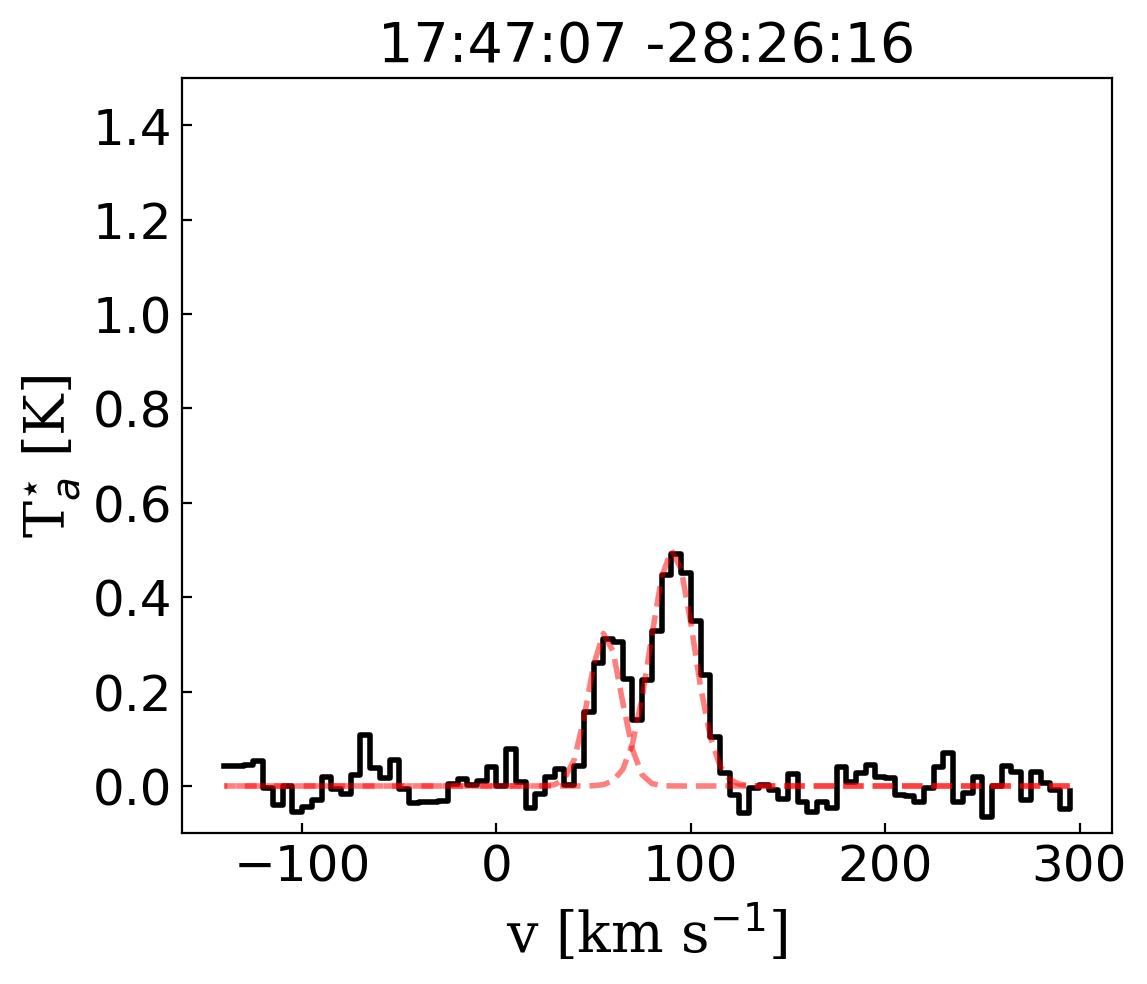}}} &\\
  \end{tabular}
  \caption{Examples of the {\sc Gausspy+} decomposition of the SiO spectra. Black line shows the SiO spectra at the same position as in Figure \ref{fig:SiO_C18O_spectra} and the red lines show the individual Gaussian components decomposed with {\sc GaussPy+}.}
  \label{fig:SiO_decomposition_gausspy}
\end{center}
\end{figure*}

Figure \ref{fig:SiO_decomposition_gausspy} shows example spectra from the SiO data cube and the results of the decomposition with {\sc GaussPy+}. These seven positions are the same as listed in Table \ref{tab:Column_density}. \par

\begin{figure*}
	\includegraphics[width=16cm]{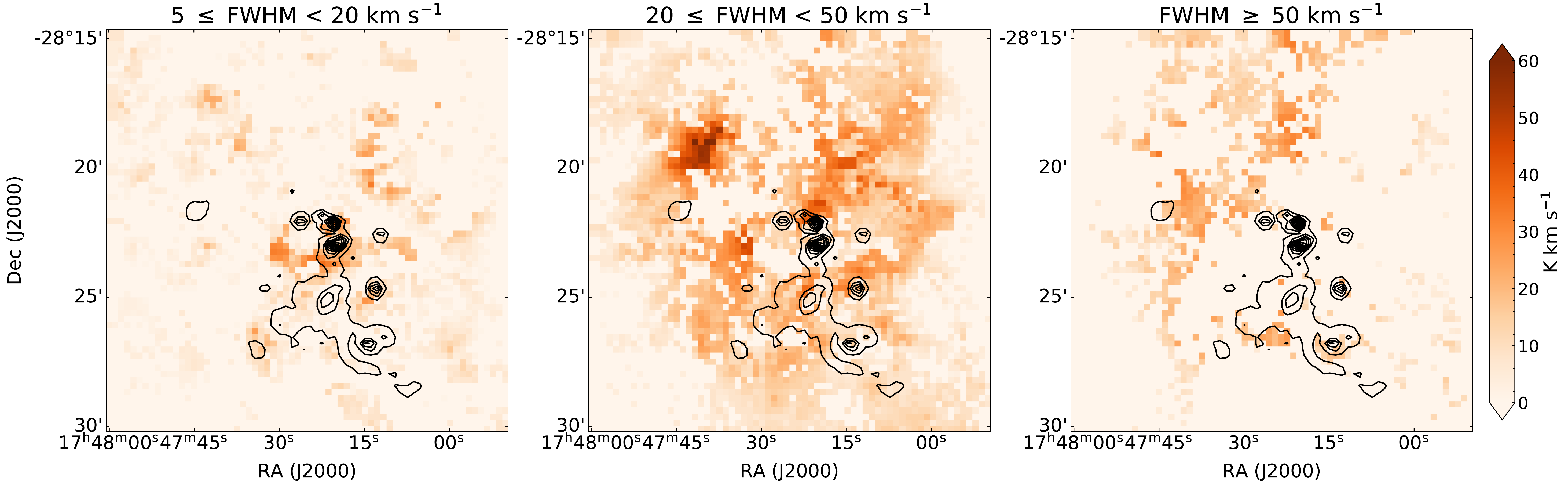}
    \caption{Integrated intensity maps of the SiO Gaussian components from the decomposition with {\sc GaussPy+}. The different panels show Gaussian components with different line widths (5 km s$^{-1} \leq$ FWHM $<$ 20 km s$^{-1}$, 20 km s$^{-1} \leq$ FWHM $<$ 50 km s$^{-1}$ km s$^{-1}$ and FWHM $\geq 50$ km s$^{-1}$). The black contours show the 20 cm radio continuum emission obtained by \protect\cite{Yusef2004} to mark the location of the HII regions in the field. Contour levels are [0.08 0.23 0.38 0.53 0.68 0.83 0.98 1.13 1.28 1.43 1.58] Jy/beam.}
    \label{fig:SiO_velocity_width_decomposition}
\end{figure*}

Figure \ref{fig:SiO_velocity_width_decomposition} shows the integrated intensity maps of the Gaussian components from the decomposition with {\sc GaussPy+}. The different panels show Gaussian components with different line widths (5 km s$^{-1} <$ FWHM $\leq$ 20 km s$^{-1}$, 20 km s$^{-1} <$ FWHM $\leq$ 50 km s$^{-1}$ km s$^{-1}$ and FWHM $>$ 50 km s$^{-1}$). All three components are extended across the whole region. The black contours show the 20 cm radio continuum emission obtained by \cite{Yusef2004}.

\begin{figure*}
\begin{center}
  \begin{tabular}{c c}
    \hspace{-0.3cm}{\resizebox{70mm}{!}{\includegraphics{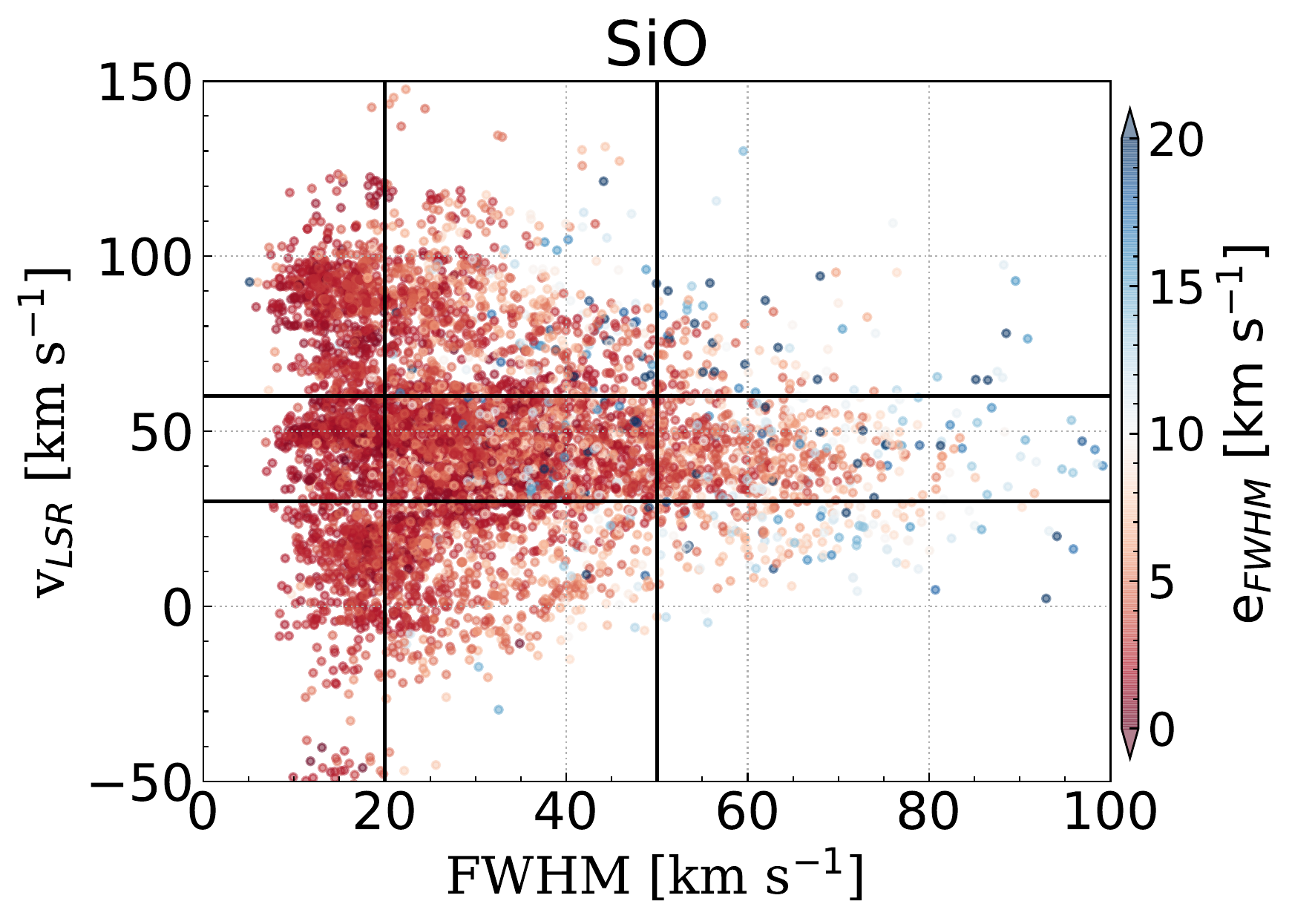}}} & 
    \hspace{-0.3cm}{\resizebox{70mm}{!}{\includegraphics{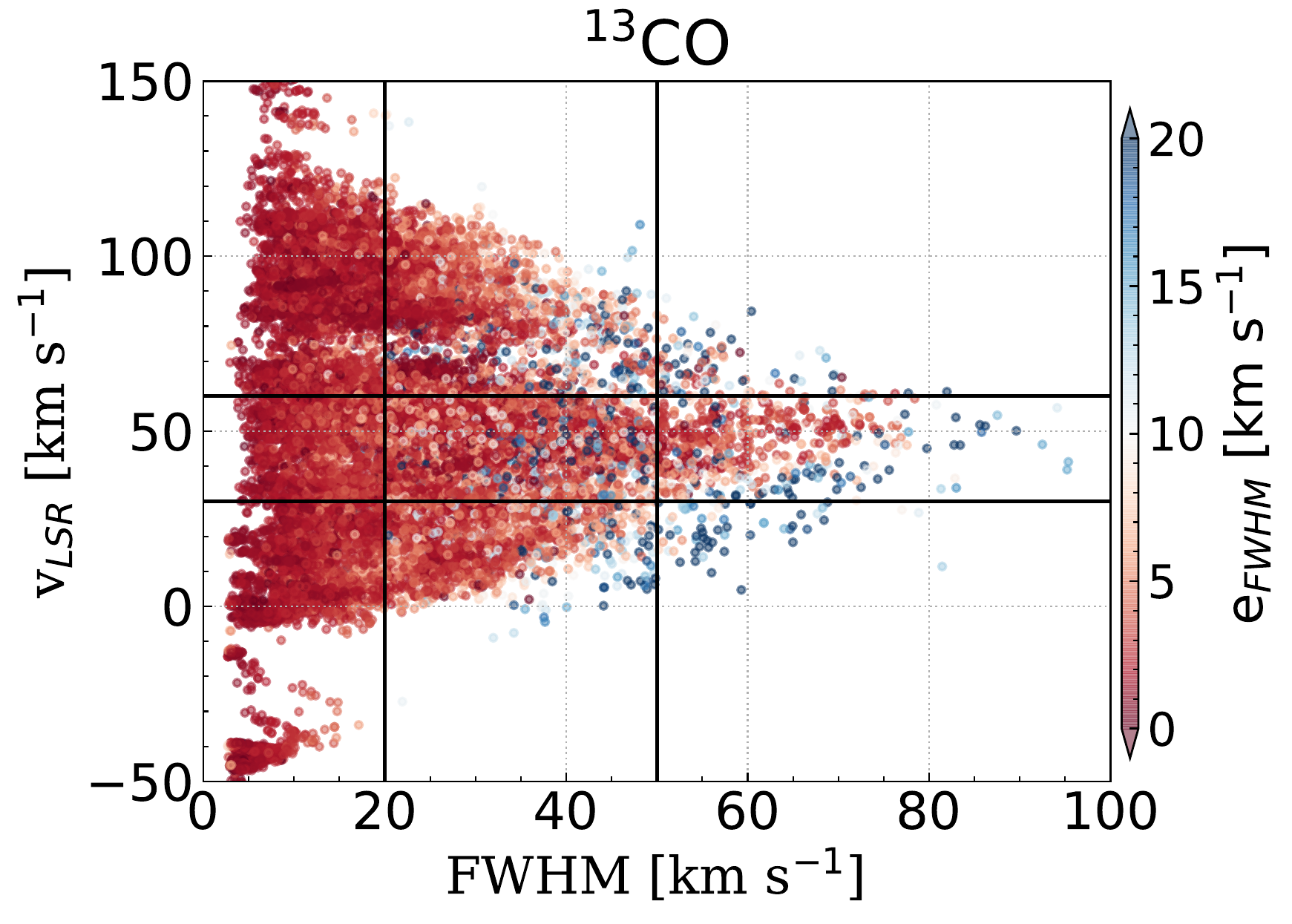}}} \\
  \end{tabular}
  \caption{Line width (FWHM) vs. velocity ($v_{LSR}$) plots for SiO and $^{13}$CO Gaussian components. Black lines mark $v_{LSR}$ = 30 and 60 km s$^{-1}$ and FWHM = 20 and 50 km s$^{-1}$. The color coding of the markers shows the uncertainty of the FWHM in the Gaussian decomposition. Most of the broad lines are between $v_{LSR} \sim$ 30 and 60 km s$^{-1}$ for both molecules.}
  \label{fig:fwhm_vlsr_gausspy}
\end{center}
\end{figure*}

To compare our results from the Gaussian decomposition of the SiO data to the ambient gas in Sgr B2, we also decomposed the $^{13}$CO data cube with {\sc GaussPy+}. Since the spectral properties of the SiO and the $^{13}$CO data are very similar (see description in Section \ref{observation}), we used the same $\alpha$ parameters for the $^{13}$CO decomposition as for the SiO decomposition. For the $^{13}$CO we find 20807 Gaussian components with similar average properties to the SiO decomposition. In particular, the average line velocity is 54 $\pm$ 5 km s$^{-1}$ and the FWHM distribution is similarly log-normal with a peak at $\sim$ 10 km s$^{-1}$ and a mean of 21 $\pm$ 2 km s$^{-1}$. Figure \ref{fig:fwhm_vlsr_gausspy} compares the line widths of the Gaussian components of the SiO and the $^{13}$CO data to their central velocity. Most of the broad lines (FWHM $>$ 50 km s$^{-1}$) are between $v_{LSR} \sim$ 30 and 60 km s$^{-1}$ for both molecules. This indicates that the broad line emission is most likely tracing mixed gas from the cloud-cloud collision.\par

\bsp	
\label{lastpage}
\end{document}